\documentclass[nofootinbib,superscriptaddress,eqsecnum,tightenlines,11pt]{revtex4}

\usepackage{hyperref}
\usepackage{graphicx}
\usepackage{amsmath,amssymb,amsfonts,amsthm,stmaryrd,mathtools}
\usepackage{color}
\usepackage{multirow,bigdelim}
\usepackage{dsfont}
 
\allowdisplaybreaks[1]

\makeatletter
\newcounter {subsubsubsection}[subsubsection]
\renewcommand\thesubsubsubsection{\thesubsubsection .\@alph\c@subsubsubsection}
\newcommand\subsubsubsection{\@startsection{subsubsubsection}{4}{\z@}%
                                     {-3.25ex\@plus -1ex \@minus -.2ex}%
                                     {1.5ex \@plus .2ex}%
                                     {\centering\normalfont\small\textit}}
\newcommand*\l@subsubsubsection{\@dottedtocline{3}{10.0em}{4.1em}}
\newcommand*{\subsubsubsectionmark}[1]{}
\makeatother

\newtheorem{Definition}{Definition}[section]
\newtheorem{Lemma}{Lemma}[section]

\def\be#1\ee{\begin{align}#1\end{align}}

\def\ba{\begin{eqnarray}}
\def\ea{\end{eqnarray}}

\def\eps{\varepsilon}

\def\tr{\text{tr}}
\def\la{\langle}
\def\ra{\rangle}
\def\de{\mathrm{d}}

\def\f{\frac}
\def\lb{\big\lbrace}
\def\rb{\big\rbrace}
\def\SU{\mathrm{SU}}
\def\U{\mathrm{U}}

\def\SL{\mathrm{SL}}
\def\su{\mathfrak{su}}

\def\openone{\mathds{1}}

\def\nn{\nonumber}
\def\q{\qquad}

\def\i{\mathrm{i}}

\def\bX{\mathbf{X}}
\renewcommand{\qed}{$\hfill\blacksquare$}

\begin{document}

\title{A new realization of quantum geometry}

\author{Benjamin Bahr}
\affiliation{II Institute for Theoretical Physics, University of Hamburg, Luruper Chaussee 149, 22761 Hamburg, Germany}
\author{Bianca Dittrich}
\affiliation{Perimeter Institute for Theoretical Physics,\\ 31 Caroline Street North, Waterloo, Ontario, Canada N2L 2Y5}
\author{Marc Geiller}
\affiliation{Institute for Gravitation and the Cosmos \& Physics Department,\\ Penn State, University Park, PA 16802, U.S.A.}

\begin{abstract}
We construct in this article a new realization of quantum geometry, which is obtained by quantizing the recently-introduced flux formulation of loop quantum gravity. In this framework, the vacuum is peaked on flat connections, and states are built upon it by creating local curvature excitations. The inner product induces a discrete topology on the gauge group, which turns out to be an essential ingredient for the construction of a continuum limit Hilbert space. This leads to a representation of the full holonomy-flux algebra of loop quantum gravity which is unitarily-inequivalent to the one based on the Ashtekar--Isham--Lewandowski vacuum. It therefore provides a new notion of quantum geometry. We discuss how the spectra of geometric operators, including holonomy and area operators, are affected by this new quantization. In particular, we find that the area operator is bounded, and that there are two different ways in which the Barbero--Immirzi parameter can be taken into account. The methods introduced in this work open up new possibilities for investigating further realizations of quantum geometry based on different vacua.
\end{abstract}

\maketitle

\tableofcontents

\newpage

\section{Introduction}

\noindent Following Einstein's insight that gravity is encoded into the geometry of spacetime, quantum gravity aims at providing realizations of quantum geometry. In this task, one key technical and conceptual challenge is to reconcile the regularizations required in quantum field theory with the diffeomorphism symmetry which underlies general relativity. Indeed, a number of approaches employ discretizations as regulators, which is the case for instance of Regge calculus \cite{regge}, and in this class of theories, where one attempts to represent geometrical data on a triangulation\footnote{As opposed to encoding the geometry directly into the combinatorial structure of the triangulation.}, diffeomorphism symmetry is generically broken \cite{bahrdittrich09a,dittrichsteinhaus,dittrichkaminskisteinhaus}.

In light of this, a very important result is therefore the construction of a continuum notion of quantum geometry, which was achieved in the context of loop quantum gravity \cite{lqg1,lqg2,lqg3} (LQG hereafter) by Ashtekar, Isham, and Lewandowski \cite{ali1,ali2,ali3,ali4}. The so-called Ashtekar--Lewandowski (AL) representation provides a Hilbert space representation of the kinematical algebra of observable of the full \textit{continuum} theory. This kinematical algebra of observable encodes the intrinsic and extrinsic geometry of a spatial hypersurface into holonomies measuring curvature (of the Ashtekar--Barbero connection \cite{barbero,immirzi}) and fluxes measuring the spatial geometry. Most importantly, this kinematical setup allows to deal successfully with spatial diffeomorphisms. Indeed, there exists a fully kinematical Hilbert space describing geometry in a coordinate-dependent way, and on which the diffeomorphisms act unitarily. The fact that the diffeomorphisms act unitarily also allows to define a Hilbert space of spatially diffeomorphism-invariant states, and, quite noticeably, this task has so far only been achieved in the (Ashtekar) connection formulation of general relativity \cite{ashtekardiffeos}. The kinematical Hilbert space allows for a rigorous definition of the quantum dynamics in the form of Hamiltonian constraints \cite{lqg3,thiemannH1,thiemannH2}, and one can hope to construct the physical Hilbert space, which would incorporate spacetime diffeomorphism-invariant states, from the solutions to these constraints.

In the present article, building upon the earlier work \cite{newvac,fluxC}, we construct an alternative realization of a continuum quantum geometry, which is unitarily inequivalent to the Ashtekar--Lewandowski representation. We hope that this new framework will make the description of states describing configurations with macroscopic geometry much easier. The reason for this expectation is that the new representation which we are constructing here supports states which are peaked on an almost everywhere flat connection (we therefore call this representation the BF representation, since the BF vacuum is peaked on a flat connection). Curvature has only distributional support on defects, which brings us much nearer to Regge's proposal \cite{regge} of approximating general relativity by a very dense gas of defects in an otherwise flat geometry\footnote{This is indeed realized in three spacetime dimension. For four spacetime dimensions, the kind of geometries described here do not correspond to the piecewise flat geometries considered by Regge \cite{dittrichryan1}, and one rather deals with a generalized class of geometries \cite{dittrichryan2,twisted,FGZ}.}.

As explained in more detail in the overview given in section \ref{overview}, our construction leads to a continuum Hilbert space which supports arbitrarily many excitations in the form of defects. In fact, both the AL representation and the BF representation can be interpreted in this way. For the AL representation, it is the defects themselves which generate (non-degenerate) geometry, and therefore a macroscopic geometry corresponds to a highly excited state\footnote{This renders the construction of smooth semi-classical geometries involved. For this reason, another representation has been proposed by Koslowski in \cite{KS1}, and investigated in \cite{KS2,KS3,varadarajan1,varadarajan2,varadarajan3} by various authors. In this representation, one considers a constant background flux describing a fixed geometry, and unitary equivalence with the AL representation is lost because diffeomorphism-invariance is replaced by a suitable notion of diffeomorphism-covariance.}. For the BF representation however, the states have (almost everywhere) maximal uncertainty in spatial geometry since they are peaked on the conjugated variable, namely on flat connections. There are numerous proposals to approximate the dynamics of general relativity by mostly flat geometries with defects, either in the classical theory \cite{regge,thooft}, or in the context of LQG \cite{FGZ,pullin,lewandowski2surface,bianchi}. Another related class of approaches is the combinatorial quantization of flat connections in three spacetime dimensions\footnote{These methods are restricted so far to three spacetime dimension, and consider the space of flat connections itself as phase space. Here we will rather consider the cotangent space to the space of flat connections. These approaches can however be related, at least in three spacetime dimensions \cite{MeusburgerNoui2}. For a four-dimensional attempt, we refer the reader to \cite{aldo,aldo2}.} \cite{Fock:1998nu,Alekseev:1994pa,Alekseev:1994au,Alekseev:1995rn,Meusburger:2003ta,Meusburger:2003hc,MeusburgerNoui}.

In this work, we construct for the first time a Hilbert space carrying a representation of a continuum observable algebra and supporting states which are peaked on almost everywhere flat connections. This continuum construction has important consequences. In particular, it requires the compactification (or the exponentiation) of the fluxes, and with it the introduction of a discrete topology for the holonomy group parameters. This in turn changes the properties of, e.g., the spectra of observables like the area operator.  

Our construction of the continuum Hilbert space is done via a so-called inductive limit. As is explained in more detail in section \ref{overview}, such a constructions allows one to keep the cake and to eat it too. For most purposes, it is sufficient to deal with a discretization. However, all the Hilbert spaces describing states restricted to such discretizations are embedded into a continuum Hilbert space. The properties of observables are therefore changed further due to redefining (as compared to the AL representation) the way in which the observables on Hilbert spaces associated to discretizations are related to the observables on the continuum Hilbert space. For the BF representation, this relation is based on a geometric coarse-graining of the fluxes, which allows to address for example the staircase problem appearing for geometric operators in the AL representation.

This work is divided in two main parts. We start in section \ref{sec:configuration space} by providing a characterization of the (cotangent) space of flat connections on manifolds with a fixed number of defects, where the positioning of the defects is prescribed by a choice of triangulation. We then define in section \ref{sec:inner product} an inner product (compatible with the inductive limit construction which comes afterwards) which leads to a discrete topology on the group. This discrete topology on the group implies that dealing with gauge invariance is far less trivial than in the AL case. We bypass this difficulty by working with the space of almost gauge-invariant wave functions, and employ a group averaging for the remaining global adjoint action. In section \ref{sec:quantization algebra}, we construct a representation of holonomies and (exponentiated) fluxes which are supported by the triangulation. This already allows us to discuss the spectra of these operators, and also that of the area operator, which we do in section \ref{sec:area}.

In the second main part of this work, we construct the continuum Hilbert space via an inductive limit. This first requires the definition of embeddings that map Hilbert spaces associated to coarser triangulations into Hilbert spaces associated to finer triangulations. We therefore develop a number of technical results in section \ref{sec:refinement},  which allow in particular to obtain a convenient splitting of the observables into coarser observables and finer observables. In section \ref{sec:inductive limit}, we describe the construction of the inductive limit Hilbert space. We then discuss how to relate the operators on the various Hilbert spaces in order to define a notion of continuum operators in sections \ref{sec:extop} and \ref{contop}. Finally, we close with a discussion of our results in section \ref{sec:discussion}. The appendices include a more detailed elaboration on the configuration space of flat connections, a number of proofs of theorems needed in the main text, as well as the construction of the (intertwiner-) spin representation for the wave functions, which replaces the usual spin network states of the AL representation.  
 
\section{Overview of the construction}
\label{overview}

\noindent In order to construct the \textit{continuum} Hilbert space, we proceed in the same way as for the AL representation, namely by considering an inductive limit of Hilbert spaces. Let us explain why this construction is very natural in a background-independent context. This will also allow us to explain how the new realization of quantum geometry presented here can be seen as one member of a potentially much bigger class of quantum geometries \cite{timeevol}.

We usually organize Hilbert space representations into a vacuum and excitations, where the excitations are generated by applying the observable algebra to the vacuum state. The vacuum states we are dealing with for both the AL and the new representation are states which are sharply peaked on a certain configuration, as opposed to the Gaussian states of free field theory for which one needs to specify a background metric. For the AL representation, one has a vacuum which is sharply peaked on a totally degenerate spatial geometry, i.e. in which all flux observables annihilate the vacuum. For the representation constructed in this work, the vacuum state is very different: it is sharply peaked on a (globally and locally) flat connection.

In fact, such states which are sharply peaked on a flat connection can be described by (first class) constraints arising from a topological field theory (these constraints play also an important role in the classical theory \cite{fluxC,FGZ}). For the new representation presented here, this topological field theory is given by BF theory \cite{horowitz}, which is also the theory underlying the construction of spin foam models \cite{baez,oriti-thesis,perez-review}. We will therefore refer to the new representation as the BF representation. The topological field theory for the AL representation is a trivial one, and coincides with the strong coupling limit of Yang--Mills theory, whereas BF theory arises from a zero coupling limit \cite{BFYangMills}.  

The topological field theory in itself does however not provide room for local excitations, and therefore does not give a representation of the observable algebra which is supposed to be generating such excitations. In order to add excitations to the theory, one needs to allow for the presence of defects, e.g. in the form of a violation of the flatness constraints in the BF case, or of a violation of the flux constraints in the AL case. Therefore, the notion of what constitutes a ``defect'' strongly depends on the choice of vacuum. In the case of the AL representation, a ``defect'' actually means an excitation which allows to have at least some of the spatial geometrical operators with non-vanishing values. This would rather be an ``anti-defect'' of geometry if one understands a degenerate geometry as being highly defected.

Defects play also an important role in condensed matter applications \cite{DefectsCM}. In this case, one wishes to describe localized excitations which are separated by very large distances so that they do not interact\footnote{One can however describe the process in which two defects merge into a bigger one. This gives the fusion rules for the charges attached to the defects.} with each other. Therefore, the precise values of the distances, i.e. the metric geometry, does not actually matter. This justifies a posteriori the use of the vacuum (or the physical state) of a topological field theory: it ensures that only relations of topological nature between the defects actually matter (for instance how two line defects wind around each other). Furthermore, it explains why frameworks for describing defects are applicable to quantum gravity, since the metric properties are then not encoded in a background, but in the defect configurations themselves.

A second property which one usually assigns to defects is a discreteness for their ``charge'' labels. In condensed matter, this arises since one needs finite energy gaps between the vacuum and the excitations described by the defects, so that there is a notion of adiabatically moving the defects around. Interestingly, there is a similar discreteness of the charges describing the defects in the framework of quantum geometry. This arises however due to a different reason, namely in the way in which the continuum limit is constructed.

To construct the continuum limit, one starts with a family of Hilbert spaces where each Hilbert space describes a fixed number of defects. For a diffeomorphism-covariant description, one does not only fix the number of defects, but also their positions. This is achieved by embedding a discrete structure, e.g. a graph or a triangulation, into the spatial manifold. Defects are then confined to this discrete structure, for instance line defects to the links of the graph or to the edges of the triangulation.

The degrees of freedom are given by the values of the charges for the defects. The values can also describe a vanishing charge, which is then equivalent to having no defect. This is an important point since it allows to embed states realized in a Hilbert space based on e.g. a coarse triangulation into a Hilbert space based on a refinement of this triangulation.

For the continuum theory, one however wishes to allow for an arbitrary number of defects (at arbitrary positions). One can construct such a continuum Hilbert space as an inductive limit of Hilbert spaces. A key point in this construction is the notion of refinement for the discrete structures carrying the defects on the one hand, and for the states on the other hand. Let us consider the example of line defects confined to a graph embedded into the manifold. One possible refinement of a graph is to add links to it. The corresponding state for the refined graph is then defined by transposing over all the charge labels from the coarse state, and by assigning the charge labels describing a vanishing defect for the new links. In other words, one puts the new additional degrees of freedom into a vacuum state \cite{timeevol}.

In order to define the inductive Hilbert space structure, the refinement operator acting on the states needs to be isometric (with respect to the inner product defined on the Hilbert spaces with a fixed discrete structure). In particular, the vacuum state where all the charges describe vanishing defects should have a finite norm. Moreover, what happens (in both the BF and the AL representation) is that the charges label an orthonormal basis for the Hilbert spaces associated to a fixed discrete structure. Therefore, even if the charge labels are continuous (as it is the case with the BF representation, where the labels are elements of the (Lie) group), we will have to equip the space of charge values with a discrete topology.

If we see the charge labels as momentum space, then a discrete topology on momentum space corresponds to a compactification of the configuration space. This compactification can be understood in the following way for the two representations:
\begin{itemize}
\item In the AL representation, the configuration space associated to a fixed discrete structure is given by group holonomies along the links of embedded graphs, and therefore corresponds to $G^L$, where $G$ is a compact Lie group (usually $\SU(2)$), and $L$ denotes the number of allowed defects, which in this case coincides with the number of links of the graph. The inductive limit of Hilbert spaces amounts to a projective limit for the configuration space, in which compactness is preserved (see for instance \cite{lqg3}). The space of charge labels is constructed through the group Fourier transform and given by representation labels (and magnetic indices). This space does indeed naturally carry a discrete topology (for a compact group $G$).

\item In the BF representation, the configuration variables are actually the fluxes\footnote{We will however mostly work in the holonomy representation. A spin representation is described in the appendices \ref{spinrepsu2} and \ref{intertwinerspin}.}, which correspond to the dual of the Lie group. We therefore need to consider a compactification of this space. In fact, for the group $G=\U(1)$ the (Pontryagin) dual is given by the Abelian group ${\mathbb Z}$, which can be Bohr-compactified. This in turn leads to a discrete topology on $G=\U(1)$. For $G= \SU(2)$, it is however easier to start with a discrete topology, and to then derive the compactification of the dual space via a (generalized) group Fourier transform. This is described in appendix \ref{spinrepsu2}.
\end{itemize}
 
Therefore, we see that the inductive limit construction leads to a compactification of the underlying configuration spaces. This also explains the difficulty to define a projective limit Hilbert space (in the AL representation) for a non-compact group like $\SL(2,\mathbb{C})$, since this latter cannot easily be compactified\footnote{For the Abelian group $\mathbb{R}$, one can use the Bohr compactification \cite{okolov}. It might actually be easier to compactify the dual of $\SL(2,\mathbb{C})$, which would then allow for a BF representation.}.
 
The compactification of the fluxes has important repercussions for the properties, and in particular the spectrum, of the observables. Due to the discrete topology on the (Lie) group, the fluxes cannot act anymore as (Lie) derivatives like in the AL representation. Instead, one has to replace the fluxes with translation operators labelled by a translation group element $\mu$. Due to the compactification, the spectrum of these translation operators is bounded. However, depending on the choice of $\mu$, the spectrum can be either discrete or continuous. The continuous spectrum arises if the action of the translation operator on the group is ergodic. We will provide a more detailed discussion of these features in section \ref{sec:area}.

A (Bohr) compactification is also employed in another context in loop quantum cosmology (LQC) \cite{AshBojLewan}. In this case, one compactifies $\mathbb{R}$, which describes the momentum conjugated to the scale factor (this momentum $p_a\sim\dot{a}$ is proportional to the connection degree of freedom). This then leads to a discrete topology for the scale factor. Note that in LQC (point) holonomies are compact while the conjugated momentum discrete, which is the reversed situation as compared to the BF representation.  
 
It is also well-known that compactifications lead to non-separable Hilbert spaces. Let us comment more on this issue, and in particular on the fact that there are now two different sources of non-separability.
\begin{itemize}
\item The first source of non-separability arises from the inductive limit Hilbert space construction based on \textit{embedded} discrete structures. This can be interpreted as introducing a discrete topology for the underlying spatial manifold. This is also mirrored in the interpretation of defects in condensed matter as charges which lie at large distance from one another. Here, we declare for instance any point (in the case in which the defects are point-like) of the underlying manifold to be separated from any other point (mathematically, these points define open sets), which gives rise to the discrete topology.

One can however expect that this kind of non-separability will be lifted (to a large extent) if one constructs the spatial diffeomorphism-invariant Hilbert space. Indeed, diffeomorphisms act by changing the embeddings of the discrete structures, and therefore the positioning of the defects. For a diffeomorphism-invariant description only the relative\footnote{Of course there are many subtleties in the precise notion of this concept. In the AL representation, the defects are described by embedded graphs. One usually works within a semi-analytic category \cite{fleischhack,LOST} of graphs. After factoring out the diffeomorphisms, one not only has knotting information of the graphs, but also moduli parameters at the vertices of the graphs.} (topological) positions of the defects matter.

\item In the BF representation, we encounter a second kind of non-separability, which results from the discrete topology on the space of charge labels. This non-separability will not be lifted by implementing spatial diffeomorphism symmetry. There is however the possibility that it can be lifted for the physical Hilbert space if one includes a positive cosmological constant (at least in the Euclidean theory). In this case, the large spin $j$ labels in the spin representation might become irrelevant: for instance, the area eigenvalues first grow with growing spin, and then start (depending on the value of $\mu$) to oscillate. Therefore, very large spins are in a sense not needed since they lead to similar area eigenvalues as the smaller spins.

One can also introduce a quantum deformation of the group at root of unity, i.e. work with $\SU(2)_q$. In this case, the cutoff on the spin is implemented explicitly. The charge labels are then described by the Drinfeld center of the $\SU(2)_q$ module category, which has only finitely many objects \cite{balsam}. In three spacetime dimensions, the corresponding vacuum is given by the Turaev--Viro state sum model \cite{TV} and describes (Euclidean) gravity with a positive cosmological constant. Future work \cite{bdtoappear} will be concerned with constructing a continuum Hilbert space representation based on quantum groups.
\end{itemize}

\section{Setup}
\label{sec:setup}

\noindent In this section we briefly summarize the setup of our construction, and present the notations and notions which are used throughout the rest of the article. We refer the reader to the companion paper \cite{fluxC} for more details.

Let $\Sigma$ be a $d$-dimensional closed (i.e. compact and without boundary) manifold, which we choose to be real, analytic, orientable, and path-connected. We will be interested in particular in the cases $d=2$ and $d=3$, which we treat simultaneously. The manifold $\Sigma$ can be thought of as a spatial boundary, and our aim is to construct states corresponding to $d$-dimensional boundary quantum geometries on it. For this, we consider a principal $G$-bundle $P:\Sigma\rightarrow G$, where the gauge group $G$ is an arbitrary compact Lie group whose Lie algebra will be denoted by $\mathfrak{g}$. If $G$ is simply-connected (which is the case for $\SU(2)$, the gauge group of interest in this work), since we consider the case $d\leq3$ it is always possible to pick a global trivialization of $P$, which we do for simplicity. The space $\mathcal{A}$ of smooth $G$ connections is then affine over $\Omega^1(\Sigma,\mathfrak{g})$, the space of $\mathfrak{g}$-valued one-form fields over $\Sigma$, and the group of local gauge transformations is given by $\mathcal{G}=\text{Map}(\Sigma,G)$.

On the phase space of LQG, a connection $A\in\mathcal{A}$ is canonically conjugated to a $\mathfrak{g}^*$-valued vector density $E$ (the electric field) of weight +1. If we write $A=A^i_a\tau_i\de x^a$ and $E=E^a_i\tau^i\partial_a$, where $\{x^a\}_{a=1,\dots,d}$ are coordinates on $\Sigma$ and $\{\tau_i\}_{i=1,\dots,\dim\mathfrak{g}}$ forms a basis of $\mathfrak{g}$, the non-vanishing Poisson brackets are given by
\be\label{firstequ}
\lb A^i_a(x),E^b_j(y)\rb
=\delta^b_a\delta^i_j\delta^d(x,y).
\ee
Following the logic of LQG, we are going to consider holonomies of the connection and fluxes of the electric field as our elementary variables. However, instead of assigning this data to arbitrary oriented embedded (dual) graphs and $(d-1)$-dimensional submanifolds, as is done in the usual construction of the holonomy-flux $\star$-algebra, here we adopt a simplicial point of view and focus rather on the triangulation itself. We now explain the type of discrete data which is relevant for our construction.

We consider a fiducial Riemannian metric on the manifold $\Sigma$, and the infinite set of embedded (or geometric) triangulations $\Delta$ defined by this metric. In both $d=2$ and $d=3$ dimensions, these embeddings refer to the vertices of the triangulations. The edges and (in $d=3$) the triangles of a given triangulation are defined respectively as geodesics and minimal surfaces with respect to the auxiliary metric. We furthermore assume that the triangulations are fine enough so that these geodesics and minimal surfaces are unique.

To each triangulation $\Delta$ we assign a dual graph $\Gamma$, which is the one-skeleton of the cellular complex dual to $\Delta$. This dual graph is such that its nodes $n\in\Gamma$ are dual to the $d$-dimensional simplices $\sigma^d$ of $\Delta$, while its links $l\in\Gamma$ are dual to the $(d-1)$-dimensional simplices $\sigma^{d-1}$ of $\Delta$. For each link we choose a fiducial orientation, and denote by $l(0)$ and $l(1)$ the source and target nodes.

To each pair given by a triangulation and its dual graph we then assign a root, which is a choice of a preferred $d$-dimensional root simplex $\sigma^d_r$, or equivalently (by duality) a choice of a preferred root node $r$. This in turn defines a rooted geometric triangulation. In order to describe the behavior of this root under the refining operations, we further equip the root simplex with a flag structure, which consists simply in a choice of nondisjoint lower-dimensional simplices $\sigma^0\subset\dots\subset\sigma^d_r$.

With these structures, we can finally consider the infinite set of flagged rooted geometric triangulations of a given manifold $\Sigma$. On this set, we now choose the refinement operations to be the Alexander moves. These Alexander moves refine a given triangulation by subdividing a $k$-dimensional simplex, where $k\in\llbracket1,d\rrbracket$, with the addition of a new (embedded) vertex. As explained in \cite{fluxC}, if the root of a triangulation is subdivided by a move, the flag structure defines in a canonical manner a new root for the refined triangulation.

A geometric triangulation $\Delta'$ is said to be finer than a geometric triangulation $\Delta$ if it can be obtained form $\Delta$ by a (finite) sequence of refining Alexander moves, and if its vertices after the moves have the same coordinates as those of $\Delta$. We will denote this relation by $\Delta\prec\Delta'$.

In oder to unambiguously specify paths $\gamma$ in a graph $\Gamma$ and to obtain a characterization of the fundamental cycles of this graph (and to perform a gauge-fixing), it is convenient to introduce (maximal spanning) trees.

A maximal or spanning tree $\mathcal{T}$ in a graph $\Gamma$ is a connected subgraph of $\Gamma$ which does not contain any closed cycles and includes all the nodes of $\Gamma$. In the rest of the paper we use `tree' as short form for maximal connected tree. The links composing a tree are called branches and denoted by $b$, while the links of the graph that are not in the tree are called leaves and denoted by $\ell$. Although the leaves correspond to the links in $\Gamma\backslash\mathcal{T}$, by abuse of language we will refer to the leaves of the tree. The set of leaves in a tree will be denoted by $\mathcal{L}$. A rooted tree with root $r$ is a tree where a preferred node is identified and called the root. There is a one-to-one correspondence between the leaves of a tree and the fundamental cycles of the underlying graph. For a rooted tree, one way to characterize these cycles is to associate them with paths $\gamma_\ell$ labeled by the leaves in $\mathcal{L}$. A path $\gamma_\ell$ starts at the root, goes along the branches of the tree until the source node of the leaf $\ell$, then goes along $\ell$, and back to the root along branches. The cardinality of the set of fundamental cycles is independent of the choice of tree. Explicitly, if we denote by $|n|$ the number of nodes in the graph $\Gamma$ and by $|l|$ the number of links, then any given tree has $|n|-1$ branches and $\text{card}(\mathcal{L})=|\ell|=|l|-|n|+1$ leaves (and fundamental cycles).

\newpage
\section*{\large PART 1. QUANTIZATION ON A FIXED TRIANGULATION}

\noindent As explained in the introduction, to construct the full continuum quantum theory we will proceed in two steps. In this first part, we are going to define the quantum theory on a fixed triangulation. For this, we will first focus on the classical connection configuration space, then on the inner product and Hilbert space structure built upon it, and finally on the quantization of the algebra of holonomies and fluxes. This will enable us in particular to discuss the properties of the area operator. In the second part we will then show the existence of a continuum (inductive limit) Hilbert space.

\section{Connection configuration space}
\label{sec:configuration space}

\noindent In this section, we focus on the connection degrees of freedom and describe the configuration space\footnote{Notice that we use here the term ``configuration space'' in a slightly unusual way. Indeed, we already have at our disposal the configuration space $\mathcal{A}$ of smooth $G$ connections on the spatial manifold $\Sigma$. In the AL framework, the quantum theory is built over a suitable generalization of $\mathcal{A}$ which is the space $\overline{\mathcal{A}}$ of generalized connections. In the present context however, as we shall see, by configuration space we mean the space of classical states which reflect the geometrical properties of the states of our (yet-to-be built) quantum theory, which turns out to be the moduli space of flat connections on the spatial manifold $\Sigma$ with certain defects.} over which we are going to build the quantum theory.

The BF vacuum is a state locally and globally peaked on flat connections. This property can be encoded in the triviality of the holonomies of the connection around the faces of the dual graph $\Gamma$ (more precisely, along the fundamental cycles of the graph). Excitations over this vacuum locally shift the holonomies away from the identity, thereby creating curvature defects which can be thought of as being carried by the $(d-2)$-dimensional simplices (the hinges) of the triangulation. For this reason, we are interested in connections which are flat appart from possible conical singularities along the hinges. If we denote by $\Delta_{(d-2)}$ the $(d-2)$-skeleton of the triangulation $\Delta$, we are therefore led to considering  the space of flat connections on $\Sigma\backslash\Delta_{(d-2)}$.

As is well-known \cite{atiyah,goldman}, the moduli space of flat connections modulo gauge transformations on a compact and connected manifold $\Sigma$ is a finite-dimensional orbifold which admits several different (although equivalent) mathematical descriptions. In particular, it is common to describe the quotient 
\be
\mathcal{A}_0
\coloneqq\lb A\in\mathcal{A}\,\big|\,F(A)=0\rb/\mathcal{G}
\ee
in terms of holonomies (i.e. group homomorphisms) as
\be\label{without defects}
\mathcal{A}_0=\text{Hom}\big(\pi_1(\Sigma),G\big)/G,
\ee
where $G$ acts on the space $\text{Hom}$ of homomorphisms by diagonal adjoint action. This defines the space of locally flat connections for a compact and connected manifold $\Sigma$ without defects. Details about the construction of the first fundamental group $\pi_1(\Sigma)$ can be found in appendix \ref{appendix:pi1}.

In the present work, since we are interested in flat connections on the defected manifold $\Sigma\backslash\Delta_{(d-2)}$, we first have to worry about the fact that this latter is not compact. In order to avoid the inconvenience of working with infinite triangulations, one can resort to the following construction: First, choose a flat metric on the simplices of $\Delta$ (which can be the auxiliary metric on $\Sigma$) and let, for some small $\eps>0$, $B_\eps(x)$ be the open ball of radius $\eps$. Then, the set
\be
\Delta_{(d-2)}^\eps
\coloneqq\bigcup_{x\in\Delta_{(d-2)}}B_\eps(x),
\ee
which represents a ``blown-up'' version of $\Delta_{(d-2)}$, is open in $\Sigma$, and if $\eps$ is small enough one has that $\Sigma\backslash\Delta_{(d-2)}$ is homotopy-equivalent to the new manifold $\overline{\Sigma}\coloneqq\Sigma\backslash\Delta_{(d-2)}^\eps$. In fact, one has that $\Sigma\backslash\Delta_{(d-2)}^\eps$ is a deformation-retract of $\Sigma\backslash\Delta_{(d-2)}$. As such, these two spaces have isomorphic fundamental groups, and we can define our configuration space to be
\be
\mathcal{A}_\Delta
\coloneqq\text{Hom}\big(\pi_1\big(\overline{\Sigma}\big),G\big)/G.
\ee
The question is now how to characterize $\pi_1\big(\overline{\Sigma}\big)$. To this end we note that the graph $\Gamma$, which is the one-skeleton of the simplicial complex dual to the triangulation $\Delta$ of $\Sigma$, is a deformation-retract of $\overline{\Sigma}$. As a consequence, their first fundamental groups are isomorphic, i.e. $\pi_1\big(\overline{\Sigma}\big)\simeq\pi_1(\Gamma)$. We are therefore led to considering the first fundamental group $\pi_1(\Gamma)$ of the graph $\Gamma$ dual to the triangulation $\Delta$.

The first fundamental group $\pi_1(\Gamma)$ of a graph $\Gamma$ is a free group, and is generated by a set of fundamental cycles of the graph \cite{hatcher} (we refer the reader to appendix \ref{appendix:pi1defects} for more details). As noted in section \ref{sec:setup}, a set of fundamental cycles can be obtained by choosing a maximal spanning tree $\mathcal{T}$ (which here in addition will be rooted) in $\Gamma$. The links of $\Gamma$ which are not part of the tree are called leaves and denoted by $\ell$. The set $\mathcal{L}$ of leaves with respect to a given choice of tree $\mathcal{T}$ defines a set of fundamental cycles $\gamma_\ell$. Such a fundamental cycle $\gamma_\ell$ starts at the root, goes along the branches of the tree until the source node of the leaf $\ell$, then goes along $\ell$, and back to the root along branches.

By definition, a set of fundamental cycles obtained by a choice of tree does freely generate the first fundamental group $\pi_1(\Gamma)$. In other words, if we denote by $\gamma_\ell$ the cycle associated to the leaf $\ell$ and choose a numbering $1,\dots,|\ell|$ of the leaves, we have that
\be
\pi_1(\Gamma)
=\big\langle\gamma_1,\dots,\gamma_{|\ell|}\big\rangle.
\ee
Therefore, the space $\text{Hom}\big(\pi_1(\Gamma),G\big)\simeq\text{Hom}\big(\pi_1\big(\overline{\Sigma}\big),G\big)$ is parametrized by the set of group-valued maps $\mathcal{L}\rightarrow G$, and by associating to each cycle $\gamma_\ell$ a group element $g_\ell$ we can write that
\be
\text{Hom}\big(\pi_1\big(\overline{\Sigma}\big),G\big)
=\big\{g_1,\dots,g_{|\ell|}\,\big|\,g_i\in G\big\}.
\ee
From this space, we then obtain the space of locally flat connection by dividing out by the diagonal adjoint action of the group on $G^{|\ell|}$, i.e.
\be
\mathcal{A}_\Delta
\simeq\big\{g_1,\dots,g_{|\ell|}\,\big|\,g_{\ell_i}\in G\big\}/G.
\ee
As we will now explain, this space in turn agrees with the space of gauge-invariant (or gauge-fixed) holonomies on the graph $\Gamma$.

Since the $|\ell|$-tuple  $\{g_1,\dots,g_{|\ell|}\}$ determines the holonomies along the fundamental cycles of the graph, it can be used to reconstruct the holonomies along all the cycles (starting and ending at the root) in $\Gamma$. A (suitable) space of functions of these cycle holonomies therefore gives the space of functions of the graph connection which are invariant under gauge transformations at all the nodes except the root. Let us explain how this comes about. A graph connection is given by an assignment of group elements $h_l$ to the links $l$ of the graph, and the action of a gauge transformation with parameter $u_n\in G$ on $h_l$ factorizes at the nodes and gives
\be
u_n\triangleright h_l
=u^{-1}_{l(1)}h_lu_{l(0)}.
\ee
Given a closed cycle $\gamma=l_n\circ\dots\circ l_1$ with $l_1$ starting at the root and $l_n$ ending at the root, the cycle holonomy $g_\gamma=h_n\dots h_1$ is therefore invariant under gauge transformations at all the nodes except at the root. Instead, under gauge transformations at the root, it transforms with the adjoint action
\be\label{rootgaugetrafo1}
G^{|n|}\ni\{u_n\}\triangleright g_\gamma
=u^{-1}_r g_\gamma u_r.
\ee

By choosing a tree, the space of holonomies which are invariant under non-root gauge transformations can again be characterized by the association of group elements to the leaves of the graph. As explained above, for ever leaf $\ell$ we can consider paths $\gamma_\ell$ describing the fundamental cycles by starting at the root, going along the tree until the source of $\ell$, then going along $\ell$, and back to the root along the tree. To every leaf we can therefore associate a group element
\be\label{hcycle}
g_\ell
\coloneqq\overrightarrow{\prod_{e\subset\gamma_\ell}}h_e^{\pm1}
=g_{r\ell(1)}^{-1}h_\ell g_{r\ell(0)}.
\ee
Any other arbitrary cycle holonomy can be obtained by composing the holonomies $g_\ell$ of the fundamental cycles. The fundamental cycle holonomies therefore parametrize the space of graph connections which are invariant under non-root gauge transformations.

Furthermore we can use the tree to perform a (partial) gauge-fixing. Given any graph connection $\{h_l\} \in G^{|l|}$ there exist a unique gauge transformation 
$\{u_n\}\in G^{|n|-1}$, acting on all nodes except the root, such that $h_b\doteq\openone$ for all the links $b$ which are branches of the tree. Here we use $\doteq$ to denote an equality which holds with a particular gauge-fixing. In this gauge, the holonomy \eqref{hcycle} along a fundamental cycle reduces to the holonomy associated to the leaf $\ell$, i.e. $g_\ell\doteq h_\ell$. Thus $\{h_\ell\}\in G^{|\ell|}$ parametrizes the space of  almost gauge-invariant graph connections, that is connections invariant under all gauge transformations except the ones at the root. 

We can therefore conclude that $G^{|\ell|}$ describes the space of almost gauge-invariant connections on $\Gamma$, as well as the space of almost gauge-invariant locally flat connections on the manifolds with defects $\overline{\Sigma}$, which we will denote by $\mathcal{A}_r$.

\section{Inner product }
\label{sec:inner product}

\noindent In this section we introduce the inner product and Hilbert space structure on a fixed triangulation. For this, we first focus on states which are gauge-invariant everywhere appart from the root, and then use the technique of group averaging in order to account for these residual gauge transformations.

\subsection{Inner product for almost gauge-invariant states}

\noindent Our goal is to construct a Hilbert space $\mathcal{H}_\Delta$ for which the space of almost gauge-invariant graph connections $\mathcal{A}_r$ serves as the configuration space\footnote{Notice that we are not considering here a phase space (i.e. symplectic) structure on $\mathcal{A}_r$, as it would be the case for example in (at least one approach to) the quantization of Chern--Simons theory.}. The reason for working with $\mathcal{A}_r$ instead of the full gauge-invariant space $\mathcal{A}_\Delta$ is that the configuration space $G^{|\ell|}$ is much simpler to handle than the orbifold $G^{|\ell|}/G$. We will therefore first consider $\mathcal{A}_r$, and then impose invariance under the residual gauge transformations at the root via a group averaging procedure.

As discussed in section \ref{sec:configuration space}, the configuration space $\mathcal{A}_r$ is parametrized by a set of group elements in $G^{|\ell|}$, and the identification of the leaves $\ell$ implicitly means that one has made a choice of tree $\mathcal{T}$. We therefore consider some space $\mathcal{F}(G^{|\ell|})$ of functions as a candidate for our Hilbert space $\mathcal{H}_\Delta$, and consider states of the form
\be
|\psi\rangle
\equiv\psi\{g_\ell\}
\coloneqq\psi(g_1,\dots,g_{|\ell|}).
\ee
Here we can interpret the group elements $g_\ell$ as giving the holonomies associated to the fundamental cycles $\gamma_\ell$ of the graph $\Gamma$. Alternatively, if we use the tree $\mathcal{T}$ to gauge-fix the branch holonomies to the identity, since in this gauge we have $g_\ell\doteq h_\ell$, we can interpret the group elements $g_\ell$ as the (edge) holonomies associated to the leaves.

Let us now propose an inner product on the space $\mathcal{F}(G^{|\ell|})$. Our choice for this inner product is motivated by the inductive limit Hilbert space construction which will be carried out in the second main part of this work. As we will see, this construction requires to isometrically embed Hilbert spaces based on coarser triangulations $\Delta$ into Hilbert spaces based on finer triangulations $\Delta'$. Recall that for a finer triangulation $\Delta'$ with $\Delta'\succ\Delta$, the associated manifold $\Sigma\backslash\Delta'_{(d-2)}$ has more defects than the manifold $\Sigma\backslash\Delta_{(d-2)}$, and the dual graph $\Gamma'$ has more independent cycles than the dual graph $\Gamma$. This means that the space $\mathcal{A}'_r$ of almost gauge-invariant holonomies is isomorphic to $G^{|\ell'|}$ with $|\ell'|>|\ell|$. Now, the coarser configuration space $\mathcal{A}_r$ can be regained from the finer configuration space $\mathcal{A}'_r$ by imposing constraints $\mathcal{C}_I\{g_{\ell'}\}\stackrel{!}{=}\openone$ (where the index $I$ has range $1,\dots,|\ell'|-|\ell|$). These constraints impose that the ``finer'' holonomies, i.e. holonomies of cycles around the additional defects in $\Sigma\backslash\Delta'_{(d-2)}$, are trivial (this will be explained in much more detail in section \ref{sec:refinement}).

The presence of these constraints means that states in the finer Hilbert space $\mathcal{H}_{\Delta'}$ which arise as the embedding of states in the coarser Hilbert space $\mathcal{H}_{\Delta}$ are sharply peaked on vanishing ``finer'' holonomies. These states need to have a finite norm as otherwise the embedding would not be isometric (and the inner product would not be cylindrically consistent, which would prevent the inductive limit construction of the continuum Hilbert space from existing). Furthermore, the motivation for this work is to construct a representation based on a vacuum state which is peaked on locally and globally flat connections, and this vacuum state should of course also have a finite norm.

If the gauge group $G$ is a finite group, the inner product provided by the discrete measure (which agrees in this case with the Haar measure) leads to the required finiteness. However, if $G$ is a Lie group (even a compact one), the states of interest will have infinite norm. We therefore need to consider an appropriate regularization procedure.

Our regularization procedure goes as follows. First, we start with an auxiliary inner product $\langle\cdot|\cdot\rangle_\text{aux}$, together with a family $\psi^\epsilon$ of regulator states having a finite norm with respect to this auxiliary inner product for finite $\epsilon$. Second, we need a family $\psi^\epsilon_\openone$ of regulator states approaching the (locally and globally) flat vacuum state $\prod_\ell\delta(\openone,g_\ell)$ when $\epsilon\rightarrow0$. With this, we can then define the inner product
\be
\langle\psi_1|\psi_2\rangle
=\lim_{\epsilon\rightarrow0}\f{\langle\psi^\epsilon_1|\psi^\epsilon_2\rangle_\text{aux}}{\langle\psi^\epsilon_\openone|\psi^\epsilon_\openone\rangle_\text{aux}}.
\ee
The use of such an inner product, which employs the vacuum state as a reference vector, was suggested already in \cite{newvac} and \cite{benjaminBF}. As we are going to show, it leads to a discrete mesure on the space of locally flat connections, which in turn will make the proof of cylindrical-consistency of the inner product rather simple to obtain. It is interesting to notice that other proposals using a non-discrete measure on the space of flat connections have appeared earlier in the literature \cite{bonzom-smerlak1,bonzom-smerlak2}, but so far do not allow for a clear understanding of the cylindrical-consistency properties of the inner product (as opposed to the present framework) \cite{benjaminBF}.

To simplify the discussion, let us now consider that the configuration space is given by a single copy of the (compact and semi-simple) gauge group $G$, and choose the auxiliary inner product to be the one induced by the Haar measure on $G$. Furthermore, let us use a heat kernel regularization in order to regularize delta-peaked states. For $\alpha\in G$, we can then define $\epsilon$-regulated families of states converging to the delta function peaked on $\alpha$ as
\be
\psi_\alpha^\epsilon(g)
\coloneqq\sum_\rho d_\rho\exp\big(-\epsilon\text{Cas}(\rho)\big)\chi_\rho(g\alpha^{-1}).
\ee
Here we have chosen representatives $\rho$, given by representation matrices $D^\rho(\cdot)$ and with dimension $d_\rho$, for the equivalence classes of unitary irreducible representations of $G$. $\text{Cas}(\rho)$ denotes the quadratic Casimir (the eigenvalue of the Laplacian on the group manifold) associated to the representation $\rho$, and $\chi_\rho(\cdot)=\tr D^\rho(\cdot)$ is the character of $\rho$. Since the representation matrices satisfy $\langle D^\rho_{mn}|D^{\rho'}_{kl}\rangle_\text{aux}=(d_\rho)^{-1}\delta_{\rho,\rho'} \delta_{mk}\delta_{nl}$, we can readily compute
\be
\f{\langle\psi^\epsilon_\alpha|\psi^\epsilon_\beta\rangle_\text{aux}}{\langle\psi^\epsilon_\openone|\psi^\epsilon_\openone\rangle_\text{aux}}
=\f{\sum_\rho d_\rho\exp\big(-2\epsilon\text{Cas}(\rho)\big)\chi_\rho(\alpha\beta^{-1})}{\sum_\rho d^2_\rho\exp\big(-2\epsilon\text{Cas}(\rho)\big)}
=\f{\psi^{2\epsilon}_\openone(\alpha\beta^{-1})}{\psi^{2\epsilon}_\openone(\openone)}.
\ee
Using the asymptotic expansion of the heat kernel on compact Lie groups (see for instance \cite{heatkernel,bahrthiemannK}), which takes the form
\be
\psi^{\epsilon}_\openone(g)
\stackrel{\epsilon\rightarrow0}{\sim}\f{1}{(4\pi\epsilon)^{(\dim G)/2}}\exp\left(-\f{|g|^2}{4\epsilon}\right),
\ee
where $|g|$ is the geodesic distance in $G$ between $g$ and the unit element $\openone$, we finally get the following limit of vanishing regulator for the inner product:
\be\label{limitIP}
\f{\langle\psi^\epsilon_\alpha|\psi^\epsilon_\beta\rangle_\text{aux}}{\langle\psi^\epsilon_\openone|\psi^\epsilon_\openone\rangle_\text{aux}}
\ \stackrel{\epsilon\rightarrow 0\vphantom{\f{1}{2}}}{\longrightarrow}\ \delta(\alpha,\beta),
\ee
Here $\delta(\alpha,\beta)=1$ if and only if $\alpha=\beta$ and is vanishing otherwise. This delta symbol should therefore be treated as a Kronecker delta with continuous parameters. This is similar to what happens in the case of the Bohr compactification of Abelian groups, in particular for example in loop quantum cosmology (LQC) \cite{AshBojLewan}. However, in LQC the Bohr compactification typically corresponds to the configuration space (i.e. to the holonomies, which are proportional to the momentum $\dot{a}$ conjugate to the scale factor $a$), whereas here, in the case of the Abelian group $\U(1)$, we are compactifying the dual group $\mathbb{Z}$ (i.e. the space of fluxes), which in turn leads to a discrete topology on $\U(1)$ itself.

Let us now turn to the basis states for the Hilbert space $\mathcal{H}_\Delta$. Introducing the collection $\{\alpha_\ell\}=\{\alpha_1,\dots,\alpha_{|\ell|}\}\in G^{|\ell|}$ of group elements, the basis states $\psi_{\{\alpha_\ell\}}$ are of the form
\be\label{basisvectors}
\psi_{\{\alpha_\ell\}}\{g_\ell\}
=\prod_{\ell\in\mathcal{L}}\delta(\alpha_\ell,g_\ell),
\ee
and are orthonormal in the above-introduced inner product:
\be\label{Eq:HSInnerProduct}
\big\langle\psi_{\{\alpha_\ell\}}\big|\psi_{\{\beta_\ell\}}\big\rangle
=\prod_{\ell\in\mathcal{L}}\delta(\alpha_\ell,\beta_\ell).
\ee
Note that the Hilbert space $\mathcal{H}_\Delta$, although being associated to a finite triangulation, is therefore non-separable. More precisely, the Hilbert space is the closure of states of the form
\be
\psi\{g_\ell\}
=\sum_{\{\alpha_\ell\}}f\{\alpha_\ell\}\psi_{\{\alpha_\ell\}}\{g_\ell\},
\ee
where the sum runs over a \textit{finite} set of values $\{\alpha_\ell\}\in G^{|\ell|}$ and $f\{\cdot\}$ is a function on this finite set.

Although we are here developing a flux formulation of LQG, it is worth noticing that we work still in the holonomy representation for the arguments of the wave functions. In appendix \ref{appendix:spin}, we introduce a dual spin representation based on a suitable discrete generalization of the (group) Fourier transform. This spin representation can also be used to express the measure functional, as explained in appendix \ref{appendix:spin}.

\subsection{Invariance under change of tree}
\label{change of tree}

\noindent The construction which we have carried out so far depends on a specific choice of maximal tree $\mathcal{T}$ with (fixed) root node $r$. However, it is not difficult to see that the inner product is independent of the choice of tree, which means that the construction of $\mathcal{H}_\Delta$ is also independent of this choice.

In order to see this explicitly, let us choose another tree $\widetilde{\mathcal{T}}$ in $\Gamma$ with the same root $r$. The trees $\mathcal{T}$ and $\widetilde{\mathcal{T}}$ define two different isomorphisms between $\mathcal{A}_r$ and $G^{|\ell|}$, which means that the same connection in $\mathcal{A}_r$ corresponds to two different sets $\{g_\ell\}_{\ell\in\mathcal{L}}$ and $\{g_{\tilde{\ell}}\}_{\tilde{\ell}\in\widetilde{\mathcal{L}}}$ of group elements. The explicit isomorphism between the two sets of group elements can be constructed as follows. For both sets $\mathcal{L}$ and $\widetilde{\mathcal{L}}$ of leaves, the corresponding sets $\{\gamma_\ell\}_{\ell\in\mathcal{L}}$ and $\{\gamma_{\tilde{\ell}}\}_{\tilde{\ell}\in\widetilde{\mathcal{L}}}$ of cycles are fundamental. Therefore, a given cycle $\gamma_\ell$ can be expressed as a word in the cycles $\{\gamma_{\tilde{\ell}}\}_{\tilde{\ell}\in\widetilde{\mathcal{L}}}$ (and their inverses), and the other way around\footnote{More precisely, we have for instance $\gamma_{\tilde{\ell}}=\overrightarrow{\prod}_{\ell\in\gamma_{\tilde{\ell}}}\gamma_\ell^{\pm1}$, where the orientation (and ordering) of $\gamma_\ell$ in the product is according to the orientation (and ordering) in which the leaves $\ell$ appear in $\gamma_{\tilde{\ell}}$.}. We denote these relations by
\be
\gamma_\ell
=W_\ell\{\gamma_{\tilde{\ell}}\},\q
\gamma_{\tilde{\ell}}
=W_{\tilde{\ell}}\{\gamma_{\ell}\}.
\ee
Because of this, the corresponding group holonomies for the same flat connection in $\mathcal{A}_r$ can therefore be expressed as
\be\label{Eq:Isomorphism}
g_\ell
=W_\ell\{g_{\tilde{\ell}}\},\q
g_{\tilde{\ell}}
=W_{\tilde{\ell}}\{g_{\ell}\}.
\ee

Let us now consider a state $\psi\{g_\ell\}$ in the representation based on the choice of tree $\mathcal{T}$. A change of tree $\mathcal{T}\rightarrow\widetilde{\mathcal{T}}$ induces a gauge transformation $\mathbf{G}$ of the function $\psi$ on $G^{|\ell|}$, which we denote by $\tilde{\psi}=\mathbf{G}(\psi)$. The gauge-transformed state is then given by
\be
\tilde{\psi}\{g_{\tilde{\ell}}\}
=\mathbf{G}(\psi)\{g_{\tilde{\ell}}\}
=\psi\big\{W_\ell\{g_{\tilde{\ell}}\}\big\}.
\ee
In particular, for the basis states $\psi_{\{\alpha_\ell\}}$ we have that
\be
\mathbf{G}\big(\psi_{\{\alpha_\ell\}}\big)\{g_{\tilde{\ell}}\}
=\psi_{\{\alpha_\ell\}}\big\{W_\ell\{ g_{\tilde{\ell}}\}\big\}
=\prod_\ell\delta\big(\alpha_\ell,W_\ell\{g_{\tilde{\ell}}\}\big)
=\prod_{\tilde{\ell}}\delta\big(W_{\tilde{\ell}}\{\alpha_\ell\},g_{\tilde{\ell}}\big)
=\psi_{\{W_{\tilde{\ell}}\{\alpha_\ell\}\}}\{g_{\tilde{\ell}}\}.
\ee
Here we have used the fact that the words $W_\ell$ (or their inverses $W_{\tilde{\ell}}$) define a bijective map $G^{|\ell|}\rightarrow G^{|\tilde{\ell}|}$, which means that $\alpha_\ell=\beta_\ell\,\forall\ \ell$ if and only if $W_{\tilde{\ell}}\{\alpha_\ell\}=W_{\tilde{\ell}}\{\beta_\ell\}\,\forall\ \tilde{\ell}$.

With these ingredients, we can finally compute the inner product using the representation based on the choice of tree $\widetilde{\mathcal{T}}$, which leads to
\be
\big\langle\mathbf{G}(\psi_{\{\alpha_\ell\}})\big|\mathbf{G}(\psi_{\{\beta_\ell\}})\big\rangle
=\big\langle\psi_{\{W_{\tilde{\ell}}\{\alpha_\ell\}\}}\big|\psi_{\{W_{\tilde{\ell}}\{\beta_\ell\}\}}\big\rangle
=\prod_\ell\delta(\alpha_\ell,\beta_\ell).
\ee
This is identical to the inner product between states expressed with respect to the choice of tree $\mathcal{T}$, which shows that the inner product \eqref{Eq:HSInnerProduct} is independent from the choice of tree, or in other words invariant under gauge transformations not acting at the root. We are now going to deal with these residual gauge transformations.

\subsection{Gauge transformations at the root}
\label{sec:raq}

\noindent The states in ${\mathcal{H}}_\Delta$ which we have constructed correspond to states on almost gauge-invariant connections $\mathcal{A}_r$. Under the gauge transformation \eqref{rootgaugetrafo1} at the root $r$, the group elements in $\{\alpha_\ell\}$ become conjugated as
\be\label{Eq:GaugeTransformationAtRoot}
\{\alpha_\ell\}^u
=\{\alpha_1\,\dots,\alpha_{|\ell|}\}^u
\equiv\{u\alpha_1u^{-1},\dots,u\alpha_{|\ell|}u^{-1}\},
\ee
where the group element parametrizing the gauge transformation is $u=u_r$. Due to the form of the inner product \eqref{Eq:HSInnerProduct}, it is clear that the gauge-averaged states which result from integrating the action \eqref{Eq:GaugeTransformationAtRoot} over the gauge group $G$ have infinite norm. Therefore, the Hilbert space $\mathcal{H}^u_\Delta$ of gauge-invariant states has to be constructed differently, and one way to do so is by using the technique of refined algebraic quantization (RAQ) \cite{RAQ}.

For this, let us denote by $\mathcal{D}_\Delta$ the dense subspace of finite linear combinations of $\psi_{\{\alpha_\ell\}}$ with\footnote{Even though this makes explicit use of a maximal tree $\mathcal{T}$ in $\Gamma$, it is easy to see that $\mathcal{D}_\Delta$ is actually invariant under a change of tree.} $\{\alpha_\ell\}\in G^{|\ell|}$. Instead of being the image of a projector on ${\mathcal{H}}_\Delta$, the Hilbert space $\mathcal{H}^u_\Delta$ of fully gauge-invariant functions is going to be obtained as a subspace of the algebraic dual $\mathcal{D}_\Delta'$ of $\mathcal{D}_\Delta$. These various spaces are organized in a so-called Gel'Fand triple
\be
\mathcal{D}_\Delta\hookrightarrow\mathcal{H}_\Delta\hookrightarrow\mathcal{D}_\Delta',
\ee
where the last inclusion is given by the Riesz representation theorem, using the inner product on $\mathcal{H}_\Delta$. The elements in $\mathcal{D}_\Delta'$ can be thought of as distributional states. 

Instead of a projector onto gauge-invariant states, RAQ employs a \textit{rigging map}, i.e. a linear function $\eta_\Delta:\mathcal{D}_\Delta\rightarrow\mathcal{D}_\Delta'$ which maps to the gauge-invariant distributional states in the sense that
\be
\eta_\Delta\big[\psi_{\{\alpha_\ell\}}\big]
=\eta_\Delta\big[\psi_{\{\alpha_\ell\}^u}\big]
\ee
for all $u\in G$. We choose this rigging map to be
\be\label{Eq:RiggingMap}
\eta_\Delta\big[\psi_{\{\alpha_\ell\}}\big]\psi
\coloneqq\int_{G/\text{stab}\,\alpha_\ell}\de\mu_\text{d}(u)\big\langle\psi_{\{\alpha_\ell\}^u}\big|\psi\big\rangle
=\sum_{u\in G}\big\langle\psi_{\{\alpha_\ell\}^u}\big|\psi\big\rangle,
\ee
where $\text{stab}\,\alpha_\ell$ is the stabilizer of $\alpha_\ell$ under the adjoint action of $G$, and where we use the discrete measure $\mu_\text{d}$ on $G$. Because an element $\psi\in\mathcal{D}_\Delta$ is a finite linear combination of basis elements $\psi_{\{\alpha_\ell\}}$, the integral in \eqref{Eq:RiggingMap} is finite since only finitely many elements in the sum are non-vanishing.
 
On the image of $\eta_\Delta$, we can now define the pre-inner product form
\be\label{Eq:PreInnerProduct}
\big\langle\eta_\Delta\big[\psi_{\{\alpha_\ell\}}\big]\big|\eta_\Delta\big[\psi_{\{\beta_\ell\}}\big]\big\rangle
\coloneqq\eta_\Delta\big[\psi_{\{\alpha_\ell\}}\big]\psi_{\{\beta_\ell\}}
=\left\{\begin{array}{l}
\text{1 if $\{\beta_\ell\}=\{\alpha_\ell\}^u$ for some $u\in G$},\\[5pt]
\text{0 otherwise.}
\end{array}\right.
\ee
Since \eqref{Eq:PreInnerProduct} is sesquilinear and non-negative, the image of non-zero vectors which have vanishing norm in $\eta_\Delta$ can be completed with respect to it and we can construct the Hilbert space as the closure
\be
\mathcal{H}^u_\Delta
\coloneqq\overline{\eta_\Delta\big(\mathcal{D}_\Delta\big)}.
\ee
Because the discrete measure $\mu_\text{d}$ is translation-invariant on $G$, the kernel of $\eta_\Delta$ is generated by elements of the form $\psi_{\{\alpha_\ell\}}-\psi_{\{\alpha_\ell\}^u}$. In other words, gauge-equivalent states are mapped to the same distributional state. As a result, an orthonormal basis of states in $\mathcal{H}^u_\Delta$ is given by states which are labelled by gauge-orbits $[\{\alpha_\ell\}]$ in $G^{|\ell|}$. The inner product in $\mathcal{H}^u_\Delta$ is such that two such states have zero norm precisely if two of these orbits do not coincide, i.e.
\be
\big\langle\psi_{[\{\alpha_\ell\}]}\big|\psi_{[\{\beta_\ell\}]}\big\rangle
=\delta\big([\{\alpha_\ell\}],[\{\beta_\ell\}]\big).
\ee
It should be noted that the isomorphisms \eqref{Eq:Isomorphism} commute with the gauge transformations \eqref{Eq:GaugeTransformationAtRoot} at the root, which implies that the resulting Hilbert space $\mathcal{H}^u_\Delta$ is of course independent of the choice of tree.

Finally, note that there exist in principle alternative versions of the rigging map \eqref{Eq:PreInnerProduct} which include an $[\{\alpha_\ell\}]$-dependent factor. With these other choices, the resulting inner product will simply change by state normalization.

\section{Quantization of the algebra of observables}
\label{sec:quantization algebra}

\noindent Now that we have introduced the Hilbert space, the inner product, and a basis of states on a fixed triangulation, we are ready to discuss the quantization of the algebra of holonomy-flux observables. This will then enable us to discuss, in the next section, the construction of the area operator and the associated notion of quantum geometry.

\subsection{Phase space on a fixed triangulation}

\noindent In the previous section, we have focused on the configuration space of connections associated to a triangulation. In order to obtain the phase space of LQG, we need to supplement the holonomies encoding the connection degrees of freedom with their conjugated fluxes of the electric field. This will allow us to discuss in which sense the quantum operators which we are going to define provide a representation of the holonomy-flux algebra.

As discussed in \cite{fluxC}, the finite-dimensional (almost) gauge-invariant phase space $\mathcal{M}_\Delta$ associated to a triangulation $\Delta$ is spanned by cycle holonomies $g_\ell$ and rooted fluxes $\bX_\ell$ associated to the leaves. These variables, which form the rooted holonomy-flux algebra $\mathfrak{A}^r$, are defined as $g_\ell\coloneqq g_{r\ell(1)}^{-1}h_\ell g_{r\ell(0)}$ and $\bX_\ell\coloneqq g_{r\ell(0)}^{-1}X_\ell g_{r\ell(0)}$, where $g_{r\ell(0)}$ is the holonomy along the tree going from the root to the source node of the leaf $\ell$. The flux $X_\ell$ is an element of the Lie algebra of $G$. In the case where $G=\SU(2)$ for example, the Poisson brackets between these variables are given by
\be
\lb\bX^i_\ell,\bX^j_{\ell'}\rb=\delta_{\ell,\ell'}{\eps^{ij}}_k\bX^k_\ell,\q
\lb\bX^k_\ell,g_{\ell'}\rb=\delta_{\ell,\ell'}g_\ell\tau^k-\delta_{\ell^{-1},\ell'}\tau^kg_{\ell'},\q
\lb g_\ell,g_{\ell'}\rb=0,
\ee
where the $\tau^k$ denote the generators of $\su(2)$. This is nothing but the Poisson structure on $(T^*G)^{|\ell|}$, where $T^*G\simeq G\times\mathfrak{g}^*$.

Note that the same Poisson algebra (however with Dirac brackets), between the holonomies $h_\ell$ and fluxes $X_\ell$ associated to the leaves, can be obtained by starting from the gauge-covariant phase space associated to the graph $\Gamma$ and performing a gauge-fixing\footnote{This agrees with our choice of notation. We denote by $g$ holonomies associated to paths along several links, and by $h$ holonomies associated to single links. Therefore, the gauge-fixed data is given by elements $(h_\ell,X_\ell)$, while at the almost gauge-invariant level we have $(g_\ell,\bX_\ell)$.}. The gauge-covariant phase space is parametrized by holonomies $h_l$ and fluxes $X_l$ associated to all the links of the graph, and the Poisson structure is that of $(T^*G)^{|l|}$. By choosing a tree, one can gauge-fix all the branch holonomies to be $h_b=\openone$. The associated Dirac brackets, involving $h_\ell$ and $X_\ell$ only, do then agree with the Poisson brackets and lead again to the Poisson structure on $(T^*G)^{|\ell|}$.

Since we are working in the connection representation, we can expect that the holonomies will be quantized as multiplication operators and the conjugated fluxes as derivative operators. While it is true that the holonomies will act by multiplication, the fluxes can however not be realized as derivative operators. This is due to the peculiar Hilbert space topology which we have chosen in order to accommodate the flat state as a normalizable state an to allow for the construction of an inductive limit Hilbert space. States in this Hilbert space, seen as functions on $G^{|\ell|}$, are generically not continuous. This effect is analogous to what happens in LQC due to the Bohr compactification of the configuration space (remember however, from the discussion of section \ref{overview}, the difference between LQC and the construction based on the BF vacuum presented here). The way around this difficulty is to introduce exponentiated derivative operators, or, in other words, translations. Indeed, we are going to see that the quantized exponentiated fluxes act as left or right translations on the appropriate copy of the group. In addition to considering the exponentiated version of an elementary flux variable, we will also have to discuss the action of exponentiated parallel-transported fluxes. It turns out that the parallel transport will appear as an adjoint action of certain holonomies on the group element parametrizing the translation.

To be more precise, the exponentiation of the fluxes is to be understood as exponentiating the symplectic flow generated by the fluxes. Indeed, if one sees the fluxes as representing vector fields (acting as derivatives) on the phase space (see for instance \cite{lqg3}), the exponentiated fluxes correspond to integrating these vector fields to a finite flow. Let us illustrate this for one copy of the group. The symplectic flow induced by an elementary flux $X_l$ associated to a link $l$ is given by
\be\label{flux1}
\exp\big(\sigma_i\lb X^i_l,\cdot\rb\big)f(h_l)
\coloneqq\sum_k\f{1}{k!}\big\{\sigma_iX^i_l,f(h_l)\big\}_k
=R_l^{\exp(\sigma_i\tau^i)}f(h_l)
\coloneqq f\big(h_l\exp(\sigma_i\tau^i)\big),
\ee
where the iterated Poisson brackets are defined by $\{f,g\}_k=\big\{f,\{f,g\}_{k-1}\big\}$ and $\{f,g\}_0=g$. For the link with a reversed orientation, we have that
\be
\exp\big(\sigma_i\lb X^i_{l^{-1}},\cdot\rb\big)f(h_l)
\coloneqq R_{l^{-1}}^{\exp(\sigma_i\tau^i)}f(h_l)
=L_l^{\exp(\sigma_i\tau^i)}f(h_l)
=f\big(\exp(-\sigma_i\tau^i)h_l\big).
\ee
In an abuse of notation, we will denote the group element parametrizing this flow by $\sigma\coloneqq\exp(\sigma_i\tau^i)$. 

So far we have considered the flow induced by an elementary flux which is not parallel-transported. Note that if we understand this description as arising from a gauge-fixing, then the flux $X_\ell$ in the gauge-fixed version does indeed correspond to the rooted (parallel-transported along the tree) flux $\bX_\ell$. Hence, the (exponentiated) action of $X_\ell$ on $h_\ell$ agrees with the (exponentiated) action of $\bX_\ell$ on the cycle holonomy $g_\ell$.

We might however encounter further parallel transport for the fluxes, not necessary along the tree. In the almost gauge-invariant description, such a parallel transport will be along a closed loop $\gamma$, and will appear through the adjoint action of a product  $g_\gamma$ of cycle holonomies $g_{\tilde{\ell}}$, which in the gauge-fixed description is just given by the corresponding product $h_\gamma$ of leaf holonomies $h_{\tilde{\ell}}$. Let us assume for now that $h_\ell$ is not appearing in this product. We then have for the symplectic flow generated by $h^{-1}_\gamma X_\ell h_\gamma$ the following action:
\be
\exp\big(\sigma_i\{h^{-1}_\gamma X^i_\ell h_\gamma,\cdot\}\big)f(h_\ell)
=f(h_\ell h_\gamma\sigma h_\gamma^{-1}).
\ee
If $h_\gamma$ includes a leaf holonomy $h_\ell$ (such that it cannot be reduced by using the inversion formula $X_{\ell^{-1}}=-h_\ell X_\ell h^{-1}_\ell$) the evaluation of the flow does become very involved (since the flux $X_\ell$ is meeting in the iterated Poisson brackets the element $h_\ell$ in the parallel transport). We will exclude this case and not allow such a parallel transports to happen.

In summary, we are going to quantize the parallel-transported fluxes $g_\gamma^{-1}\bX_\ell g_\gamma $ in their exponentiated form as parallel-transported (right) translations, i.e.
\be
g_\gamma^{-1}\bX_\ell g_\gamma\ \rightarrow\ R_\ell^{g_\gamma\sigma g_\gamma^{-1}},
\ee
and the parallel-transported fluxes associated to inverted links as left translations, i.e.
\be
g_\gamma^{-1}\bX_{\ell^{-1}}g_\gamma\ \rightarrow\ L_\ell^{g_\gamma\sigma g_\gamma^{-1}}.
\ee
In all cases, we assume that the parallel transport $g_\gamma$ does not contain the holonomy $h_\ell$ (except in the case that the right translation can be rewritten into a left one and vice versa). Note that although a representation of the fluxes as derivative operators is not available, it is still possible to approximate derivative operators in terms of translations generated by the exponentiated fluxes. This is what we will use in the next section in order to discuss the area operator.

We have here considered only the fluxes associated to the leaves. It is however possible to reconstruct the fluxes associated to the branches starting from the ones associated to the leaves by solving the Gau\ss~constraints. The corresponding translation operators are then products of translation operators associated to the leaves. This construction is explained in appendix \ref{appendix:branchX}.

\subsection{Quantization of the holonomies}

\noindent Let us consider the space $\mathcal{F}$ of continuous functions $f:G^{|\ell|}\rightarrow\mathbb{C}$ over $|\ell|$ copies of the gauge group. Being continuous functions on a compact space, these functions are bounded, and in the holonomy representation the holonomy operators acting as multiplication of states by elements of $\mathcal{F}$ are therefore bounded operators. Examples of such functions are matrix elements $D^{\rho}_{mn}(\cdot)$ for a unitary irreducible representation $\rho$. Although states of the form $\psi= \prod_\ell D^{\rho_\ell}_{m_\ell n_\ell}$ are not normalizable in our inner product (if $G$ is not a finite group), such functions are allowed to be used as multiplication operators, i.e. lead to densely defined operators. Another example is given by functions of the form
\be
f\{g_\ell\}
=\sum_if_i\prod_\ell\delta\big(g_\ell,\alpha^{(i)}_\ell\big),
\ee
where the index $i$ runs over a finite set, and $\alpha^{(i)}_\ell$ is the $\ell$-th entry in the $i$-th element $\{\alpha_\ell\}^i$ of a finite subset of $G^{|\ell|}$. Again, we denote by $\delta(g,\alpha)$ the Kronecker Delta as defined below equation \eqref{limitIP}.

The action of an holonomy operator on a state in $\mathcal{H}_\Delta$ is given by
\be\label{holop1}
\widehat{f}\psi\{g_\ell\}
=f\{g_\ell\}\psi\{g_\ell\}.
\ee
The holonomy operators are straightforward to diagonalize, and the eigenvectors agree with the basis vectors $\psi_{\{\alpha_\ell\}}$. Therefore, such an operator has a discrete spectrum given by the image of $f$ in $\mathbb{C}$ (or in $\mathbb{R}$ if one takes the real or imaginary parts of $f$). Note that the spectrum can appear as continuous, but always comes equipped with a discrete spectral measure.

\subsection{Quantization of the fluxes}
\label{sec:Qfluxes}

\noindent As discussed in the beginning of this section, we do not quantize the fluxes themselves, but rather their exponentiated version, which leads to an action as right translation operators. To the flux associated to the $i$-th leaf $\ell_i$ we associate the operator
\be
R_{\ell_i}^\sigma\psi\{g_\ell\}
=\psi(g_1,\dots,g_i\sigma,\ldots,g_{|\ell|}).
\ee
A (first order) difference operator can then be defined as
\be
D^\sigma_{\ell_i}
\coloneqq\frac{-\i}{2|\sigma|}\left(R_{\ell_i}^\sigma-R_{\ell_i}^{\sigma^{-1}}\right),
\ee
with $|\sigma|$ denoting the geodesic distance in $G$ between $\sigma$ and $\openone$.

The translation operators are isometric since the discrete measure employed for the inner product on $\mathcal{H}_\Delta$ is invariant under translations. They are also invertible, i.e.
\be
\big(R_\ell^\sigma\big)^{-1}
=\big(R_\ell^\sigma\big)^\dagger
=R_\ell^{\sigma^{-1}},
\ee
and therefore unitary. It is quite instructive to discuss the spectrum of such a translation operator, and to compare it to the spectrum of a flux operator in the AL representation. This will be the focus of the next subsection.

Let us now also discuss the action of parallel-transported exponentiated fluxes, which, as discussed above, are quantized as operators acting in the following way:
\be
R_{\ell_i}^{g_\gamma\sigma g_\gamma^{-1}}\psi\{g_\ell\}
=\psi(g_1,\dots,g_ig_\gamma\sigma g_\gamma^{-1},\dots,g_{|\ell|}).
\ee
These operators are also isometric (and unitary) because of the invariance of the discrete measure under translations. The action of a parallel-transported exponentiated flux on a basis state is given by
\be
R_{\ell_i}^{g_\gamma\sigma g_\gamma^{-1}}\psi_{\{\alpha_\ell\}}\{g_\ell\}
&=\psi_{\{\alpha_\ell\}}(g_1,\dots,g_ig_\gamma\sigma g_\gamma^{-1},\dots,g_{|\ell|})\nn\\
&=\delta\big(\alpha_i\alpha_\gamma\sigma\alpha^{-1}_\gamma,g_i\big)\prod_{\ell\neq i}\delta(\alpha_\ell,g_\ell)\nn\\
&=\psi_{\{\{\alpha_\ell\}_{\ell\neq i};\alpha_i\alpha_\gamma\sigma\alpha^{-1}_\gamma\}}\{g_\ell\},
\ee
where $\alpha_\gamma$ denotes the same product in terms of $\alpha_\ell$ as $g_\gamma$ in terms of the fundamental cycle holonomies $g_\ell$ (remember that we do not allow the link $\ell_i$ to appear in the parallel transport, as otherwise the last equality would not hold).

\subsection{Spectrum of the translation operator}\label{spectrans}

\noindent We are now going to discuss the spectrum of the translation operator. For simplicity, let us focus on one single copy of the group, and consider the Hilbert space $\mathcal{H}=L^2(G,\mu_\text{d})$, where $\mu_\text{d}$ denotes the discrete measure. This Hilbert space being non-separable, the spectral analysis is a priori not straightforward to carry out. Fortunately, since the translation operator on $\SU(2)$ and $\U(1)$ is invertible, it leaves separable subspaces of $\mathcal{H}$ invariant, and its spectral analysis can therefore be done on each separable subspace separately. For a fixed $\alpha\in G$, consider the invariant subspace
\be
\mathcal{H}_{\alpha}^\sigma
=\overline{\big\{(R^\sigma)^n\psi_\alpha=\psi_{\alpha\sigma^{-n}}\,\big|\,n\in\mathbb{Z}\big\}},
\ee
which is the closure of the space of states obtained by applying an arbitrary integer power of the translation operator to a basis state $\psi_\alpha$. We can transform the matrix representation of the group with the adjoint transformations $\alpha'=u\alpha u^{-1}$ and $\sigma'=u\sigma u^{-1}$, in such a way that $\sigma'=\text{diag}(e^{\i\phi},e^{-\i\phi})$ is diagonal with $\phi\in[0,2\pi)$, and restrict ourselves to considering the translation operator for $\U(1)$.

Let us therefore turn to the spectral analysis of the translation operator $R^\phi$ on $\U(1)$ equipped with the discrete topology. We consider basis functions $\psi_\alpha:[0,2\pi)\rightarrow\mathbb{C}$ given by $\psi_\alpha(\theta)=\delta(\alpha,\theta)$, which we declare to be orthonormal in the Kronecker delta inner product $\langle\psi_\alpha|\psi_\beta\rangle=\delta(\alpha,\beta)$, where the arguments are understood modulo $2\pi$. These functions form a basis of the non-separable Hilbert space
\be
\mathcal{H}
=\overline{\text{span}\big\{\psi_\alpha\,\big|\,\alpha\in[0,2\pi)\big\}},
\ee
a typical element of which is given by a function of the form
\be
\psi(\theta)
=\sum_if(\alpha_i)\psi_{\alpha_i}(\theta),
\ee
where $i$ takes values in some finite subset of $\mathbb{N}$, and $\theta\in[0,2\pi)$ is the $\U(1)$ angle. The inner product between two such functions is simply given by
\be
\langle\psi_1|\psi_2\rangle
=\int\de\mu_\text{d}(\theta)\,\overline{\psi_1(\theta)}\psi_2(\theta)
=\sum_i\overline{f_1(\alpha_i)}f_2(\alpha_i).
\ee
The action of the translation operator on elements of $\mathcal{H}$ is given by
\be
R^\phi\psi(\theta)
=\psi(\theta+\phi),
\ee 
and its action on basis functions by\footnote{Let us also mention what happens for the parallel-transported translations $R_{\ell_i}^{g_\gamma\sigma g^{-1}_\gamma}$. On basis vectors $\psi_{\{\alpha_\ell\}}$, the group elements $g_\gamma=W\{g_\ell\}$ evaluate to $\alpha_\gamma=W\{\alpha_\ell\}$ (where $\alpha_{\ell_i}$ does not appear). The discussion is therefore exactly the same, since the class angle of $\sigma$ is invariant under conjugation.}
\be
R^\phi\psi_\alpha
=\psi_{\alpha-\phi}.
\ee
As mentioned above, since this operator is invertible it leaves separable subspaces of $\mathcal{H}$ invariant. The spectral analysis can therefore be done separately on each separable subspace.

Let us consider the sub-Hilbert space which is the closure of the space of states generated by applying an arbitrary integer power of the translation operator to the basis $\psi_\alpha$, i.e.
\be
\mathcal{H}_\alpha^\phi
\coloneqq\overline{\big\{R^{n\phi}\psi_\alpha=\psi_{\alpha-n\phi}\,\big|\,n\in\mathbb{Z}\big\}},
\ee
where the label $\alpha$ in $\psi_\alpha$ is to be understood in $\mathbb{R}$ modulo $2\pi$. The properties of this subspace and of the translation operator acting on it will depend on whether the angle $\phi$ parametrizing the translation is rational or not. We now discuss these two cases separately.
\begin{enumerate}
\item \underline{$\phi$ is a rational angle:}\\
This case corresponds to $\phi\in[0,2\pi)\cap2\pi\mathbb{Q}$, which means that $\phi=2\pi p/q$ with $q,p\in\mathbb{N}$ such that $\text{gcd}(q,p)=1$. Then we have that $(R^\phi)q=\openone$, and the repeated action of $R^\phi$ on a basis state $\psi_\alpha$ generates finite-dimensional (in fact $q$-dimensional) subspaces
\be
\mathcal{H}_\alpha^\phi
\coloneqq\big\{\psi_{\alpha-n\phi}\,\big|\,\text{$\alpha\in[0,2\pi/q)$ and $n=0,\dots,p-1$}\big\}.
\ee
In terms of these finite-dimensional spaces, the whole Hilbert space decomposes as
\be
\mathcal{H}
=\bigoplus_{\alpha\in[0,2\pi/q)}\mathcal{H}^\phi_\alpha.
\ee
It is straightforward to diagonalize the translation operator on the finite-dimensional spaces $\mathcal{H}_\alpha^\phi$. In particular, the normalized eigenvectors are of the form
\be\label{eigen1}
v_{\alpha,\kappa}
=\f{1}{\sqrt{q}}\sum^{q-1}_{n=0}e^{\i n\kappa\phi}\psi_{\alpha+n\phi},
\ee
and the associated eigenvalues are
\be
\text{spec}(R^\phi)
=\big\{e^{\i\kappa\phi}\,\big|\,\kappa=0,\dots,q-1\big\}.
\ee
Let us now Fourier transform these eigenvectors using the (group) Fourier transform introduced in appendix \ref{appendix:spin} (this will also help illustrate the case of an irrational angle $\phi$). In the case of $\U(1)$, the momentum representation is defined with the discrete measure by
\be
\psi(k)
=\int\de\mu_\text{d}(\theta)\,e^{-\i k\theta}\psi(\theta),
\ee
where $k\in\mathbb{Z}$. The  inner product expressed in $k$-space (and which actually defines the Bohr compactification on $\mathbb{Z}$) is
\be\label{inp}
\la\psi_1|\psi_2\ra
=\lim_{T\rightarrow\infty}\f{1}{(2T+1)}\sum_{|k|\leq T}\overline{\psi_1(k)}\psi_2(k).
\ee
One would expect the functions $v_\kappa(k)=\delta(\kappa,k)$ to be eigenvectors of the translation operator. However, these functions have vanishing norm in the inner product \eqref{inp}. Let us therefore consider the $k$-representation of the eigenvectors \eqref{eigen1}. With $\psi_{\alpha}(k)=e^{-\i k\alpha}$, we have 
\be
v_{\alpha,\kappa}(k)
=\f{1}{\sqrt{q}}\sum_{n=0}^{q-1}e^{\i n\kappa\phi}\psi_{\alpha+n\phi}(k)
=e^{-\i k\alpha}\f{1}{\sqrt{q}}\sum_{n=0}^{q-1}e^{\i n\phi(\kappa-k)}
=e^{-\i k\alpha}\sqrt{q}\,\delta_{(q)}(\kappa,k),
\ee
where we have used $\phi=2\pi p/q$, the summation representation of the periodic Kronecker delta $\delta_{(q)}(\kappa,k)$ (with arguments modulo $q$), and the fact that $\delta_{(q)}(\kappa,k)=\delta_{(q)}(p\kappa,pk)$ since $q$ does not divide $p$. Therefore, instead of vectors $v_\kappa(k)\sim\delta(\kappa,k)$, which have vanishing norm, we have eigenvectors $v_\kappa(k)\sim\sqrt{q}\,\delta_{(q)}(\kappa,k)$ with finite norm. Letting $q\rightarrow\infty$ illustrates the fact that we have to expect generalized eigenvectors in the case of an irrational angle.

\item\underline{$\phi$ is an irrational angle:}\\
This case corresponds to $\phi\in[0,2\pi)\cap2\pi(\mathbb{R}\backslash\mathbb{Q})$. The angle $\alpha$ then labels an equivalence class $[\alpha]$ of angles defined by the equivalence relation
\be\label{eclass}
\text{$\alpha\sim\alpha'$ if and only if $\exists\,n\in\mathbb{Z}$ such that $\alpha'=\alpha+n\phi$}.
\ee
Let us introduce the set $I\coloneqq[0,2\pi)/\sim$ of all such equivalence classes, as well as the (countably) infinite-dimensional Hilbert spaces
\be
\mathcal{H}_\alpha^\phi
\coloneqq\overline{\big\{\psi_\text{$(\alpha-n\phi)$ mod $2\pi$}\,\big|\,n\in\mathbb{Z}\big\}}.
\ee
The whole Hilbert space $\mathcal{H}=L^2\big(\U(1),\mu_\text{d}\big)$ then decomposes as
\be
\mathcal{H}
=\bigoplus_{[\alpha]\in I}\mathcal{H}_\alpha^\phi.
\ee
We prove in appendix \ref{appendix:spectrum} that the spectrum of the translation operator $R^\phi$ on each subspace $\mathcal{H}_\alpha^\phi$ consists of the whole circle, i.e.
\ba
\text{spec}(R^\phi)
=\U(1),
\ea
and is continuous (in the sense that the spectral measure is continuous). The generalized eigenvectors are given by
\be\label{irreigen}
w_{\alpha,\rho}
=\sum_{n\in\mathbb{Z}}e^{in\rho}\psi_{\alpha+n\phi},
\ee
with generalized eigenvalues $e^{\i\rho}$ where $\rho\in[0,2\pi)$. Choosing a different representative $\alpha'\in[\alpha]$ in the equivalence class $[\alpha]$ defined in \eqref{eclass} just leads to a phase factor for $w_{\alpha',\rho}$. In $k$-space, where again $\psi_{\alpha}(k)=e^{-\i k\alpha}$, the generalized eigenvectors are given by
\be
w_{\alpha,\rho}(k)
=\sum_{n\in\mathbb{Z}}e^{\i n\rho}\psi_{\alpha+n\phi}(k)
=e^{-\i k\alpha}\sum_{n\in\mathbb{Z}}e^{\i n(\rho-k\phi)}
=2\pi e^{-\i k\alpha}\delta(\rho-k\phi),
\ee
where $\delta(\rho-k\phi)$ is now the delta \textit{function} with respect to the continuous measure $\de\theta$ on $\U(1)$. We can in fact understand these generalized eigenvectors via a regularization procedure which uses the variable $T$ in the inner product \eqref{inp} as a cutoff. One considers the inner product \eqref{inp} with fixed $T$ and eigenvectors
\be\label{cutoffvec}
w^T_{\alpha,\kappa}(k)
=C(T)e^{-\i k\alpha}\delta(\kappa,k),
\ee
where $\kappa\in\mathbb{Z}\cap[-T,T]$, and $C(T)$ is an appropriate normalization factor. To keep the norm of $w^T$ finite when $T\rightarrow\infty$, the normalization factor has to grow to infinity.
\end{enumerate}

Let us finish this subsection by comparing the spectra of the translation operator for the different cases discussed above, with the spectra of the translation operator defined on the Hilbert space $\mathcal{H}_\text{Haar}=L^2\big(\U(1),\de\theta\big)$. These can be obtained by evaluating the function $\exp(i\phi\,\cdot)$ on the spectrum $\mathbb{Z}$ of the momentum operator $-i\partial_\theta$. We then have to discuss separately the two cases for the angle $\phi$.
\begin{enumerate}
\item\underline{$\phi$ is a rational angle:}\\
If $\phi=2\pi p/q$ with $q,p\in\mathbb{N}$ such that $\text{gcd}(q,p)=1$, then we have a discrete spectrum $\text{spec}(R^\phi)=\{e^{\i \kappa\phi}\,|\,\kappa=0,\dots,q-1\}$, and eigenvectors $v_\kappa(k)=\delta(\kappa,k)$ in the $k$-space representation which now have finite norm.
\item\underline{$\phi$ is an irrational angle:}\\
If $\phi$ is irrational, then we have a ``discrete'' spectrum $\text{spec}(R^\phi)=\overline{\{e^{\i\kappa\phi}\,|\,\kappa\in\mathbb{Z}\}}=\U(1)$, in the sense that the spectral measure is discrete since it derives from that of the momentum operator with discrete spectrum. Furthermore, the eigenvectors are again given by $v_\kappa(k)=\delta(\kappa,k)$.
\end{enumerate}

We therefore see that the difference between the two quantizations expresses itself in a rather subtle way, namely in the different nature of the spectral measure for the translation operator with irrational angle. Now that we have an understanding of these subtleties at the level of the translation operator, we can turn to the more interesting case of the area operator.

\section{Area operator}
\label{sec:area}

\noindent In this section we are going to introduce the area operator and discuss some of its properties. For this, we focus on the case with $d=3$ spatial dimensions. The length operator in $d=2$ spatial dimensions exhibits similar features, but it is more instructive to study the three-dimensional case in order to grasp all of the subtleties and to understand the role of the Barbero--Immirzi parameter.

\subsection{Definition}

\noindent Recall that the fluxes arise as the smearing of the densitized triad fields, which themselves encode the information about the intrinsic geometry, and in particular the spatial metric tensor. The area of a surface can therefore naturally expressed in terms of the fluxes. Let us fix a surface $S$, which we can think of as being built from minimal triangular surfaces spanned by piecewise geodesic arcs. It is therefore naturally a triangulated surface $\Delta_S$, and we denote its elementary triangles by $t$. The area of this triangulated surface is given by
\be\label{Classical-area}
\text{Ar}_{\Delta_S}
=\beta_\text{BI}\sum_{t\in\Delta_S}\bigg|\sum_{i=1,2,3}\bX^i_t\bX^i_t\bigg|^{1/2}
\coloneqq\sum_{t\in\Delta_S}\text{Ar}_t,
\ee
where the flux $\bX^i_t$ associated to a triangle $t\in\Delta_S$ is going to be defined below. Here we have introduced the Barbero--Immirzi (BI) parameter\footnote{
In \eqref{firstequ} we wrote the LQG variables as a connection one form $A^i_a$ and its conjugated electrical field variable $E^b_j$. The Barbero--Immirzi  parameter appears in the Ashtekar--Barbero connection through $A^i_a=\Gamma^i_a+\beta_\text{BI}K^i_a$, where $\Gamma^i_a$ is the spin connection and $K^i_a$ the extrinsic curvature (contracted with a triad). The electric field result from a rescaling by $1/\beta_\text{BI}$ of the densitized triad fields. Therefore, the geometric fluxes are given by the fluxes $\bX$ multiplied by $\beta_\text{BI}$.} $\beta_\text{BI}$, and set $8\pi G_\text{N}=1$.

Before quantizing this expression, let us collect a couple of remarks which already highlight at the classical level some key differences between the area operator in the BF representation and in the AL one.
\begin{itemize}
\item The subdivision of the surface $S$ into a set $\Delta_S$ of triangles $t$ is an essential ingredient of the definition. The choice of $\Delta_S$ provides a ``scale'' on which the area will be measured, which implies in particular that two different triangulations\footnote{Note that in this picture the triangulation is part of the definition of operator itself, and not of the state.} of $S$ may a priori lead to different results. This should not be seen as a disadvantage of the framework. On the contrary, it addresses the so-called staircase (or ``coast of Britain'') problem for fractal geometries, which is that since the measurement of a given length depends on a choice of measuring scale, for fractal-like geometries one can in principle get infinite results in the limit where the measuring scale approaches zero. In the present framework, the triangles $t$ play the role of the smallest available rulers with which one can measure the geometry. As usual in loop quantum gravity, ``smallest'' should here be understood in relative terms, since the metric properties are encoded in the states themselves. What actually matters is rather the fineness of the triangulation $\Delta_S$ as compared to the fineness of the states onto which the area operator is applied.

Since one can choose $\Delta_S$ to be much coarser than the triangulations underlying the states, the BF-based representation provides a natural way of defining coarse-grained geometric quantities. This possibility does a priori not exist in the AL representation, which leads to problems when trying to construct suitable semi-classical states approximating smooth geometries (see for instance the discussion in \cite{1308.5648}). 

Because of its behavior under refinement operations, the area operator introduced here is similar to the one proposed by Livine and Terno in \cite{gr-qc/0603008}. This latter is an alternative to the usual area operator in the AL Hilbert space \cite{RovelliSmolin,AshtekarArea}. The usual area operator \cite{RovelliSmolin,AshtekarArea} involves a sum over the Casimir operators for each spin network edge piercing the surface. Under refinement of the states, the sum is extended in order to include the Casimir operators associated to the additional edges. The alternative suggestion made in \cite{gr-qc/0603008} is to first couple the representation spaces associated to the spin network edges going through a piece of surface (which here is defined to be a triangle in $\Delta_S$) to a total angular momentum vector space. One should then considers the $\SU(2)$ Casimir with respect to this total angular momentum. Therefore, under refinement one still considers just one Casimir operator, but now for a different total angular momentum vector space arising from more recoupling steps.

The difference with our new proposal is that here we have to replace the angular momentum operators with exponentiated operators (due to the compactification of the Lie algebra of $\SU(2)$), while the behavior under refinement of states is similar. Indeed the cylindrical-consistency of the area operator (i.e. its consistent behavior under refinement) in the BF-based representation is ensured by the recoupling prescription. This follows from the discussion in section \ref{extflux}, where we define the action of  exponentiated flux operators on refined states, which we summarize in the next two points.

\item The area $\text{Ar}_{\Delta_S}$ given by \eqref{Classical-area} can be defined on all the triangulations $\Delta$ which support $\Delta_S$, i.e. in which the triangles $t$ of $\Delta_S$ arise as the possible coarse-graining of (unions of) triangles $\sigma^2\in\Delta$. Because of this, the fluxes $\bX^i_t$ have to be thought of as composed objects. One possibility is to define these as ``integrated fluxes'' associated to the triangles $t$ following the definition of \cite{fluxC} (the triangle $t$ is then to be understood as a co-path $\pi$ in \cite{fluxC}), i.e.
\be\label{Xt}
\bX_t
\coloneqq g^{-1}_{rl_1(0)}\left(\sum_{l_i\in t}g^{-1}_{l_1(0)l_i(0)}X_{l_i}g_{l_1(0)l_i(0)}\right)g_{rl_1(0)}.
\ee
Here $l_i$ labels all the links which are dual to the triangles of $\Delta$ composing $t$, the source node $l_1(0)$ of the link $l_1$ serves as a root for the surface $t$, and $g_{l_1(0)l_i(0)}$ is the parallel transport from this source node to the source node of the link $l_i$. This parallel transport takes place along a surface tree (defined in detail in \cite{fluxC}), which is as close to the surface as possible (and below the oriented surface). This means that $\bX_t$ (and therefore the area operator itself) depends on this choice of surface tree (for every triangulation $\Delta$ on which one wishes to define $\bX_t$). Note that in \eqref{Xt} we also have to consider in general fluxes associated to links that are not necessarily leaves. Such branch fluxes can however be expressed in terms of fluxes associated to the leaves, as explained in appendix \ref{appendix:branchX}.

One can come up with alternative definitions for $\bX_t$  by changing the prescription for the parallel transports. For instance, one can extend the triangulation $\Delta_S$ to a $d$-dimensional triangulation $\Delta(S)$ in such a way that each $t$ is a boundary triangle (with outward orientation) of a $d$-simplex in $\Delta(S)$. We can then allow for a parallel transport which is not necessarily as close to the surface as possible, but rather confined to be inside the coarse $d$-simplex in $\Delta(S)$ which bounds $t$. This definition corresponds to the notion of coarse flux observables which we will define in section \ref{splitting algebra}. The area operator will then depend on this additional data $\Delta(S)$, and on a definition of parallel transport in each of its simplices.

\item With the definition(s) of $\bX_t$ discussed above, we can write down a consistent area operator which can be applied to states based on different triangulations. Consistency means that the area operator will give the same result for states connected by the refinement operations which we define in section \ref{sec:refinement}. This allows for the area operator to be defined on a continuum (inductive limit) Hilbert space.

Given a triangulation $\Delta$ supporting $\Delta_S$ or $\Delta(S)$, we need to specify the parallel transport for the definition of the fluxes $\bX_t$. 
This will then give a full definition of the area operator $\text{Ar}_{\Delta_S}$ or $\text{Ar}_{\Delta(S)}$ for the triangulation $\Delta$. The area operator can also be defined for all states based on a triangulation which are coarser than $\Delta$. To this end, one just needs to refine the coarser states to states defined on $\Delta$ (in a manner defined in the second part of this work). However, the extension to states defined on triangulations $\Delta'$ finer than $\Delta$ is less trivial. The basic problem is to define a notion of parallel transport which leads to consistent results for arbitrary fine triangulations. This prescription corresponds therefore to a continuum prescription, and what one needs to specify is a method for projecting this prescription to a given triangulation $\Delta'$. We outline in section \ref{contop} a prescription for parallel transport in the continuum which can be projected in a consistent manner to triangulations. Using this method, we can define the area operator as a continuum operator.

\item Since only the exponentiated fluxes can be quantized, we have to approximate the differential operators by difference operators. Such approximations to the gauge-invariant expressions of the form $\tr(\bX_t\bX_t)$ are however not completely gauge-invariant anymore (this concerns only the gauge transformations at the root). One could therefore introduce a further change of parallel transport for the (exponentiation of the fluxes) $\bX_t$, so that the fluxes associated to the triangles $t$ are first transported to some node in $\Delta(S)$, and only then to the root. This makes the definition of the area operator more local in its dependence on data near or far away from the surface. By first transporting the fluxes along the surface, we will have a more local dependence. Data off of the surface appear only via the global parallel transport from the chosen node in $\Delta(S)$ to the root.
 \end{itemize}

Let us now discuss the quantization of the area operators. As mentioned above, we have to approximate the fluxes with difference operators. To this end, we choose a parameter $\mu>0$, along with a basis $\tau^i\in\su(2)$, and define $\mu_i^\pm\in\SU(2)$ as
\be
\mu^\pm_i
\coloneqq\exp(\pm\mu\tau^i).
\ee
We then define a quantization of the quantity $\sum_i\bX^i_l\bX^i_l$ as 
\be\label{areaelement}
\sum_i\bX^i_l\bX^i_l\ \rightarrow\ \f{-\hbar^2}{\mu^2}\sum_i\left(R^{\mu^+_i}_l+R^{\mu^-_i}_l-2\cdot\openone\right)
\ee
This quantization covers the case in which $\Delta_S$ coincides with a triangulated surface of the triangulation of $\Delta$. In this case, the fluxes $\bX_t$ appearing in the expression \eqref{Classical-area} for the area are given by the fluxes $\bX_l$. The quantized area operator is then given by
\be\label{areaoperator}
\text{Ar}^\mu_{\Delta_S}
=\f{\beta_\text{BI}\hbar}{\mu}\sum_{l\in\Delta_S}\bigg|-\sum_i\left(R^{\mu^+_i}_l+R^{\mu^-_i}_l-2\cdot\openone\right)\bigg|^{1/2}.
\ee
Before discussing the spectrum of this operator, let us gather a few remarks concerning its definition.

\begin{itemize}
\item In section \ref{contop}, we discuss the way in which the translation operators can be defined on the continuum Hilbert space. This can be applied in order to obtain a consistent continuum definition of the area operator, as discussed above for the classical expression.

\item As mentioned above, the quantized area operators are not invariant under the frame rotations at the root anymore. We therefore discuss later on in this section strategies for defining a fully gauge-invariant version of the area operator.

\item Interpretation of the parameter $\mu$: The approximation of the flux derivative operator by difference operators is similar to the approximation of the extrinsic curvature (or the momentum conjugated to the scale factor) by (point) holonomies in loop quantum cosmology (LQC) \cite{AshBojLewan}. In fact, in LQC the configuration space of the scale factor is also quantized with a discrete topology, which leads to an evolution equation which is discrete in $\mu$ steps. In the case of LQC, it is necessary to adjust the step parameter $\mu$ to the value of the scale factor \cite{improvedmu}. In our case, $\mu$ parametrizes a translation on the group, which encodes the values of the curvature defects. One could attempt\footnote{We thank Abhay Ashtekar for pointing out this possibility.} to introduce a dependence on the class angle of the group element which is being translated, which would lead to a curvature-dependent discretization step. This would in turn change the properties (and the spectrum) of the area operator. We will discuss such state-dependent $\mu$ parameters for the fully gauge-invariant area operator.

\item The Barbero--Immirzi parameter: In \eqref{areaoperator}, we have introduced the BI parameter by treating it as a factor which multiplies the difference operator. There exists however an alternative (quantization) ambiguity, which is to define the group elements by which one translates as
\be\label{altversion}
\tilde{\mu}^\pm_i
\coloneqq\exp\big(\pm\beta_\text{BI}\mu\tau^i\big),
\ee
and to omit the prefactor $\beta_\text{BI}$ in \eqref{areaoperator}. In this case (at least for $\U(1)$) the BI parameter can redefine the values of the parameter $\mu$ which lead to discrete and continuous spectra respectively. This alternative changes also the bound on the spectrum of the area operator: whereas in the version \eqref{areaoperator} it multiplies the bound, version \eqref{altversion} has a bound which is independent of the BI parameter. Also, at least in the case of $\U(1)$, if the Barbero--Immirzi parameter is included using \eqref{altversion}, the spectrum of the area operator does then not depend on this parameter as long as $\beta_\text{BI}\mu$ is irrational.
\end{itemize}

\subsection{Spectrum of the area operator}
\label{sec:spectrumAr}

\noindent Let us consider the area operator associated to one triangle $t$, and focus on a triangulation $\Delta$ where this triangle $t$ appears and is not further refined. With this choice, the triangle $t$ is dual to a link $l$, and we have that
\be\label{arqu}
\text{Ar}^\mu_{l}
=\f{\beta_\text{BI}\hbar}{\mu}\bigg|-\sum_i\left(R^{\mu^+_i}_l+R^{\mu^-_i}_l-2\cdot\openone\right)\bigg|^{1/2}.
\ee
Note that the (squared) area operator is a linear combination of bounded operators, and therefore is itself bounded. The bound does however grow with $1/\mu$. This boundedness results from the compactification of the dual of the Lie group, and is similar to the bound on curvature which appears in LQC due to the compactification of the momentum variable conjugated to the scale factor. This bound on the area is also similar to the quantum group deformation (at root of unity), where only a finite range of representations $j$ are admissible. On the other hand, it is not so clear that the spectrum of the area operator is still discrete, as it is in the AL representation. Let us now discuss the spectra for the two gauge groups of interest.

\subsubsection{Abelian group $\U(1)^3$}

\noindent We start by discussing the case $\U(1)^3$, where the issue with gauge transformations at the root does not arise. Furthermore, in this case the translation operators commute, so we can use directly the results of the spectral analysis of the translation operators themselves. Again, the properties of the area operator will depend on whether the parameter $\mu$ is rational or irrational.

\begin{enumerate}
\item\underline{$\mu$ is a rational angle:}\\
This case corresponds to $\mu\in[0,2\pi)\cap2\pi\mathbb{Q}$, which means that $\mu=2\pi p/q$ with $q,p\in\mathbb{N}$ such that $\text{gcd}(q,p)=1$. Then we have eigenvectors of the form
\be
v_{\alpha_1,\kappa_1}\otimes v_{\alpha_2,\kappa_2}\otimes v_{\alpha_3,\kappa_3},
\ee
where $v_{\alpha,\kappa}$ is given by \eqref{eigen1}. This leads to a discrete spectrum
\be
\f{\beta_\text{BI}\hbar}{\mu}\times\bigg\{\Big|\sum_i\big(2-2\cos(\kappa_i\mu)\big)\Big|^{1/2}\,\big|\,(\kappa_1,\kappa_2,\kappa_3)\in({\mathbb Z}_q)^3\bigg\}
\ee
for the $\U(1)^3$ area operator. 

\item\underline{$\mu$ is an irrational angle:}\\
This case corresponds to $\mu\in[0,2\pi)\cap2\pi(\mathbb{R}\backslash\mathbb{Q})$. We can then apply the spectral theorem on the separable invariant subspaces discussed in section \ref{spectrans}. Thus, we have generalized eigenvectors
\be
w_{\alpha_1,\rho_1}\otimes w_{\alpha_2,\rho_2}\otimes w_{\alpha_3,\rho_3},
\ee
where $w_{\alpha,\rho}$ is given by \eqref{irreigen}, and a $\rho_i$ labels an eigenvalue $e^{\i\rho}=e^{\i\kappa_i\mu}$ of the translation operator $R^{\mu_i}$. In order to make the similarity with the discrete case more transparent, we can define implicitly $\kappa_i\in\mathbb{Z}$ as the integer for which $\kappa_i\mu-\rho=0$ mod $2\pi$ holds (remember that $w_{\alpha_i,\rho_i}$ can indeed be understood as being concentrated on $k_i=\kappa_i$ in the $k$-representation). In this sense, the spectrum is given by the closure of the set
\be
\f{\beta_\text{BI}\hbar}{\mu}\times\bigg\{\Big|\sum_i\big(2-2\cos(\kappa_i\mu)\big)\Big|^{1/2}\,\big|\,(\kappa_1,\kappa_2,\kappa_3)\in({\mathbb Z}_q)^3\bigg\}.
\ee
If we choose $\mu$ as well as the $\kappa_i$ to be small, one finds a very good approximation to the usual case, i.e. spectral values $\sim\sqrt{\kappa_1^2+\kappa_2^2+\kappa_3^2}$. For growing $\kappa_i$, the bound sets in and one has an oscillatory behavior which fills densely all the allowed range, which is given by 
\be
\f{\beta_\text{BI}\hbar}{\mu}\times\big[0,\sqrt{12}\big].
\ee
\end{enumerate}

\subsubsection{Non-Abelian group $\SU(2)$}

\noindent Let us now turn to the more complicated case of the gauge group $\SU(2)$. Here, the translation operators $R^{\mu_i}$ do not commute, so we cannot apply the results from the spectral analysis of these operators straightforwardly. However, we can define as before subspaces which are left invariant by the area operator, by taking the closure of the space of states generated by applying arbitrary integer powers of $R^{\mu_i}$ to a given basis state $\psi_\alpha$.

If the parameters $\mu_i$ are elements of a finite subgroup of $\SU(2)$, this results in a finite-dimensional subspace with discrete spectrum. These are however only the subgroups corresponding to the Platonic solids, and thus no regular family of finite groups that generates rotations in all three directions.

Therefore, we can rather expect to have a continuous spectrum, at least in the sense of having generalized eigenvectors (it still could happen that there are normalizable eigenvectors, as explained below). We can nevertheless diagonalize the area operator by employing the spin representation laid out in appendix \ref{spinrepsu2}. This uses a (discrete generalization of the) group Fourier transform, so that the states are wave functions of the form $\psi(j,M,N)=\psi_{MN}(j)$, where $j$ is the spin representation label and $M,N$ are magnetic indices. The translation operators act in this representation as
\be
(R^\mu\psi)(j)
=D^j (\mu)\cdot\psi(j),
\ee 
and therefore leaves the $j$ label and one of the magnetic indices $N$ invariant. The difference with the usual case is that $\psi(j)\sim\delta_{j,j'}$ is a state with vanishing norm, since the inner product is given by
\be\label{inpL}
\la\psi_1|\psi_2\ra
=\lim_{\Lambda\rightarrow\infty}\f{1}{\mathcal{N}(\Lambda)}\sum_{j=0}^\Lambda\tr\big((\psi_1^\dagger(j)\psi_2(j)\big),
\ee
with a normalization factor $\mathcal{N}(\Lambda)\sim\Lambda^3$ (see \eqref{nfactor}). In order to consider the (so far non-existant) subspaces associated to one fixed spin $j$, we introduce a fixed cutoff $\Lambda_\text{fix}$ on the spins $j$, and define an inner product on the space of functions $\psi_{MN}(j)$ as in \eqref{inpL} but without taking the limit. The associated Hilbert spaces $\mathcal{H}(\Lambda)$ are then finite dimensional (and just correspond to (rescalings of) $\oplus_{j\leq\Lambda}V_j\otimes V_j^*$). This enables us to consider vectors of the form
\be\label{jN}
\psi^{j'N'}_{MN}(j) 
=C(\Lambda)a_M(j,N)\delta_{j,j'}\delta_{N,N'},
\ee
where $C(\Lambda)$ is a normalization factor which grows to infinity if the regulator is taken away, i.e. in the limit $\Lambda\rightarrow\infty$. 

\begin{center}
\begin{figure}[h]
\includegraphics[scale=0.5]{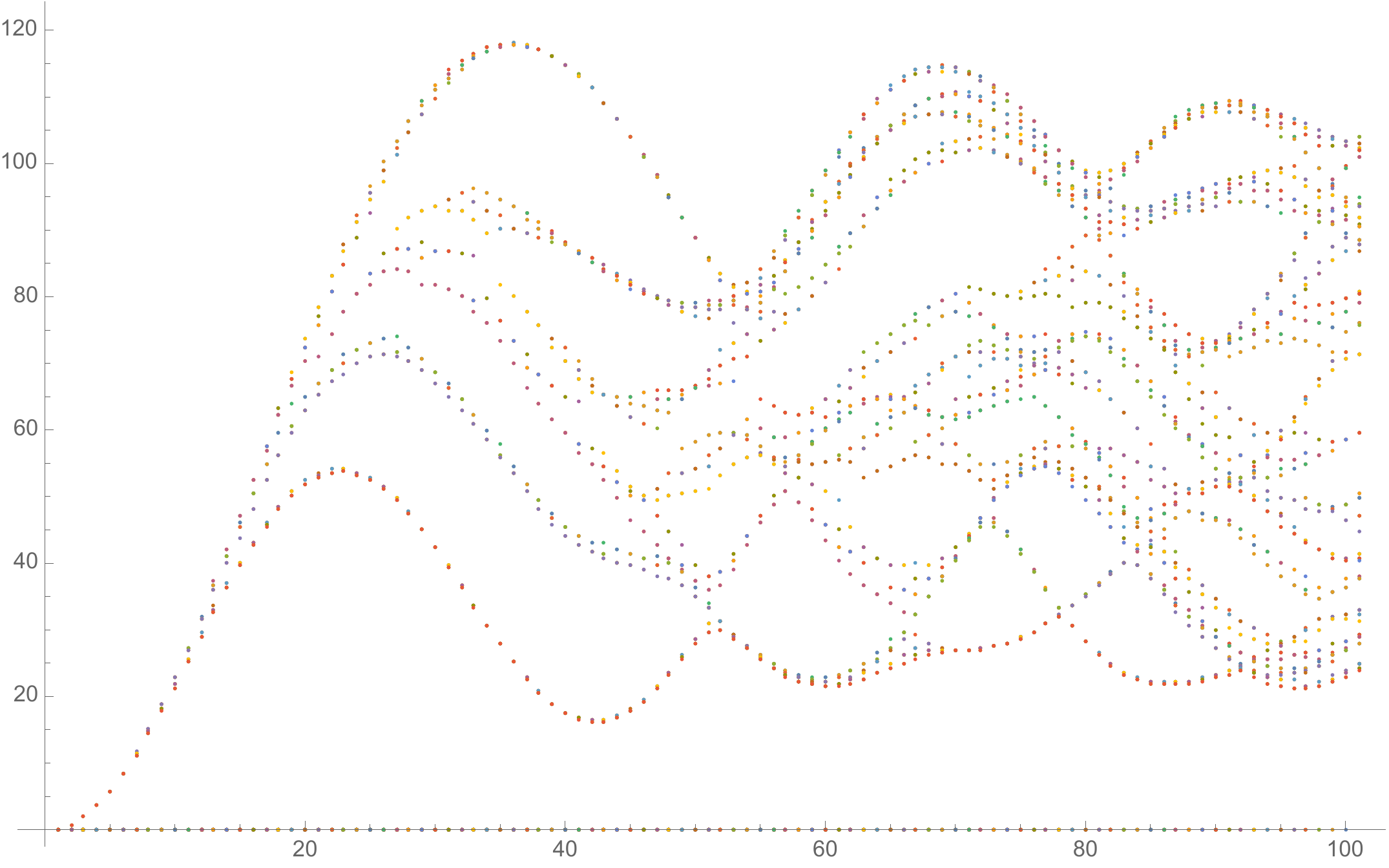}
\caption{Eigenvalues ($y$ axis) of the squared area operator as a function of the spin $j$ ($x$ axis), for $\mu=0.3$, $\beta_\text{BI}=1$, and $\hbar=1$.}
\label{fig:Area2}
\end{figure}
\end{center}

The action of the square of the area operator on a subspace spanned by the vectors \eqref{jN}, for fixed $j'=j$ and any fixed $N'=-j',\dots,+j'$, is given by the matrix\footnote{Note that this operator is not invariant under gauge transformations at the root.}
\be
A_{MN}^{j'}(\mu)
=-\f{\beta_\text{BI}^2\hbar^2}{\mu^2}\sum_i\left(D^{j'}_{MN}(\mu_i)+D^{j'}_{MN}(\mu^{-1} _i)-2\delta_{M,N}\right).
\ee
We now have to diagonalize these matrices and examine the eigenvalues of the area operator on the cutoff Hilbert space $\mathcal{H}(\Lambda)$. The general behavior of the eigenvalues is similar to the case of the gauge group $\U(1)^3$ treated above, but with an important difference: for small $\mu$ and $j$, the matrix $A^j(\mu)$ is nearly diagonal with almost equal eigenvalues, approximating well the Casimir eigenvalues $j(j+1)$. For growing spin, the approximate degeneracy of the eigenvalues is lifted, and the eigenvalues spread more and more. Moreover, the eigenvalues are bounded, and an oscillatory behavior sets in, as can be seen on figure \ref{fig:Area2}. However, it is not clear whether the oscillatory behavior will cover the entire allowed interval.

\begin{center}
\begin{figure}[h]
\includegraphics[scale=0.5]{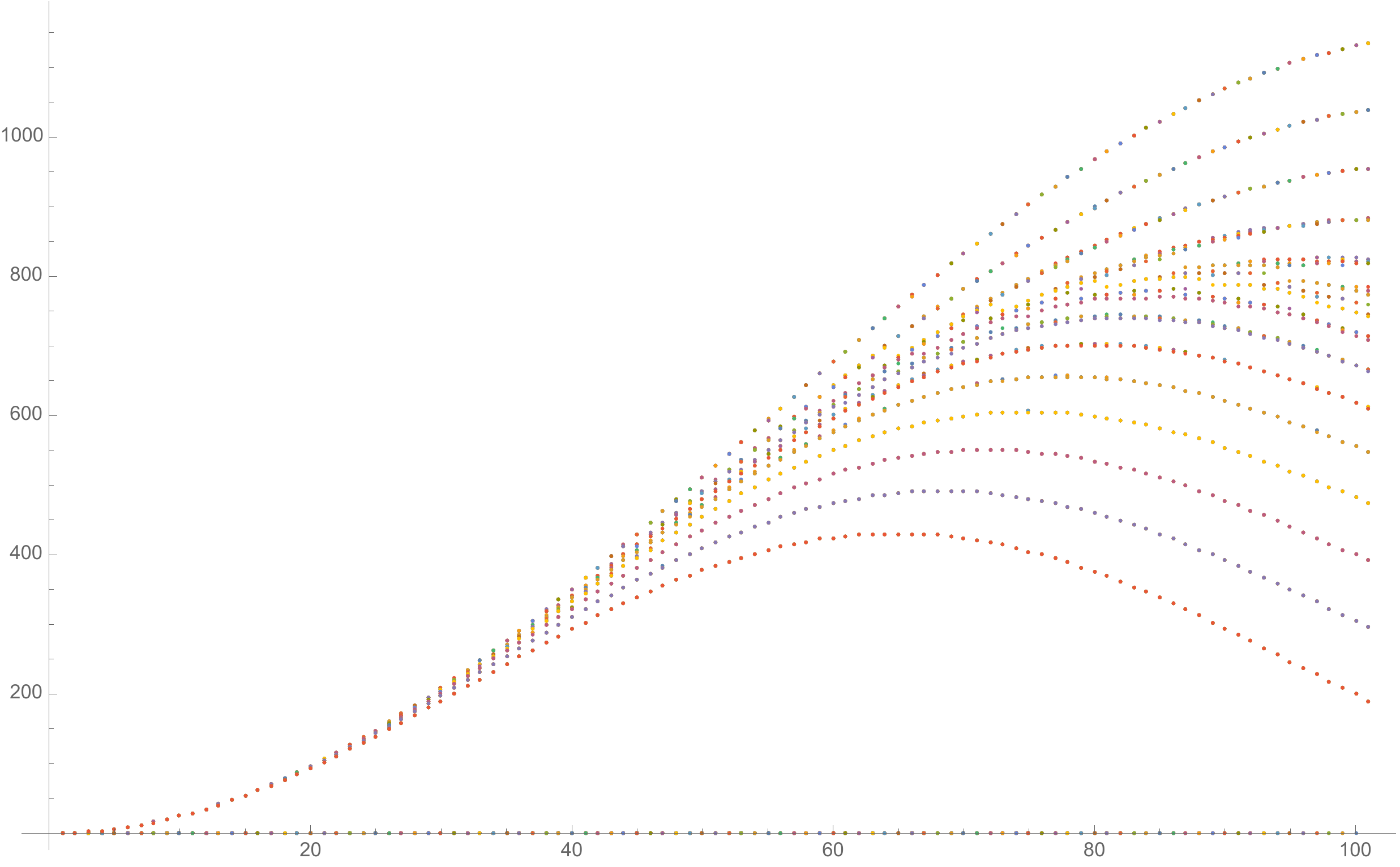}
\caption{Eigenvalues ($y$ axis) of the squared area operator as a function of the spin $j$ ($x$ axis), for $\mu=0.1$, $\beta_\text{BI}=1$, and $\hbar=1$.}
\label{fig:Area1}
\end{figure}
\end{center}

So far, we have discussed the area operator associated to one single link. Under refinement, this link can be subdivided into several links, and in this case the translation operators $R^\mu_l$ have to be replaced by a product of (possibly parallel-transported) translation operators associated to the finer links $l'$. However the essential features do not change: using the intertwiner-spin representation laid out in appendix \ref{intertwinerspin}, one can couple\footnote{This recoupling does also absorb a parallel transport for the translations. Note that the group element corresponding to this parallel transport can be considered to be fixed, since it result from links on which the translations do not act.} the representation spaces associated to the finer links into a total spin $j$. The discussion then proceeds as before, with the difference that one now has higher multiplicities for the eigenvalues caused by the recoupling procedure. As before, the spectrum is bounded by $\beta_\text{BI}\hbar\sqrt{12}/\mu$.

Moreover, we have only considered so far the case of a basic triangle appearing in the triangulation of the surface $S$ underlying the definition of the area operator for this surface. These basic area operators will usually commute with each other. In order to obtain the spectrum of the full area operator, one needs to add up the (square root) of each of these contributions of the basic triangle (squared) area operators. The bound on the spectrum, which is given by $\beta_\text{BI}\hbar\sqrt{12}/\mu$, is therefore multiplied by the number of triangles appearing in the surface $S$. Although the basic area operators are bounded, we can achieve a large area by subdividing a given area in many pieces. This exemplifies even more the ``coast of Britain" paradox: the measurement scale, which here is the subdivision of the surface into triangles, determines the maximal area of this surface.

To summarize our discussion on the spectrum of the area operator, let us emphasize that the compactification of the flux space, i.e. of the dual to the Lie group, naturally leads to a compactification of the area spectrum for the area operator associated to a basic triangle. Note that the subdivision of the surface $S$ is inherent to the definition of the are operator associated to this (triangulated) surface, and not to the triangulation on which the states are defined. Therefore, by adding up the area operators associated to many triangles, we can increase the bound on the spectrum of the composed operator. Furthermore, we have seen that it is possible to account for the Barbero--Immirzi parameter $\beta_\text{BI}$ in two different ways, which does also influence the bound on the spectrum.

Since the spectrum associated to a given basic area operator is bounded, it will wind around in the compact range of values specified by the bound $(\beta_\text{BI}\hbar/\mu)\times[0, \sqrt{12}]$. For $\U(1)$, with a rational angle $\mu$ this winding happens in a periodic way and results in a discrete spectrum, while for $\mu$ irrational the winding is ergodic and gives a continuous spectrum. For $\SU(2)$, we can also diagonalize the area operator in the spin representation. These are indeed proper eigenvalues if one introduces a cutoff Hilbert space which only admits spins smaller than or equal to $\Lambda_\text{fix}$. However, it is not clear how the spectrum changes if the cutoff is removed. The spectrum associated to small spins $j$ (and sufficiently small $\mu$) approximates the usual area spectrum $\sim\sqrt{j(j+1)}$ quite well. For larger spin values $j$ (or larger $\mu$), an oscillatory behavior appears. One can expect that this leads to a dense filling of at least a part of the space of allowed values. It will therefore be interesting to see what happens for the fully gauge-invariant area operator, which we now discuss.

\subsection{Fully gauge-invariant area operator}

\noindent The lifting of the degeneracy of the eigenvalues in figure \ref{fig:Area1} and figure \ref{fig:Area2} is due to the non-invariance of the area operator under frame rotations (i.e. gauge transformations) at the root. Let us therefore discuss possibilities for defining a fully gauge-invariant area operator.

One possible starting point is to take the action of the area operator in the spin representation, and to consider the following (normalized) trace of the area operator:
\be\label{areag}
A^j_\tr(\mu)
=\f{1}{d_j}\sum_{M}A_{MM}^{j}(\mu)
=\f{6\beta_\text{BI}^2\hbar^2}{\mu^2}\left[1-\f{1}{d_j}\f{\displaystyle\sin\left(d_j\f{\mu}{2}\right)}{\displaystyle\sin\left(\f{\mu}{2}\right)}\right].
\ee
If \eqref{areag} would give a well-defined operator (on either the kinematical Hilbert space or on the Hilbert space of fully gauge-invariant wave functions), and also if we took the limit $\Lambda_\text{fix}\rightarrow\infty$, we could read off the spectrum from the representation in \eqref{areag}. There, the oscillatory behavior of the sine function is suppressed by a factor of $1/d_j$, which leads to a discrete spectrum for sufficiently small spins $j$.

To answer the question whether \eqref{areag} can give an operator on the kinematical Hilbert space $\mathcal{H}_\Delta$, we can consider the (kernel of the) operator in the holonomy representation, namely
\be
(A_\tr^\mu\psi)(h')
=\int\de\mu_\text{d}(h)A^\mu_\tr(h',h)\psi(h),
\ee
with
\be\label{Akernel}
A^\mu_\tr(h',h)
&=\lim_{\Lambda\rightarrow\infty}\f{1}{\mathcal{N}(\Lambda)}\sum_{j\leq\Lambda}d_j\chi_j(h'h^{-1})A^j_\tr(\mu)\nn\\
&=\f{c}{\mu^2}\delta(h',h)-\f{c}{\mu^2}\lim_{\Lambda\rightarrow\infty}\f{1}{\mathcal{N}(\Lambda)}\sum_{j\leq\Lambda}\chi_j\big(\theta(h'h^{-1})\big)\chi_j\big(\theta(\mu)\big),
\ee
where $c=6\beta_\text{BI}^2\hbar^2$, and $\theta(\alpha)\in[0,\pi]$ is the class angle of the group element $\alpha$. The sum in the second term gives the (Dirac) delta function (without the factor of $1/\mathcal{N}(\Lambda)$) on the equivalence classes of the adjoint action, i.e.
\be\label{sumL}
\sum_{j\in{\mathbb{N}}/2}\chi_j(\alpha)\chi_j(\beta)
=\int\de\mu_\text{Haar}(g)\delta\big(\alpha g\beta^{-1}g^{-1}\big),
\ee
and one could in principle interpret the area operator as a translation operator acting on these equivalence classes. However, the sum \eqref{sumL} with a cutoff $\Lambda$ scales rather as
\be
\sum_{j\leq\Lambda}\chi_j( \alpha)\chi_j(\beta)
\sim\left\{\begin{array}{cl}
\Lambda&\text{if $\theta(\alpha)\approx\theta(\beta)\neq0$ or $\pi$},\\[5pt]
\Lambda^3&\text{if $\theta(\alpha)\approx\theta(\beta)=0$ or $\pi$.}
\end{array}\right.
\ee
Because of this, the kernel \eqref{Akernel} would be equal to the Kronecker delta for $h\neq\pm\openone$ (since the second term is suppressed in the limit $\Lambda\rightarrow\infty$), but would be vanishing for $h,h'=\pm\openone$. The expression \eqref{areag} does therefore not define a useful area operator on the kinematical Hilbert space. Nevertheless, it could be possible to define matrix elements in the inner product of the fully gauge-invariant Hilbert space, as defined in section \ref{sec:raq}, via the RAQ procedure. This would lead to an averaging over the orbits of the adjoint action (acting on the entry $h'$ in \eqref{Akernel}) with respect to the discrete measure on the group. The difficulty is however to exchange the sum and the limit in $\Lambda$ in \eqref{Akernel} with this discrete measure integral in a well-defined way.

In summary, the derivation of a fully gauge-invariant area operator from the kinematical version is rather involved (this difficulty is however lifted if one changes to a quantum group at root of unity, as will be discussed in section \ref{sec:discussion}). An alternative way to proceed is offered by the interpretation of the kernel \eqref{Akernel} with respect to the continuous topology: it would then define a translation operator for the class angle.

Indeed, the usual area operator is given as a Laplace operator on the group if we work with $L^2(G,\de\mu_\text{Haar})$. We could therefore consider the Laplace operator restricted to class functions, which is
\be\label{lap}
\text{L}
=\f{1}{4\sin^2 \theta}\partial_\theta(\sin^2\theta\partial_\theta),
\ee
and approximate this differential operator by a difference operator. This (discretization) procedure does however involve lots of ambiguities, and one would rather wish for a more natural connection with the area operator defined on the kinematical Hilbert space.

\begin{center}
\begin{figure}[h]
\includegraphics[scale=0.5]{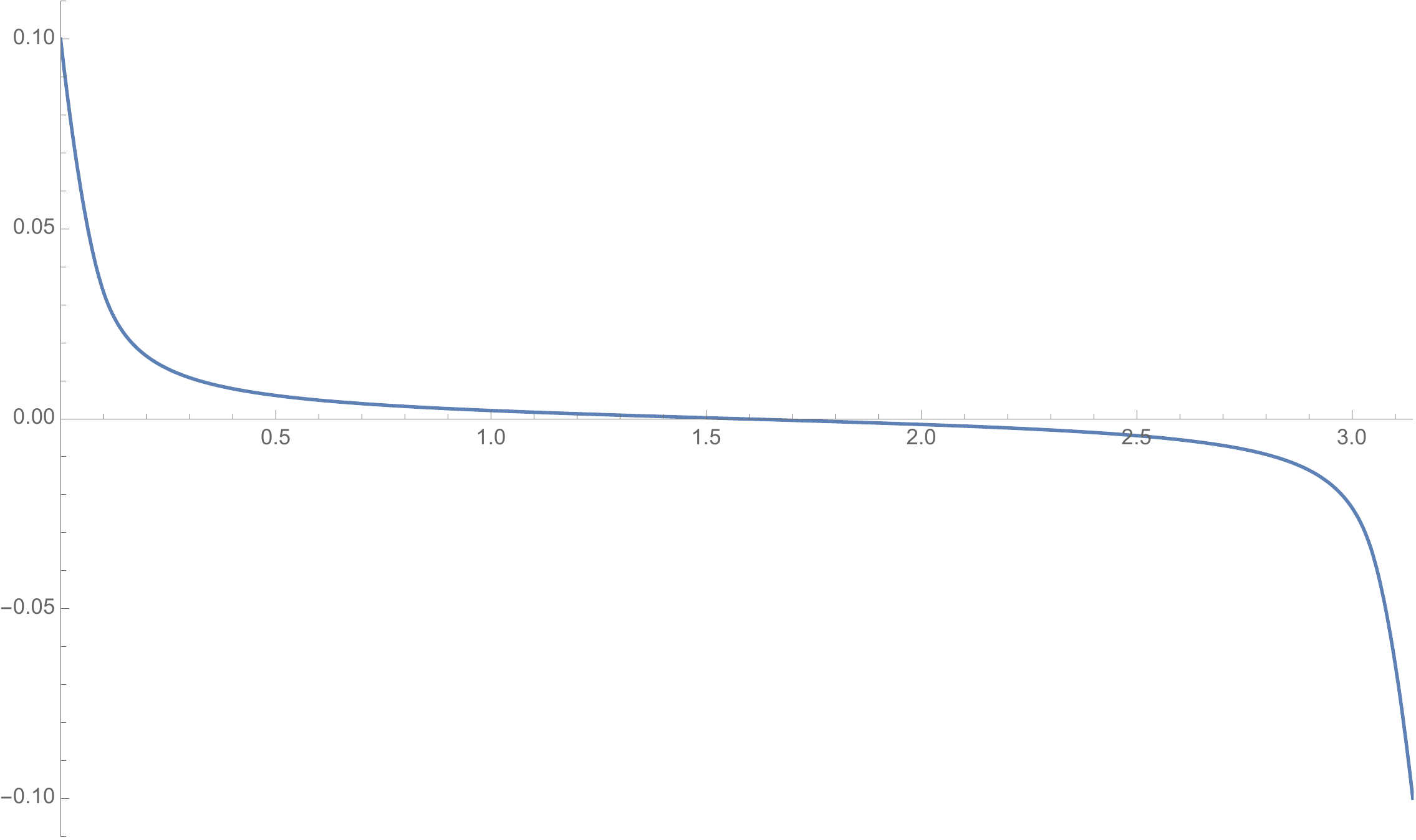}
\caption{Difference $R^\mu(\theta)-\theta$ between the translated and untranslated class angles, with translation parameter $\mu=0.2$, as a function of the untranslated class angle $\theta$.}
\label{fig:theta1}
\end{figure}
\end{center}

Let us therefore study how the translation operators $R^{\mu_i^\pm}$, which are instrumental for the area operator \eqref{arqu} on the kinematical Hilbert space, act on the class angle of the argument $g$ being translated in the wave function. For this, let us use the parametrization
\be
g=\openone\cos\theta(g)-\i\vec{n}(g)\cdot\vec{\sigma}\sin\theta(g),\q
\mu_i^\pm=\openone\cos\f{\mu}{2}\mp\i\vec{n}_i\cdot\vec{\sigma}\sin\f{\mu}{2},
\ee
where $\vec{\sigma}=(\sigma_1,\sigma_2,\sigma_3)$ are the Pauli sigma matrices, $\vec{n}(g)$ is a unit vector, and $(\vec{n}_i)_k=\delta_{ik}$. With this, the class angle of the product $g\mu_i^\pm$ is given by
\be\label{classa}
\theta(g\mu_i^\pm)
=\arccos\left(\cos\theta(g)\cos\f{\mu}{2}\mp\big(\vec{n}(g)\big)_i\sin\theta(g)\sin\f{\mu}{2}\right).
\ee
The problem is now that this class angle depends through $\vec{n}(g)$ on the full group element $g$, and not only on its class angle. To get rid of this dependency, we can integrate over $u\in G$ the adjoint action $g\rightarrow ugu^{-1}$, which happens to be equal to a rotation of the vector $\vec{n}(g)$ over the sphere $\mathbb{S}^2$. This will lead to the same averaged class angles for the product of $g$ with $\mu^+_i$ and $\mu_i^-$, and for all $i=1,2,3$. With the parametrization
\be
n_1=\sin\psi\cos\phi,\q
n_2=\cos\psi,\q
n_3=\sin\psi\sin\phi,
\ee
where $\phi\in[0,2\pi)$ and $\psi\in[0,2\pi]$, the integration is easiest to perform for $\mu_2$. Choosing the normalized invariant measure on the sphere, we obtain
\be
R^\mu\big(\theta(g)\big)
&\coloneqq\f{1}{4\pi}\int_0^{2\pi}\de\phi\int_0^{2\pi}\theta(g\mu_2^+)\sin\psi\,\de\psi\nn\\
&=\f{1}{2}\int_{-1}^{1}\arccos\left(\cos\theta(g)\cos\f{\mu}{2}-x\sin\theta(g)\sin\f{\mu}{2}\right)\de x\nn\\
&=\f{1}{\displaystyle2\sin\theta(g)\sin\f{\mu}{2}}\bigg(-\arccos\left[\cos\left(\theta(g)+\f{\mu}{2}\right)\right]\cos\left(\theta(g)+\f{\mu}{2}\right)+\sqrt{\sin^2\left(\theta(g)+\f{\mu}{2}\right)}\nn\\
&\phantom{=\f{1}{\displaystyle2\sin\theta(g)\sin\f{\mu}{2}}\bigg(}+\arccos\left[\cos\left(\theta(g)-\f{\mu}{2}\right)\right]\cos\left(\theta(g)-\f{\mu}{2}\right)-\sqrt{\sin^2\left(\theta(g)-\f{\mu}{2}\right)}\bigg).\label{theta1}
\ee 
Here one can choose the roots such that $\sqrt{\sin^2(x)}\geq0$, and furthermore 
\be
\arccos\big(\cos(x)\big)
=
\left\{\begin{array}{cl}
-x&\text{if $-\pi\leq x\leq0$,}\\[5pt]
x&\text{if $0\leq x\leq\pi$,}\\[5pt]
\pi-x&\text{if $\pi\leq x\leq2\pi$.}
\end{array}\right.
\ee
This ensures that the result \eqref{theta1} is finite for $\theta(g)=0$ (and equal to $R^\mu(0)=\mu/2$) and for $\theta(g)=\pi$ (and equal to $R^\mu(\pi)=\pi-\mu/2$), as can be seen on figure \ref{fig:theta1}.

Consider now a fully gauge-invariant Hilbert space $\mathcal{H}^u_\Delta$ as defined in section \ref{sec:raq}, with a basis of wave functions $\psi_{[\{\alpha_\ell\}]}$, and an area operator acting on a leaf $\ell$. We parametrize the orbits $[\{\alpha_\ell\}]$ under the adjoint action in such a way that the leaf $\ell$ on which the area operator is acting appears only with a class angle, i.e. $[\{\alpha_\ell\}]=\{\theta_\ell,\dots\}$. We can then define the square of the area operator as
\be\label{areainv1}
\text{Ar}_\ell^\mu\psi_{\{\theta_\ell,\dots\}}
=-\f{6\beta_\text{BI}^2\hbar^2}{\mu^2}\left(\psi_{\{R^\mu(\theta_\ell),\dots\}}-\psi_{\{\theta_\ell,\dots\}}\right).
\ee
We therefore have a translation operator acting on the class angle with a configuration-dependent translation parameter. This makes the investigation of the spectrum more complicated, and we leave this rather important question for future work. 

It should be emphasized that \eqref{areainv1} is one particular proposal for a fully gauge-invariant area operator, obtained by averaging the shift in the class angle over the orbits of the adjoint action. It is certainly possibly to define alternative versions, for instance by discretizing the action \eqref{lap} of the Laplacian on class functions.

\newpage

\section*{\large PART 2. THE CONTINUUM HILBERT SPACE}

\noindent We now turn to the second main part of this work, which concerns the construction of the continuum inductive limit Hilbert space. For this, we are going to introduce in section \ref{sec:refinement} a few technical results concerning the refinement operations. This will then enable us to establish the cylindrical-consistency of the inner product, which is required in order for the inductive limit to be constructed. We will then discuss the relationship between continuous operators and operators defined on a fixed triangulation.

\section{Refinement operations}
\label{sec:refinement}

\noindent In this section, we are going to construct refining maps which embed Hilbert spaces $\mathcal{H}_\Delta$ into Hilbert spaces $\mathcal{H}_{\Delta'}$ associated to finer triangulations. These refinement maps are needed in order to obtain the continuum Hilbert space $\mathcal{H}_\infty$ as an inductive limit from the family of Hilbert spaces $\mathcal{H}_\Delta$.

Since our description of the Hilbert spaces $\mathcal{H}_\Delta$ involves choices of trees (for describing bases for these Hilbert spaces), we are first going to discuss the notion of refined trees. This will then allow us to express the refinement maps in a compact way. It will also facilitate the expression of data on a finer phase space in terms of data on a coarser phase space, which will be used later on in order to provide extensions of the holonomy-flux observables associated to a given triangulation to a finer triangulation.

In order to help the reader, we provide here a short summary of the various notations which are used throughout the rest of this work to describe structures related to graphs, trees, and their refinements.
\begin{enumerate}
\item[]\underline{Glossary:}
\be\nn\begin{array}{ll}
\mathbf{P}&\text{map which projects paths in $\Gamma'$ to paths in $\Gamma$,}\\[2pt]
\mathbf{P}^{-1}(l)&\text{pre-image of $l$, i.e. the set of \textit{paths} in $\Gamma'$ which map under $\mathbf{P}$ to the link $l$,}\\[2pt]
\mathbf{P}^{-1}_\text{L}(l)&\text{link pre-image $l$, i.e. the set of \textit{links} in $\Gamma'$ which map under $\mathbf{P}$ to the link $l$,}\\[2pt]
\mathcal{T}&\text{tree in $\Gamma$,}\\[2pt]
\mathcal{T}'&\text{refinement in $\Gamma'$ of a tree $\mathcal{T}$ in $\Gamma$,}\\[2pt]
\mathcal{L}'_\mathcal{T}&\text{set of finer link representatives for the coarse leaves $\mathcal{L}$ in $\Gamma$,}\\[2pt]
\mathbf{P}_\text{R}&\text{restriction of $\mathbf{P}$ to  $\mathcal{L}'_\mathcal{T}$ such that $\mathbf{P}_\text{R}:\mathcal{L}'_\mathcal{T}\rightarrow\mathcal{L}$ is a bijection,}\\[2pt]
\ell'_i&\text{leaves in $\mathcal{L}'_\mathcal{T}$ with labelling induced from $\mathcal{L}$, i.e. such that $\mathbf{P}_\text{R}(\gamma_{\ell'_i})=\gamma_{\ell_i}$,}\\[2pt]
\ell'_I&\text{leaves in $\mathcal{L}'\backslash \mathcal{L}'_\mathcal{T}$.}
\end{array}
\ee
\end{enumerate}

\subsection{Projection maps for the space of paths}

\noindent Let us consider two triangulations $\Delta$ and $\Delta'$ such that $\Delta\prec\Delta'$, along with their respective dual graphs $\Gamma$ and $\Gamma'$. In \cite{fluxC}, we have described the projection maps $\mathbf{P}$ which send paths $\gamma'$ in $\Gamma'$ (which can be viewed as paths in $\Delta'$) to paths $\gamma$ in $\Gamma$ (viewed as paths in the coarser triangulation $\Delta$). Let us recall how simplices and paths in $\Delta$ and $\Delta'$ (and their dual graphs) are related by $\mathbf{P}$.

Since $\Delta'$ is a refinement of the triangulation $\Delta$, we can identify each simplex $\sigma_\Delta$ of $\Delta$ which is involved in the refinement with a union $\cup\sigma_{\Delta'}$ of simplices (of the same dimension as the simplices $\sigma_\Delta$) in the finer triangulation $\Delta'$. In other words, $\cup\sigma_{\Delta'}$ gets coarse-grained to $\sigma_\Delta$ under the inverse of the refining map. We denote this relationship by\footnote{Note that we use a slight abuse of notation here, and use the same symbol $\mathbf{P}$ to denote a map from the space of path in $\Gamma'$ to the space of path in $\Gamma$, and for denoting the relationship between simplices $\sigma_\Delta$ in $\Delta$ and the set of simplices $\mathbf{P}^{-1}(\sigma_\Delta)$ in $\Delta'$ into which the simplices $\sigma_\Delta$ are refined.} $\mathbf{P}^{-1}(\sigma_\Delta)=\cup\sigma_{\Delta'}$.

Consider now a path $\gamma'$ in $\Gamma'$. When seen as a path in $\Delta'$, this $\gamma'$ goes inside $d$-dimensional simplices and intersects some of their boundary $(d-1)$-dimensional simplices. We define the path $\gamma=\mathbf{P}(\gamma')$ in $\Delta$ in such a way that it enters and leaves a $d$-dimensional simplex $\sigma^d_\Delta$ in the same order and through the same boundary simplices $\partial\sigma^d_\Delta$ as $\gamma'$ enters and leaves the complex $\mathbf{P}^{-1}(\sigma^d_\Delta)$ with respect to the boundary complexes $\mathbf{P}^{-1}(\partial\sigma^d_\Delta)$.

An important property which follows from this definition of the projection maps $\mathbf{P}$ is their transitivity. This can be summarized in the following result (here we denote by $\mathbf{P}_{\Delta',\Delta}$ the projection map from $\Delta'$ to $\Delta$):

\begin{Lemma}[Transitivity of the projection maps]\label{lemma:Ptransitivity}
Given three triangulations $\Delta\prec\Delta'\prec\Delta''$ and the corresponding projection maps $\mathbf{P}_{\Delta',\Delta}$, $\mathbf{P}_{\Delta'',\Delta'}$, and $\mathbf{P}_{\Delta'',\Delta}$, then we have that
\be\label{Ptransitivity}
\mathbf{P}_{\Delta',\Delta}\mathbf{P}_{\Delta'',\Delta'}
=\mathbf{P}_{\Delta'',\Delta}.
\ee
\end{Lemma}

Every link $l\in\Gamma$, whether it is a leaf or a branch of a tree, has a pre-image $\mathbf{P}^{-1}(l)$ in the set of paths in $\Gamma'$ whose nature depends on the $d$-dimensional simplices dual to the end nodes $l(0)$ and $l(1)$. If the $d$-dimensional simplices dual to the end nodes of $l$ are not involved in the refinement of the triangulation, then the pre-image $\mathbf{P}^{-1}(l)$ consists of a single link $l'$. If at least one of the $d$-dimensional simplices dual to the end nodes of $l$ is being refined, then the pre-image $\mathbf{P}^{-1}(l)$ consists of a set of paths (which can a priori differ in their source and/or target nodes) and therefore contains several links.


In general, $\mathbf{P}^{-1}(l)$ will include paths composed out of several links of the finer triangulation. It will turn out to be useful to isolate the paths which include only one link. For this ,we therefore introduce the notation $\mathbf{P}_\text{L}^{-1}(l)$ for the ``link pre-image'', i.e. the subset of $\mathbf{P}^{-1}(l)$ containing all the links $l'$ such that $\mathbf{P}(l')=l$. This link pre-image will be essential for discussing the refinement of flux observables, which are based on so-called co-paths (a set of connected edges in $d=2$ and a set of connected triangles in $d=3$). Basically, $\mathbf{P}_\text{L}^{-1}(l)\cup\mathbf{P}_\text{L}^{-1}(l^{-1})$ contains all the links which are dual to the $(d-1)$-dimensional simplices appearing in the refinement of the $(d-1)$-dimensional simplex dual to the link $l$.


We are now going to use these projections of paths to characterize refined trees as, roughly speaking, trees whose fundamental cycles can be projected to the fundamental cycles determined by the coarser tree.

\subsection{Refined trees}
\label{refinementtrees}

\noindent Consider a triangulation $\Delta$, its dual graph $\Gamma$, and a finer triangulation $\Delta'$ obtained from $\Delta$ by Alexander moves. Given a tree $\mathcal{T}$ in $\Gamma$, it turns out that not all the trees $\mathcal{T}'\subset\Gamma'$ will be useful in our construction, not even if they satisfy the condition $\mathbf{P}(\mathcal{T}')=\mathcal{T}$. This is illustrated with some examples in appendix \ref{appendix:trees}. In what follows, we will therefore characterize the trees $\mathcal{T}'\subset\Gamma'$ which arise as ``natural extensions'' of $\mathcal{T}$, and call them refined trees.

\begin{Definition}[Refined tree]\label{def:refine-tree}
Consider two triangulations such that $\Delta\prec\Delta'$, and a tree $\mathcal{T}$ in the graph dual to the coarser triangulation $\Delta$. We say that a tree $\mathcal{T}'$ is a refinement of a coarser tree $\mathcal{T}$ if the following two conditions are satisfied:
\begin{itemize}
\item[$\mathrm{(\text{$i$})}$] All the paths along $\mathcal{T}'$ are mapped under $\mathbf{P}$ to paths along $\mathcal{T}$, which we write as $\mathbf{P}(\mathcal{T}')=\mathcal{T}$.
\item[$\mathrm{(\text{$ii$})}$] For every $d$-dimensional simplex $\sigma^d_\Delta\in\Delta$ being refined when going to the finer triangulation $\Delta'$, the nodes dual to any two simplices in the union $\cup\sigma^d_{\Delta'}=\mathbf{P}^{-1}(\sigma^d_\Delta)$ can be connected by a path in $\mathcal{T}'$ without ever crossing a $(d-1)$-dimensional simplex in the boundary $\mathbf{P}^{-1}(\partial\sigma^d_\Delta)$.
\end{itemize}
From these two conditions, it follows in fact that
\begin{itemize}
\item[$\mathrm{(\text{$ii'$})}$] \textit{There exists a subset $\mathcal{L}'_{\text{\tiny{$\mathcal{T}$}}}\subset\mathcal{L}'$ of the set of leaves of $\mathcal{T}'$ such that all the fundamental cycles determined by $\mathcal{L}'_{\text{\tiny{$\mathcal{T}$}}}$ can be mapped bijectively to the fundamental cycles determined by $\mathcal{L}$.}
\end{itemize}
\end{Definition}

One can select such a set $\mathcal{L}'_{\text{\tiny{$\mathcal{T}$}}}$ in the following way: For a given leaf $\ell$ in $\mathcal{L}$, consider the corresponding dual simplex $\sigma^{d-1}_\Delta(\ell)$ and the subdivision of this simplex under refinement, i.e. the set $\mathbf{P}^{-1}\big(\sigma^{d-1}_\Delta(\ell)\big)$ of simplices. Because of condition $(ii)$, all the simplices in this set are dual to leaves with respect to the refined tree $\mathcal{T}'$. One can then choose one of the leaves $\ell'$ dual to one of these simplices to be in $\mathcal{L}'_{\text{\tiny{$\mathcal{T}$}}}$ and to ``represent'' the leave $\ell$. By making this choice for all the leaves $\ell\in\mathcal{L}$, we determine the set $\mathcal{L}'_{\text{\tiny{$\mathcal{T}$}}}$.

Note that this choice specifies additional information on top of the refined tree, and in the following we implicitly assume that with the refined tree we have also chosen such a subset. We also adopt the convention according to which the orientations of the leaves $\ell'\in\mathcal{L}'_{\text{\tiny{$\mathcal{T}$}}}$ coincide with the orientations of the corresponding leaves in $\mathcal{L}$. Therefore, if necessary, we need to invert the orientations of some of the leaves $\ell'$.

One can replace condition $(ii)$ with $(ii)'$, which results in a weaker notion of refined tree, as illustrated with the examples of appendix \ref{appendix:trees}. This weaker notion still allows for the identification of constraints in section \ref{constraints}. However, condition $(ii)$ makes the splitting of (in particular the flux) observables, into coarse and fine ones (detailed in section \ref{splitting algebra}) much easier. 

There is a way to explicitly construct refined trees, which we will now explain. The construction follows two steps, and requires some choices along the way.
\begin{enumerate}
\item For each $d$-dimensional simplex $\sigma_\Delta^d$ in $\Delta$, consider the collection of $d$-dimensional simplices which build up $\sigma_\Delta^d$, i.e. the set $\mathbf{P}^{-1}(\sigma_\Delta^d)$. Choose a maximal tree dual to this piece of the triangulation, i.e. a set of edges connecting the simplices $\sigma_{\Delta'}^d\subset \mathbf{P}^{-1}(\sigma_\Delta^d)$, which meets every vertex, but which has no closed loops. By doing this for all the $d$-dimensional simplices in $\Delta$, one gets a (non-necessarily maximal) tree in $\Gamma'$.
\item For each branch in $\mathcal{T}$, take the dual $(d-1)$-dimensional simplex $\sigma_\Delta^{d-1}$ in $\Delta$, and consider the collection of $(d-1)$-dimensional simplices in $\Delta'$ which build up this simplex, i.e. $\mathbf{P}^{-1}(\sigma_\Delta^{d-1})$. Choose one of these simplices, and add its dual link to the list of branches in $\mathcal{T}'$. By doing this for all the branches in $\mathcal{T}$, one ends up with a maximal tree $\mathcal{T}'$ in $\Delta'$.
\end{enumerate}

It is easy to check that the properties $(i)$ and $(ii)$ of definition \ref{def:refine-tree} are satisfied for the resulting maximal tree $\mathcal{T}'$. In fact, it can be shown that every refined tree is the result of such a construction process. A proof of this fact is given in appendix \ref{app:Construction}.

In the following section, we are going to use different subdivisions of the set of leaves $\mathcal{L}'$ of the refined tree. We now summarize what these various leaves are.
\begin{itemize}
\item One subdivision of $\mathcal{L}'$ is into the representatives $\mathcal{L}'_{\mathcal{T}}$ for the coarse leaves and the remaining leaves $\mathcal{L}'\backslash \mathcal{L}'_{\mathcal{T}}$. This corresponds to a subdivision of the holonomy degrees of freedom into coarse holonomies and finer holonomies. The latter are constrained for configurations which arise from a refining procedure.
\item Furthermore, for each coarse leave $\ell$, we have a subset $\mathcal{L}'\supset\mathcal{S}(\ell)=\mathbf{P}^{-1}_\text{L}(\ell)\cup\mathbf{P}^{-1}_\text{L} (\ell^{-1})$ of leaves. The leaves in a given set $\mathcal{S}(\ell)$ give all the links which are dual to simplices $\sigma_{\Delta'}^{d-1}\subset\mathbf{P}^{-1}\big(\sigma^d_\Delta(\ell)\big)$ which compose the coarse simplex $\sigma^d_\Delta(\ell)$ dual to $\ell$. For a leave $\ell' \in\mathbf{P}^{-1}_\text{L} (\ell^{\pm1})$ we have $\mathbf{P}(\gamma_{\ell'})=\gamma^{\pm1}_\ell$.
\item The remaining leaves, i.e. the ones in $\mathcal{L}'\backslash\cup_\ell\mathcal{S}(\ell)$, are either links ``inside'' the coarser simplices $\sigma^d_\Delta$, or are part of $\mathcal{S}(b)\coloneqq\mathbf{P}^{-1}_\text{L}(b)\cup\mathbf{P}^{-1}_\text{L}(b^{-1})$ (which is not a subset of $ \mathcal{L}'$ since it contains exactly one branch $b'$). The links in $\mathcal{S}(\ell)$ cross a coarser $(d-1)$-dimensional simplex $\sigma^{d-1}_\Delta(b)$ which is dual to a branch $b$ of the tree $\mathcal{T}$. For all the leaves $\ell'\in\mathcal{L}'\backslash\cup_\ell \mathcal{S}(\ell)$ we have that $\mathbf{P}(\gamma_{\ell'})=r$, i.e. the corresponding cycles are mapped to the trivial cycles.
\end{itemize}

Finally, let us conclude this subsection by mentioning an additional important property satisfied by the notion of refined trees, which is that of transitivity. This can be summarized in the following lemma, which we prove in appendix \ref{proof:Ttransitivity}:

\begin{Lemma}[Transitivity of the refined trees]\label{lemma:Ttransitivity}
Let us consider three triangulations $\Delta\prec\Delta'\prec\Delta''$. Let $\mathcal{T}'$ be a refined tree for $\mathcal{T}$, and $\mathcal{T}''$ be a refined tree for $\mathcal{T}'$. Then we have that $\mathcal{T}''$ is also a refined tree for $\mathcal{T}$.
\end{Lemma}

Now that we have obtained criteria for characterizing refined trees and shown that such trees can always be constructed, we are going to use this to derive the constraints satisfied by the finer cycles of the refined graph. These constraints will in turn enable us to define the refinement maps, and then to separate the holonomy-flux algebra on the finer phase space into two mutually-commuting subalgebras.

\subsection{Constraints on the finer cycles}
\label{constraints}

\noindent With our definition of refined trees, we had to choose a subset $\mathcal{L}'_{\text{\tiny{$\mathcal{T}$}}}$ of ``coarse'' leaves in $\Gamma'$ which are representatives for the leaves in $\Gamma$. This means that if we index the leaves of $\mathcal{L}$ with small Latin letters $i,j,\dots$, then for each $\ell_i\in\mathcal{L}$ there exists a unique leaf $\ell'\in\mathcal{L}'_{\text{\tiny{$\mathcal{T}$}}}$ such that
\be\label{Pell'}
\gamma_{\ell_i}
=\mathbf{P}(\gamma_{\ell'}).
\ee 
We will denote this unique leave $\ell'\in\mathcal{L}'_{\text{\tiny{$\mathcal{T}$}}}$ which maps under $\mathbf{P}$ to $\ell_i$ as $\ell'_i$. Therefore, ${\mathbf{P}}$ restricted to $\mathcal{L}'_{\text{\tiny{$\mathcal{T}$}}}$ is a bijection. We will use the symbol ${\mathbf{P}}_\text{R}$ for the corresponding map, so that $\mathbf{P}^{-1}_\mathrm{R}(\gamma_{\ell_i})=\gamma_{\ell'_i}$.  

For the remaining leaves in $\mathcal{L}'\backslash\mathcal{L}'_{\text{\tiny{$\mathcal{T}$}}}$, there can \textit{a priori} be three possibilities. The corresponding cycles can be mapped under $\mathbf{P}$ to $(a)$ the trivial cycle of $\Gamma$ (i.e. to the root), $(b)$ to a fundamental cycle $\gamma_{\ell_i}$ or $(b')$ to the inverse $\gamma^{-1}_{\ell_i}$ of a fundamental cycle, or $(c)$ to an arbitrary non-fundamental cycle (which can in turn be described as a combination of fundamental cycles). Explicitly, if we index the leaves of $\mathcal{L}'\backslash\mathcal{L}'_{\text{\tiny{$\mathcal{T}$}}}$ with capital Latin letters $I,J,\dots$, for each $\ell'_I\in\mathcal{L}'\backslash\mathcal{L}'_{\text{\tiny{$\mathcal{T}$}}}$ we can therefore define\footnote{We define $\mathbf{P}^{-1}_\text{R}(r)=r'$, where $r$ and $r'$ stand for the trivial cycles (i.e. the roots) in $\Gamma$ and $\Gamma'$ respectivly.}
\be\label{Well'}
W_I\{\gamma_{\ell'_i}\}
\coloneqq\mathbf{P}^{-1}_\text{R}\big(\mathbf{P}(\gamma_{\ell'_I})\big)
\ee
as a word in the set $\{\gamma_{\ell'_i}\}$ of fundamental cycles determined by the leaves $\ell'_i$. However, one can show that with the definition \ref{def:refine-tree} of a refined tree option $(c)$ does not occur. This is summarized in the following result:

\begin{Lemma}\label{lemma:words}
A refined tree, together with a separation of its leaves into leaves $\ell'_i\in\mathcal{L}'_{\text{\tiny{$\mathcal{T}$}}}$ and $\ell'_I\in\mathcal{L}'\backslash\mathcal{L}'_{\text{\tiny{$\mathcal{T}$}}}$, is such that the words $W_I\{\gamma_{\ell'_i}\}\coloneqq\mathbf{P}^{-1}_\mathrm{R}\big(\mathbf{P}(\gamma_{\ell'_I})\big)$ are either trivial or one-letter words in the cycles $\gamma_{\ell'_i}$ determined by the leaves in $\mathcal{L}'_{\text{\tiny{$\mathcal{T}$}}}$ (or their inverses).
\end{Lemma}

This result is tight to condition $(ii)$ in the definition of the refined trees. Without this condition, one can obtain in \eqref{Well'} words of several letters in the cycles $\gamma_{\ell'_i}$. The proof of this lemma follows from the discussion in the previous subsection, about the partition of the refined leaves into different subsets. If the leaf $\ell'_I$ leads to the case $(a)$, we have $W_I\{\gamma_{\ell'_i}\}=r'$. In this case, the leave $\ell'_I$ will not appear in any of the sets $\mathbf{P}^{-1}_\text{L}(\ell_i)$ or $\mathbf{P}^{-1}_\text{L}(\ell^{-1}_i)$. Such a leave $\ell'_I$ does not cross any of the $(d-2)$-dimensional simplices in $\Delta'$ which coarse grain to the $(d-2)$-dimensional simplices in $\Delta$. If the leave $\ell'_I$ leads to the cases $(b)$ or $(b')$, then there exists a unique $\ell_i$ such that either $(b)$ $\ell'_I=\mathbf{P}^{-1}_\text{L}(\ell_i)$ or $(b')$ $\ell'_I=\mathbf{P}^{-1}_\text{L}(\ell^{-1}_i)$. Accordingly, we have $W_I\{\gamma_{\ell'_i}\}=\gamma_{\ell_i}$ or $W_I\{\gamma_{\ell'_i}\}=\gamma^{-1}_{\ell_i}$. 

Let us now discuss how to obtain the constraints on the finer cycles of the refined tree. From equation \eqref{Well'}, it follows that we have the composition of paths $\big(\mathbf{P}(\gamma_{\ell'_I})\big)^{-1}\circ\mathbf{P}\big(W_I\{\gamma_{\ell'_i}\}\big)=r$, where $r$ is the trivial path (i.e. the root). In terms of holonomies $g_{\ell'}$ along the paths $\gamma_{\ell'}$, this means that we have found a set of independent constraints given by\footnote{More precisely, we should choose coordinates on the group in order to express these constraints.}
\be\label{constraints1}
\mathcal{C}_I\coloneqq W_I\{g_{\ell'_i}\}g^{-1}_{\ell'_I}
\stackrel{!}{=}\openone.
\ee

Let us now show that these constraints are sufficient in order to assign trivial holonomies to all cycles of $\Gamma'$ which are mapped by $\mathbf{P}$ to trivial cycles. Consider a non-fundamental cycle in $\Gamma'$ which is mapped by $\mathbf{P}$ to the trivial cycle (i.e. to the root $r$). Any arbitrary cycle in $\Gamma'$ can be described by a word $W$ in the fundamental cycles, which are given by the union set $\{\gamma_{\ell'_i}\}\cup\{\gamma_{\ell'_I}\}$.  We can therefore write that
\be\label{step1}
\mathbf{P}\big(W\{\gamma_{\ell'_i}\,;\gamma_{\ell'_I}\}\big)
=r.
\ee
We therefore have to show that the constraints \eqref{constraints1} imply that $W\{ g_{\ell'_i}\,;g_{\ell'_I}\}=\openone$.

Since $\mathbf{P}$ is a homeomorphism and $\mathbf{P}_\text{R}$ is a bijection, using $\mathbf{P}(\gamma_{\ell'_i})=\gamma_{\ell_i}$, equation \eqref{step1} is equivalent to
\be
W\big\{\mathbf{P}(\gamma_{\ell'_i})\,;\mathbf{P}(\gamma_{\ell'_I})\big\}
=W\big\{\gamma_{\ell_i}\,;\mathbf{P}_\text{R}\big(\mathbf{P}_\text{R}^{-1}\big(\mathbf{P}(\gamma_{\ell'_I})\big)\big)\big\}
=r.
\ee
Using $\mathbf{P}^{-1}_\text{R}\big(\mathbf{P}(\gamma_{\ell'_I})\big)=W_I\{\gamma_{\ell'_i}\}$ and again $\mathbf{P}(\gamma_{\ell'_i})=\gamma_{\ell_i}$ we can now rewrite this equation into
\be\label{trivialW}
W\big\{\gamma_{\ell_i}\,;\mathbf{P}_\text{R}\big(W_I\{\gamma_{\ell'_i}\}\big)\big\}
=W\big\{\gamma_{\ell_i}\,;W_I\{\mathbf{P}_\text{R}(\gamma_{\ell'_i})\}\big\}
=W\big\{\gamma_{\ell_i}\,;W_I\{\gamma_{\ell_i}\}\big\}
=r.
\ee
The last relation is written in terms of the fundamental cycles of $\Gamma$, which generate a free group. The word $W\big\{\cdot\,;W_I\{\cdot\}\big\}$ on the right-hand side of \eqref{trivialW} can therefore be reduced to the identity by applying successively the concatenation of paths $\gamma\circ\gamma^{-1}=r$. In terms of holonomies, the constraints \eqref{constraints1} imply that $g_{\ell'_I}=W_I\{g_{\ell'_i}\}$. By using these relations in the word $W\{g_{\ell'_i}\,;g_{\ell'_I}\}$ and by following the same steps of reductions (using $gg^{-1}=\openone$) as for the cycles, we can show that the word  $W\{g_{\ell'_i}\,;g_{\ell'_I}\}$ indeed reduces to the unit element of the group.

\subsection{Splitting of the observable algebra}
\label{splitting algebra}

\noindent The notion of refined tree allows us to define a splitting of the algebra of observables on a finer triangulation, into a set of coarser observables which can be mapped to the observable algebra on a coarser triangulation, and into a set of finer observables. This split will preserve the symplectic structure and can therefore be used to define the coarse-graining of spin network states (see for instance \cite{eteradeformed,howmany,clement2}). This circumvents a problem noted in \cite{eteradeformed,fluxC} and coined ``curvature-induced torsion'', namely the fact that if one coarse-grains an entire region into a (spin network) node, the fluxes associated to the links adjacent to this note do not necessarily need to satisfy the closure (or Gau\ss) constraint anymore. This problem is avoided here by making the splitting of the algebra at the (almost) gauge-invariant level. Here we use a tree to determine the (almost) gauge-invariant algebra of observables. Therefore, every region which is coarse-grained into a node is entered by at least one tree. The flux associated to this tree is however not part of the observable algebra, so one can thus not keep track of the closure constraint after coarse-graining.

A splitting of the algebra of observables underlies also the framework of \cite{lanery1,lanery2,lanery3,lanery4} aimed at enlarging the LQG state space. However, \cite{lanery4} leaves open the treatment of the Gau\ss~constraints. It was noted there that this requires indeed to work with parallel-transported fluxes. Here we do provide a full description of the corresponding algebra and a splitting of this algebra into finer and coarser observables.

Let us recall the rooted holonomy-flux algebra $\mathfrak{A}^r$, given by holonomies $g_\ell\coloneqq g_{r\ell(1)}^{-1}h_\ell g_{r\ell(0)}$ and  fluxes $\bX_\ell\coloneqq g_{r\ell(0)}^{-1}X_\ell g_{r\ell(0)}$.  Here $g_{r\ell(0)}$ is the holonomy along the tree going from the root to the source node of the leaf $\ell$. The Poisson brackets between these variables are given by (for $G=\SU(2)$)
\be
\lb\bX^i_\ell,\bX^j_{\ell'}\rb=\delta_{\ell,\ell'}{\eps^{ij}}_k\bX^k_\ell,\q
\lb\bX^k_\ell,g_{\ell'}\rb=\delta_{\ell,\ell'}g_\ell\tau^k-\delta_{\ell^{-1},\ell'}\tau^kg_{\ell'},\q
\lb g_\ell,g_{\ell'}\rb=0,
\ee
where the $\tau^k$ denote the generators of $\su(2)$.

We are going to show that, when considering a refined triangulation $\Delta'\succ\Delta$, the algebra of (almost) gauge-invariant observables can be split into two mutually commuting (with respect to the symplectic structure) algebras $\mathfrak{A}^r_\text{c}$ and $\mathfrak{A}^r_\text{f}$, which correspond respectively to coarse and fine observables.

Let us choose a refined tree in the graph dual to the refined triangulation and, following the notation introduced above, denote by $\mathbf{P}_\text{L}^{-1}(\ell)$ the set of oriented leaves $\ell'\in\mathcal{L}'$ which satisfy the condition $\mathbf{P}(\gamma_{\ell'})=\gamma_{\ell}$. Furthermore, let us denote by $\mathbf{P}_\text{L}^{-1}(\ell^{-1})$ the set of oriented leaves in $\mathcal{L}'$ which satisfy $\mathbf{P}\big(\gamma^{-1}_{\ell'}\big)=\gamma_\ell$. The set $\mathbf{P}_\text{L}^{-1}(\ell)\cup\mathbf{P}_\text{L}^{-1}(\ell^{-1})$ therefore contains all the links dual to the fine $(d-1)$-dimensional simplices of $\Delta'$ which coarse-grain to the coarser $(d-1)$-dimensional simplex of $\Delta$ dual to $\ell$.

With this notation, we can now define on the phase space $\mathcal{M}_{\Delta'}$ the refined extension\footnote{There is here a subtle difference in notation. $\bX'_\ell$ is an extension of $\bX_\ell$, which is different from $\bX_{\ell'}$.} $\bX'_{\ell_i}$ of a flux observable $\bX_{\ell_i}$ as
\be\label{fluxrefined}
\bX'_{\ell_i}
\coloneqq\sum_{\ell'\in\mathbf{P}_\text{L}^{-1}(\ell_i)}\bX_{\ell'}+\sum_{\ell'\in\mathbf{P}_\text{L}^{-1}(\ell_i^{-1})}\bX_{(\ell')^{-1}}.
\ee
This definition is a priori not the same as that of the integrated fluxes $\bX_\pi$ introduced in \cite{fluxC} for co-paths $\pi$ of $(d-1)$-dimensional simplices in $\Delta$. For the fluxes $\bX_\pi$, the parallel transport of the elementary fluxes (associated to the $(d-1)$-dimensional simplices building up $\pi$) is defined through the notion of a shadow tree in the shadow graph of the co-path $\pi$, and therefore stays as close as possible to $\pi$. Here however, the parallel transport of the fluxes used to define \eqref{fluxrefined} is along the sub-tree which remains inside a set $\cup\sigma^d_{\Delta'}=\mathbf{P}^{-1}(\sigma^d_\Delta)$ of $d$-dimensional simplices. Fortunately, the constraints impose the flatness of the parallel transport inside the coarse simplices. One can thus replace the parallel transport for the integrated flux \eqref{fluxrefined} by a parallel transport along a surface tree without changing its action on refined states (classically and on the constraint hypersurface). For the sake of obtaining a splitting into mutually commuting subalgebras, we will stick with the definition \eqref{fluxrefined}. The changes in the parallel transport described above might lead to commutators that vanish (only) on the constraint hypersurface. 

We are now ready to split the holonomies and fluxes on $\mathcal{M}_{\Delta'}$ into coarse and fine observables. For this, remember that with our definition of refined trees there is a natural separation of the set of leaves of $\mathcal{T}'$ into two subsets $\mathcal{L}'_{\text{\tiny{$\mathcal{T}$}}}$ and $\mathcal{L}'\backslash \mathcal{L}'_{\text{\tiny{$\mathcal{T}$}}}$. The decomposition of the holonomy-flux algebra on $\mathcal{M}_{\Delta'}$ consists in assigning different observables for the leaves in these subsets. Explicitly, the coarse observables in $\mathfrak{A}^r_\text{c}$ are given by
\be\label{HFcoarse}
\left(g_{\ell'_i},\bX'_{\ell_i}\right),\q\forall\ \ell'_i\in\mathcal{L}'_{\text{\tiny{$\mathcal{T}$}}},
\ee
with the fluxes $\bX'_{\ell_i}$ defined as in \eqref{fluxrefined}. Depending on the type of leaf $\ell'_I\in\mathcal{L}'\backslash\mathcal{L}'_{\text{\tiny{$\mathcal{T}$}}}$, the fine observables in $\mathfrak{A}^r_\text{f}$ are given by
\begin{subequations}\label{HFfine}
\begin{alignat}{2}
&\left(\mathcal{C}_I=g_{\ell'_i}g_{\ell'_I}^{-1},\,\bX_{(\ell'_I)^{-1}}\right),\q&&
\text{if}\ \exists\ \ell'_i\in\mathcal{L}'_{\text{\tiny{$\mathcal{T}$}}}\ \text{with}\ \ell'_I\in\mathbf{P}_\text{L}^{-1}(\ell_i),\label{Hfine1}\\
&\left(\mathcal{C}_{I^{-1}}=g_{\ell'_i}g_{\ell'_I},\,\bX_{\ell'_I}\right),\q&&
\text{if}\ \exists\ \ell'_i\in\mathcal{L}'_{\text{\tiny{$\mathcal{T}$}}}\ \text{with}\ \ell'_I\in\mathbf{P}_\text{L}^{-1}(\ell_i^{-1}),\label{Hfine2}\\
&\left(\mathcal{C}_I=g_{\ell'_I}^{-1},\,\bX_{(\ell'_I)^{-1}}\right),\q&&
\text{if}\ \nexists\ \ell'_i\in\mathcal{L}'_{\text{\tiny{$\mathcal{T}$}}}\ \text{with}\ \ell'_I\in\mathbf{P}_\text{L}^{-1}(\ell_i^{\pm1}).\label{Hfine3}
\end{alignat}
\end{subequations}
This separation of the holonomies follows from the constraints \eqref{constraints1} and the result of lemma \ref{lemma:words}. With this separation, we now have the following result, which we prove in appendix \ref{proof:separation}:

\begin{Lemma}\label{lemma:separation}
The coarse observables \eqref{HFcoarse} and the fine observables \eqref{HFfine} form two mutually-commuting subalgebras of the rooted holonomy-flux algebra $\mathfrak{A}^r$, which we denote respectively by $\mathfrak{A}^r_\mathrm{c}$ and $\mathfrak{A}^r_\mathrm{f}$.
\end{Lemma}

\subsection{Refinement maps and tree-independence}

\noindent We can now introduce the refinement maps $\mathcal{R}_{\Delta,\Delta'}$ acting on the Hilbert spaces $\mathcal{H}_\Delta$. We have seen that given a tree $\mathcal{T}\subset\Gamma$ and a refined tree $\mathcal{T}'\subset\Gamma'$, there is a subdivision of the set $\mathcal{L}'$ of leaves of $\mathcal{T}'$ into the two disjoint sets $\mathcal{L}'_{\text{\tiny{$\mathcal{T}$}}}$ and $\mathcal{L}'\backslash\mathcal{L}'_{\text{\tiny{$\mathcal{T}$}}}$. The leaves in the first set are labelled with small Latin indices, while that of the second set carry capital Latin indices. For every leave $\ell'_I\in\mathcal{L}'\backslash\mathcal{L}'_{\text{\tiny{$\mathcal{T}$}}}$, we have a constraint 
\be\label{constraints2}
\mathcal{C}_I
\coloneqq W_I\{g_{\ell'_i}\}g^{-1}_{\ell'_I}
\stackrel{!}{=}\openone,
\ee
as introduced in \eqref{constraints1}. This constraint can be used to to define the refinement map in the following way:

\begin{Definition}\label{def:refinement}
For a state $\psi\{g_{\ell_i}\}\in\mathcal{H}_\Delta$, we define the action of the refinement map as 
\be\label{refined}
\mathcal{R}_{\Delta,\Delta'}(\psi)\{g_{\ell'_i}\,;g_{\ell'_I}\}
\coloneqq\psi'\{g_{\ell'_i}\,;g_{\ell'_I}\}
=\psi\{g_{\ell'_i}\}\prod_I\delta\left(W_I\{g_{\ell'_i}\}g^{-1}_{\ell'_I},\openone\right).
\ee
\end{Definition}

This definition means that in order to obtain the refined wave function $\psi'$, we first have to replace the variables $g_{\ell_i}$ with the variables $g_{\ell'_i}$ associated to the leaves $\ell'_i $ in the set $\mathcal{L}'_\mathcal{T}$ (this set resulted from a choice of one representative leave in the finer triangulation $\Delta'$ per leave $\ell_i$ of the coarser triangulation $\Delta$), and then multiply the state with the (discrete measure) delta functions peaked on the constraints $\mathcal{C}_I $. One can equivalently replace any of the constraints $\mathcal{C}_I=W_I\{g_{\ell'_i}\}g^{-1}_{\ell'_I}$ in the argument of the delta functions in \eqref{refA} with constraints $\mathcal{C}_{I^{-1}}=(W_I\{g_{\ell'_i}\})^{-1}g_{\ell'_I}$. In this way, we match the splitting of the observable algebra associated to $\Delta'$ into coarser and finer observables as described in section \ref{splitting algebra}.

\subsubsection{Tree-independence}

\noindent We are now going to address an important point, which is the gauge-independence of the refinement maps. In the definition \eqref{refined}, the state and its refinement are expressed with respect to a particular gauge-fixing, i.e. a choice of tree, a choice of refined tree, and a choice $\mathcal{L}'_{\text{\tiny{$\mathcal{T}$}}}$ for the set of coarse leave representatives. We are going to prove that our construction is independent from this gauge information, i.e. that gauge transformations commute with the refinement operations.


Let us consider a change $\mathcal{T}\rightarrow\widetilde{\mathcal{T}}$ of tree, a change $\mathcal{T}'\rightarrow\widetilde{\mathcal{T}}'$ of refined tree, and with it a change $\mathcal{L}'_{\mathcal{T}}\rightarrow\mathcal{L}'_{ \widetilde{\mathcal{T}}}$ of the set of coarse leave representatives. This leads to a different basis for the fundamental cycles. If we denote the two different basis and their splitting into coarse and fine cycles by $\{\gamma_{\ell'_i}\,;\gamma_{\ell'_I}\}$ and $\{\gamma_{\tilde{\ell}'_j}\,;\gamma_{\tilde{\ell}'_J}\}$, the change of basis can be written as
\be
\gamma_{\ell'_i}=W_{\ell'_i}\{\gamma_{\tilde{\ell}'_j}\,;\gamma_{\tilde{\ell}'_J}\},\q
\gamma_{\ell'_I}=W_{\ell'_I}\{\gamma_{\tilde{\ell}'_j}\,;\gamma_{\tilde{\ell}'_J}\}.
\ee
Therefore as discussed in section \ref{change of tree}, the associated gauge transformation of the refined state \eqref{refined} is given by
\be\label{refrans}
\mathbf{G}(\psi')\{g_{\tilde{\ell}'_j}\,;g_{\tilde{\ell}'_J}\}
=\psi\big\{W_{\ell'_i}\{g_{\tilde{\ell}'_j}\,;g_{\tilde{\ell}'_J}\}\big\}\prod_I\delta\left(W_I\big\{W_{\ell'_i}\{g_{\tilde{\ell}'_j}\,;g_{\tilde{\ell}'_J}\}\big\}W^{-1}_{\ell'_I}\{g_{\tilde{\ell}'_j}\,;g_{\tilde{\ell}'_J}\},\openone\right).
\ee
Our task is to show that this state is equal to the state which we would obtain by applying first a gauge transformation $\mathcal{T}\rightarrow\widetilde{\mathcal{T}}$ and then the refinement operation. We are going to study the two terms of this expression separately.

Let us first consider the delta function factors. These give only a non-vanishing value equal to $\openone$ to holonomy configurations which satisfy the constraints \eqref{constraints1}. These constraints can be expressed with respect to different choices of trees and sets $\mathcal{L}'_{\text{\tiny{$\mathcal{T}$}}}$. Let us denote the constraints in the cycle basis $\gamma_{\tilde{\ell}}$ by
\be
\widetilde{\mathcal{C}}_J
=\widetilde{W}_J\{g_{\tilde{\ell}'_j}\}g^{-1}_{\tilde{\ell}'_J}.
\ee
As shown below \eqref{constraints1}, these constraints are independent and sufficient to determine the constraint hypersurface on which every ``finer'' cycle carries a trivial holonomy. This property does not depend on the choice of cycle basis being used, and therefore we can write that
\be
\prod_I\delta\left(W_I\big\{W_{\ell'_i}\{g_{\tilde{\ell}'_j}\,;g_{\tilde{\ell}'_J}\}\big\}W^{-1}_{\ell'_I}\{g_{\tilde{\ell}'_j}\,;g_{\tilde{\ell}'_J}\},\openone\right)
=\prod_J\delta\left(\widetilde{W}_J\{g_{\tilde{\ell}'_j}\}g^{-1}_{\tilde{\ell}'_J},\openone\right).
\ee

Let us now focus on the first factor on the right-hand side of \eqref{refrans}. First, notice that due to the delta functions we can proceed to the following replacement:
\be
\psi\big\{W_{\ell'_i}\{g_{\tilde{\ell}'_j}\,;g_{\tilde{\ell}'_J}\}\big\}\ \rightarrow\ \psi\big\{W_{\ell'_i}\big\{g_{\tilde{\ell}'_j}\,;\widetilde{W}_J\{g_{\tilde{\ell}'_j}\}\big\}\big\}.
\ee
Now, the argument of the state on the right-hand side also gives the transformation law for the fundamental cycles in $\Gamma$, which can be obtained by applying $\mathbf{P}$ to the fundamental cycles in $\Gamma$'. This means that
\be
\gamma_{\ell_i'}
=W_{\ell'_i}\big\{\gamma_{\tilde{\ell}'_j}\,;\widetilde{W}_J\{\gamma_{\tilde{\ell}'_j}\}\big\}
\ee
implies for the cycles $\gamma_{\ell_i}$ and $\gamma_{\tilde{\ell}_i}$ with respect to the trees $\mathcal{T}$ and $\widetilde{\mathcal{T}}$ respectively the relation
\be
\gamma_{\ell_i}
=W_{\ell_i}\big\{\gamma_{\tilde{\ell}_j}\,;\widetilde{W}_J\{\gamma_{\tilde{\ell}_j}\}\big\}
=W_{\ell_i}\{\gamma_{\tilde{\ell}_j}\}.
\ee
Therefore, we can finally write that
\be
\mathbf{G}(\psi')\{g_{\tilde{\ell}'_j}\,;g_{\tilde{\ell}'_J}\}
=\psi\big\{W_{\ell_i}\{g_{\tilde{\ell}'_j}\}\big\}\prod_J\delta\left(\widetilde{W}_J\{g_{\tilde{\ell}'_j}\}g^{-1}_{\tilde{\ell}'_J},\openone\right),
\ee
and the right-hand side corresponds to the state which would have been obtained by applying first a gauge transformation $\mathcal{T}\rightarrow\widetilde{\mathcal{T}}$ in the coarse triangulation and then refining using a refined tree $\widetilde{\mathcal{T}}'$ and a choice of set $\mathcal{L}'_{\widetilde{\mathcal{T}}}$. This shows that the refinement commutes with gauge transformations, and that the refinement operation is independent\footnote{More explicitly, due to the independence and the sufficiency of the constraints with respect to every allowed choice of set $\mathcal{L}'_{\text{\tiny{$\mathcal{T}$}}}$, the delta function factor in \eqref{refined} leads to the same set of configurations on which the refined state does not vanish. Second, a different choice for the set of coarse leave representatives will also replace some leaves $\ell'_i$ by $\ell'_I$ in the argument of $\psi$ in \eqref{refined}. However, any $\ell'_I$ that replaces a $\ell'_i$ needs to satisfy $\mathbf{P}(\gamma_{\ell'_I})=\gamma_{\ell_i}$. Due to the delta function factors we can therefore replace back $g_{\ell'_I}$ with $g_{\ell'_i}$.} from the particular choice of coarse leave representatives.
 

\subsubsection{Transitivity}

\noindent We now prove another very important property of the refinement maps, which is that they satisfy a transitivity condition. This is needed in order for the inductive limit construction of the continuum Hilbert space to be possible. We have the following result:

\begin{Lemma}[Transitivity of the embedding maps]\label{lemma:transitivity}
The embedding maps $\mathcal{R}_{\Delta,\Delta'}:\mathcal{H}_\Delta\rightarrow\mathcal{H}_{\Delta'}$ defined in \eqref{refined} satisfy
\be
\mathcal{R}_{\Delta',\Delta''}\circ\mathcal{R}_{\Delta,\Delta'}
=\mathcal{R}_{\Delta,\Delta''}.
\ee
\end{Lemma}

In order to prove this result, it is sufficient to show that it holds for the basis vectors \eqref{basisvectors}, and for a choice of tree in $\Gamma$ and of refined tree in $\Gamma'$. The trees define an orthonormal basis \eqref{basisvectors} for the two Hilbert spaces $\mathcal{H}_\Delta$ and $\mathcal{H}_{\Delta'}$ respectively. In terms of these bases, the refinement operation \eqref{refined} is given by
\be\label{Eq:TransitivityOfRefinements}
\mathcal{R}_{\Delta,\Delta'}\psi_{\{\alpha_\ell\}}\
=\psi_{\{\alpha'_{\ell'}\}},
\ee
where $\{\alpha'_{\ell'}\}$ is determined as follows:
\begin{subequations}\label{labelref}
\begin{alignat}{2}
\alpha_{\ell'}'&=\openone\q&&\text{if $\mathbf{P}_{\Delta',\Delta}(\gamma_{\ell'})=r$,}\\[5pt]
\alpha_{\ell'}'&=\alpha_\ell\q&&\text{if $\mathbf{P}_{\Delta',\Delta}(\gamma_{\ell'})=\gamma_\ell$},\\[5pt]
\alpha_{\ell'}'&=\alpha^{-1}_\ell\q&&\text{if $\mathbf{P}_{\Delta',\Delta}(\gamma_{\ell'})=\gamma_\ell^{-1}$.}
\end{alignat}
\end{subequations}
Now, let us consider three triangulations $\Delta\preceq\Delta'\preceq\Delta''$, a tree $\mathcal{T}$, a refined tree $\mathcal{T}'$ for $\mathcal{T}$, and a refined tree $\mathcal{T}''$ for $\mathcal{T}'$. By virtue of lemma \ref{lemma:Ttransitivity}, $\mathcal{T}''$ is then also a refined tree for $\mathcal{T}$. Therefore, we have that $\mathcal{R}_{\Delta,\Delta''}\psi_{\{\alpha_\ell\}}=\psi_{\{{\alpha''_{\ell''}}\}}$ with
\begin{subequations}
\begin{alignat}{2}
\alpha_{\ell''}''&=\openone\q&&\text{if $\mathbf{P}_{\Delta'',\Delta}(\gamma_{\ell''})=r$,}\\[5pt]
\alpha_{\ell''}''&=\alpha_\ell\q&&\text{if $\mathbf{P}_{\Delta'',\Delta}(\gamma_{\ell''})=\gamma_\ell$},\\[5pt]
\alpha_{\ell''}''&=\alpha^{-1}_\ell\q&&\text{if $\mathbf{P}_{\Delta'',\Delta}(\gamma_{\ell''})=\gamma_\ell^{-1}$.}
\end{alignat}
\end{subequations}
A similar statement holds for $\mathcal{R}_{\Delta',\Delta''}$. By employing the transitivity of the projection maps $\mathbf{P}$ described in lemma \ref{lemma:Ptransitivity}, this shows the transitivity of the refinement maps $\mathcal{R}$. \qed

\section{Inductive limit Hilbert space and cylindrical-consistency of the inner product}
\label{sec:inductive limit}

\noindent With the refinement maps at hand, we can finally define the inductive limit Hilbert space $\mathcal{H}_\infty$. Such an inductive limit construction is based on a partially-ordered directed set, which is here provided by the set of triangulations $\{\Delta,\prec\}$. The refinement maps \eqref{refined}) satisfy 
\begin{subequations}
\begin{alignat}{2}
&(i)\q&&\mathcal{R}_{\Delta,\Delta}=\text{id}_\Delta\q\forall\ \Delta,\\
&(ii)\q&&\mathcal{R}_{\Delta',\Delta''}\circ\mathcal{R}_{\Delta,\Delta'}=\mathcal{R}_{\Delta,\Delta''}\q\forall\ \Delta\prec\Delta'\prec\Delta''.
\end{alignat}
\end{subequations}
The inductive limit is defined as the following disjoint union on which an equivalence relation is imposed:
\be
\mathcal{H}_\infty=\overline{\cup_\Delta\mathcal{H}_\Delta/\sim},
\ee
where $\psi_\Delta\in\mathcal{H}_\Delta$ and $\psi_{\Delta'}\in\mathcal{H}_{\Delta'}$ are equivalent, i.e. $\psi_\Delta\sim\psi_{\Delta'}$, if there exists a triangulation $\Delta''$ such that $\mathcal{R}_{\Delta,\Delta''}(\psi_\Delta)=\mathcal{R}_{\Delta',\Delta''}(\psi_{\Delta'})$. In words, the two states are equivalent if they become eventually equal under refinement.

The inductive limit carries a linear vector space structure which is inherited from the vector space structure on $\mathcal{H}_\Delta$ together with the linearity of the refinement maps. The inductive limit can also inherit the inner product from the Hilbert spaces $\mathcal{H}_\Delta$. For this, we need the refinement maps to be isometric, i.e. to satisfy the condition
\be\label{isometry}
\langle\mathcal{R}_{\Delta,\Delta'}(\psi_1)|\mathcal{R}_{\Delta,\Delta'}(\psi_2)\rangle_{\Delta'}
=\langle\psi'_1|\psi'_2\rangle_{\Delta'}=\langle\psi_1|\psi_2\rangle_\Delta,
\ee
which is also known as cylindrical-consistency for the inner product. This condition makes the inner product well-defined on the equivalence classes $[\psi]$ of states defined above. In order to compute the inner product between two states $\psi_\Delta\in\mathcal{H}_\Delta$ and $\psi_{\Delta'}\in\mathcal{H}_{\Delta'}$, one chooses a common refining triangulation $\Delta''$ and takes the inner product between the states $\mathcal{R}_{\Delta,\Delta''}(\psi_\Delta)$ and $\mathcal{R}_{\Delta',\Delta''}(\psi_{\Delta'})$ in $\mathcal{H}_{\Delta''}$. The condition \eqref{isometry} ensures that the result does not depend on the choice of the common refining triangulation $\Delta''$.

Let us therefore show that the refinement maps are indeed isometric. It is sufficient to show this for a set of basis states. In \eqref{Eq:TransitivityOfRefinements}, we have considered the following refinement of the basis states:
\be\label{Eq:TransitivityOfRefinements2}
\mathcal{R}_{\Delta,\Delta'}\psi_{\{\alpha_\ell\}}\
=\psi_{\{\alpha'_{\ell'}\}},
\ee
where $\{\alpha'_{\ell'}\}$ is determined by
\begin{subequations}\label{labelref}
\begin{alignat}{2}
\alpha_{\ell'}'&=\openone\q&&\text{if $\mathbf{P}_{\Delta',\Delta}(\gamma_{\ell'})=r$,}\\[5pt]
\alpha_{\ell'}'&=\alpha_\ell\q&&\text{if $\mathbf{P}_{\Delta',\Delta}(\gamma_{\ell'})=\gamma_\ell$},\\[5pt]
\alpha_{\ell'}'&=\alpha^{-1}_\ell\q&&\text{if $\mathbf{P}_{\Delta',\Delta}(\gamma_{\ell'})=\gamma_\ell^{-1}$.}
\end{alignat}
\end{subequations}
 

This refinement operation does indeed leave the inner product \eqref{Eq:HSInnerProduct} between basis states invariant in the sense that
\be\label{innerproductc}
\langle\psi_{\{\alpha_\ell\}}|\psi_{\{\beta_\ell\}}\rangle_\Delta
=\prod_\ell\delta(\alpha_\ell,\beta_\ell)
=\langle\mathcal{R}_{\Delta,\Delta'}\psi_{\{\alpha_\ell\}}|\mathcal{R}_{\Delta,\Delta'}\psi_{\{\beta_\ell\}}\rangle_{\Delta'},
\ee
where we have made explicit the triangulation on which the inner product is computed. This gives an inner product which is well-defined (with respect to the equivalence relation) on the inductive limit Hilbert space $\mathcal{H}_\infty$.

In equation \eqref{measureBF} of appendix \ref{spinrepsu2}, we have defined a measure functional $\mu_\text{BF}$ which reproduces the inner product \eqref{innerproductc}. This measure is also well-defined on the continuum Hilbert space $\mathcal{H}_\infty$. It can be expressed via its action on basis states as
\be
\mu_\text{BF}(\psi_{\{\alpha_\ell\}})
=\prod_{\ell}\delta(\alpha_\ell,\openone).
\ee

Finally, let us note that the construction of the fully gauge-invariant states via the RAQ procedure in section \ref{RAQpro} can also be extended to the continuum Hilbert space. These fully gauge-invariant states are then linear functionals on (a dense subspace of) the kinematical Hilbert space $\mathcal{H}_\infty$. The basic argument why this construction is cylindrical consistent is that the stabilizer group associated to a label $\{\alpha_\ell\}\in G^{|\ell|}$ for a basis state $\psi_{\{\alpha_\ell\}}$ does not change if this state is refined. This follows from the action of the refinement on the labels which is detailed in \eqref{labelref}: an $|\ell|$-tuple $\{\alpha_\ell\}\in G^{|\ell|}$ gets extended to an $|\ell'|$-tuple $\{\alpha'_{\ell'}\}\in G^{|\ell'|}$ by $(i)$ including entries equal to the identity group element, $(ii)$ copying an entry $\alpha$ a number of times, and $(iii)$ copying the inverse of an entry $\alpha^{-1}$ a number of times. Therefore, any $u\in G$ which leaves the label $\{\alpha_\ell\}$ invariant under the adjoint action does also leave $\{\alpha'_{\ell'}\}$ invariant.

\section{Extension of basic operators}
\label{sec:extop}

\noindent In addition to having a well-defined inner product on the inductive limit Hilbert space $\mathcal{H}_\infty$, we would also like to have well-defined operators $\mathcal{O}$ acting on $\mathcal{H}_\infty$ (here and in what follows we will denote operators without using hats on top of the symbols). 

Assume that we have at our disposal a continuum operator $\mathcal{O}$, i.e. a prescription for how $\mathcal{O}$ acts on states $\psi\in\mathcal{H}_\Delta$ for any triangulation $\Delta$ (possibly by mapping these states to finer Hilbert spaces). The condition of cylindrical-consistency for such an operator $\mathcal{O}$ is then
\be\label{ccO}
\mathcal{O}\circ\mathcal{R}_{\Delta,\Delta'}
=\mathcal{R}_{\Delta, \Delta'}\circ\mathcal{O}.
\ee
This condition guarantees that applying the operator to states in the same equivalence class $[\psi]\in\mathcal{H}_\infty$ gives the same result. We will comment more in section \ref{contop} on how to obtain such continuum operators.

Here, we are rather going to focus on the following question: Given an operator $\mathcal{O}_\Delta:\mathcal{H}_\Delta\rightarrow\mathcal{H}_\Delta$ and a finer triangulation $\Delta'$, we would like to define an extension $\mathcal{E}_{\Delta,\Delta'}(\mathcal{O}_{\Delta}):\mathcal{H}_{\Delta'}\rightarrow\mathcal{H}_{\Delta'}$ which acts on the finer Hilbert space $\mathcal{H}_{\Delta'}$ and satisfies
\be\label{consistency O}
\mathcal{E}_{\Delta,\Delta'}(\mathcal{O}_\Delta)\circ\mathcal{R}_{\Delta,\Delta'}
=\mathcal{R}_{\Delta,\Delta'}\circ\mathcal{O}_\Delta.
\ee
One could hope that starting with a given $\mathcal{O}_\Delta$ and extending it to all possible finer triangulations, one will end up with a description of a continuum operator $\mathcal{O}$. However, the difficulty is that the extension maps which we are going to define are in general not transitive. Transitivity can nevertheless be obtained if one has a consistent way to project a continuum operator onto the different $\mathcal{H}_\Delta$. We will discuss this option in section \ref{contop}. For this discussion it will be helpful to first understand how an operator on a fixed triangulation can be extended to a finer triangulation. This is going to be the topic of the present section.

An operator $\mathcal{O}_\Delta$ determines the properties of any extension $\mathcal{E}_{\Delta,\Delta'}(\mathcal{O}_\Delta)$ on the subspace of states which are cylindrical consistent over $\Delta$ (i.e. states which can be obtained via refinements from $\mathcal{H}_\Delta$). For instance, the spectrum of $\mathcal{O}_\Delta$ coincides with the spectrum of $\mathcal{E}_{\Delta,\Delta'}(\mathcal{O}_\Delta)$ restricted to this subspace. For a discussion of the notion of (extended) observables in the classical theory, and the relation to symplectic reductions, we refer the reader to \cite{fluxC,FGZ}.

Let us recall how the refinement maps are defined. For a state $\psi\{g_{\ell_i}\}$, the refinement map acts as
\be\label{refA}
\mathcal{R}_{\Delta,\Delta'}(\psi)\{g_{\ell'_i}\,;g_{\ell'_I}\}
\coloneqq\psi'\{g_{\ell'_i}\,;g_{\ell'_I}\}
=\psi\{g_{\ell'_i}\}\prod_I\delta\left(W_I\{g_{\ell'_i}\}g^{-1}_{\ell'_I},\openone\right).
\ee
In words, this means that we first replace the variables $g_{\ell_i}$ associated to the leaves $\ell_i\in\mathcal{L}$ with the variables $g_{\ell'_i}$ associated to the leaves $\ell'_i$ in the set $\mathcal{L}'_\mathcal{T}$, and then multipliy the state with the (discrete measure) delta functions peaked on the constraints $\mathcal{C}_I$.

We are now going to discuss separately the extension of the holonomies, of the fluxes, and of the parallel-transported fluxes.

\subsection{Holonomies}

\noindent Let us start the discussion with a holonomy operator $f\{g_{\ell_i}\}$. We define (one version of) an extension of this holonomy operator as
\be\label{exthol}
\mathcal{E}_{\Delta,\Delta'}(f\{g_{\ell_i}\})
\coloneqq f\{ g_{\ell'_i}\}.
\ee
This definition means that we just replace the arguments $g_{\ell_i}$ in $f$ with $g_{\ell'_i}$ (with $i=1,\dots,|\ell|$), and then see the resulting $f$ as a function on $G^{|\ell'|}$ which is constant in all arguments $g_{\ell'_I}$ (with $I=1,\dots,|\ell'|-|\ell|$) which parametrize the finer holonomies.

It is straightforward to see that this refinement map satisfies the extension property \eqref{consistency O}. Indeed, we have that
\be\label{refinedhol}
\mathcal{R}_{\Delta,\Delta'}\big(f\psi\big)\{g_{\ell'_i}\,;g_{\ell'_I}\}
&=f\{g_{\ell'_i}\}\psi\{g_{\ell'_i}\}\prod_I\delta\big(\mathcal{C}_{I^{\pm1}}\{g_{\ell'}\},\openone\big)\nn\\
&=\mathcal{E}_{\Delta,\Delta'}(f)\mathcal{R}_{\Delta,\Delta'}(\psi)\{g_{\ell'_i}\,;g_{\ell'_I}\}.
\ee

The prescription \eqref{exthol} for extending holonomy operators is not the only possible one. For example, the following changes do not affect the action of $\mathcal{E}_{\Delta,\Delta'}(f)$ on refined states, and therefore do not affect the extension property \eqref{refinedhol}:
\begin{enumerate}
\item One can multiply $\mathcal{E}_{\Delta,\Delta'}(f)$ with any other function on $G^{|\ell'|}$ which evaluates to the identity on the constraint hypersurface.
\item Any of the arguments $g_{\ell'_i}$ in the extended holonomy operator $\mathcal{E}_{\Delta,\Delta'}(f\{ g_{\ell_i}\})$ can be replaced by $g_{\ell'_i}$ multiplied with any word in the constraints $\mathcal{C}_I$ and $\mathcal{C}_{I^{-1}}$ (or their inverses), i.e. one can proceed to the replacement
\be
g_{\ell'_i}\ \rightarrow\ g_{\ell'_i}W\{\mathcal{C}_I\,;\mathcal{C}_{I^{-1}}\}.
\ee
\end{enumerate}
This second modification amounts to deforming the path underlying the holonomies. The original and deformed path must however be in the same homotopy class of paths on $\Sigma\backslash\Delta_{(d-2)}$. This freedom in changing paths is essential for the construction of continuum observables as discussed in section \ref{contop}.

\subsection{Fluxes}
\label{extflux}

\noindent As explained in section \ref{sec:Qfluxes}, we obtain right translation operators $R^\sigma_\ell$ as quantizations of the exponentiated (flow of the) fluxes $\bX_\ell$, and left translations as quantizations of the exponentiated (flow of the) fluxes $\bX_{\ell^{-1}}$.

In section \ref{splitting algebra}, we have defined one possible extension for the fluxes, when seen as classical phase space functions. Based on the choice of refined tree $\mathcal{T}'$ in $\Delta'$, the extended flux is given by
\be\label{fluxrefined2}
\bX'_{\ell_i}
\coloneqq\sum_{\ell'\in\mathbf{P}_\text{L}^{-1}(\ell_i)}\bX_{\ell'}+\sum_{\ell'\in\mathbf{P}_\text{L}^{-1}(\ell_i^{-1})}\bX_{(\ell')^{-1}}.
\ee
Accordingly, we can define an extension for the operators $R^\sigma_\ell$ as
\be
\mathcal{E}_{\Delta,\Delta'}(R^\sigma_\ell)
\coloneqq\prod_{\ell'\in\mathbf{P}_\text{L}^{-1}(\ell_i)}R^\sigma_{\ell'}\times\prod_{\ell'\in\mathbf{P}_\text{L}^{-1}(\ell_i^{-1})}L^\sigma_{\ell'}.
\ee

In order to see that this definition satisfies the extension property \eqref{consistency O}, first note that $\mathcal{E}_{\Delta,\Delta'}(R^\sigma_\ell)$ leaves the (Kronecker) delta functions in the constraints invariant, i.e.
\be\label{cflux1}
\mathcal{E}_{\Delta,\Delta'}(R^\sigma_\ell)\delta(\mathcal{C}_I,\openone)
=\delta(\mathcal{C}_I,\openone).
\ee
This relation follows at the classical level from the fact that the fluxes $\bX'_{\ell_i}$ Poisson-commute with either $\mathcal{C}_I$ or $\mathcal{C}_{I^{-1}}$ (they commute with both versions weakly). To see this in the quantum theory, consider for instance the case described by equation \eqref{Hfine1}. There, we have a constraint $\mathcal{C}_I=g_{\ell'_i}g^{-1}_{\ell'_I}$ associated to a leave $\ell'_I$ which is in the link pre-image $\mathbf{P}_\text{L}(\ell_i)$ of some leave $\ell_i\in\mathcal{L}$. It is therefore not included in any other link pre-image $\mathbf{P}_\text{L}(\ell_j)$ with $j\neq i$. For this reason, $\mathcal{E}_{\Delta,\Delta'}(R^\sigma_{\ell_j})$ will not affect  $\mathcal{C}_I=g_{\ell'_i}g^{-1}_{\ell'_I}$ for any $j\neq i$.  

For the case $j=i$, we have that
\be
\mathcal{E}_{\Delta,\Delta'}(R^\sigma_{\ell_i})F\big(g_{\ell'_i}g^{-1}_{\ell'_I}\big)
=\left(R^\sigma_{\ell'_i}\times R^\sigma_{\ell'_I}\right)F\big(g_{\ell'_i}g^{-1}_{\ell'_I}\big)
=F\big(g_{\ell'_i}g^{-1}_{\ell'_I}\big),
\ee
for any function $F$. Therefore, the refined fluxes leave in particular $\delta(\mathcal{C}_I,\openone)$ (and also $\delta(\mathcal{C}_{I^{-1}},\openone)$ since the argument is only changed by a conjugation) invariant. Similarly, one can go through the remaining cases. Furthermore, it is easy to see that the action of $\mathcal{E}_{\Delta,\Delta'}(R^\sigma_{\ell_i})$ on $\{g_{\ell'_j}\}_{j=1}^{|\ell|}$ equals the action of $R^\sigma_{\ell_i}$ on $\{g_{\ell_j}\}_{j=1}^{|\ell|}$. 

Since the refinement maps multiply the states with delta functions in the constraints, and then replace the arguments $\{g_{\ell_j}\}_{j=1}^{|\ell|}$ in the coarse wave function with $\{g_{\ell'_j}\}_{j=1}^{|\ell|}$ in the finer wave function, we can now conclude that the extension map for the fluxes satisfies the condition \eqref{consistency O}.

Again, we can change a given extension of a translation operator without violating the extension property. A particularly important class of changes is to include a parallel transport for the fluxes, in the form
\be
\mathcal{E}^\text{Alt}_{\Delta,\Delta'}(R^\sigma_\ell)
\coloneqq\prod_{\ell'\in\mathbf{P}_\text{L}^{-1}(\ell_i)}R^{g[\ell']\sigma(g[\ell'])^{-1}}\times\prod_{\ell'\in\mathbf{P}_\text{L}^{-1}(\ell_i^{-1})}L^{g[\ell']\sigma(g[\ell'])^{-1}}_{\ell'}.
\ee
Here $g[\ell']$ is an association to $\ell'$ of an holonomy along a cycle (starting and ending at the root) which is homotopy-equivalent to the trivial cycle in $\Sigma\backslash\Delta_{(d-2)}$. This means that $g[\ell']=\openone$ if the (coarse-graining) constraints $\{ \mathcal{C}\}_I$ are satisfied.

In order to avoid ordering issues, one can demand that for a fixed coarse leave $l_i$, none of the parallel transports $g[\ell']$ associated to leaves $\ell'$ in the set $\mathcal{S}(\ell_i)\coloneqq\mathbf{P}^{-1}_\text{L}(\ell_i)\cup\mathbf{P}^{-1}_\text{L}(\ell^{-1}_i)$ (i.e. the set of leaves that appear in the refined translation) include any of the leaves in $\mathcal{S}(l_i)$. Indeed, the parallel transport along a surface tree, as defined in \cite{fluxC}, does not include any links which cross this surface. One can therefore define an extension which matches the definition for an integrated flux (where the finer fluxes are parallel-transported along a surface tree) given in \cite{fluxC}.

The restrictions on the set of leaves which can appear in the parallel transport can however be weakened. For this, one can choose an ordering of the elementary translation operators by numbering the leaves in $\mathcal{S}(l)$, so that $\mathcal{S}(l)=\{\ell'_A\}_{A=1}^N$, and then order the product of translation operators from $A=1$ (appearing on the right) to $A=N$ (appearing on the left end). The condition on a parallel transport $g[l'_A]$ is then that it should $(a)$ be homotopy-equivalent to the trivial cycle in $\Sigma\backslash\Delta_{(d-2)}$, $(b)$ not include the link $l'_A$ itself, and $(c)$ be such that $g[l'_A]$ is left invariant by the product of translation operators associated to the leaves $\ell'_{A+1},\dots,\ell'_N$.

\subsection{Parallel-transported fluxes}

\noindent Let us now close this section by discussing the refinement of the parallel-transported translation operator $R^{g_\gamma\sigma g_\gamma^{-1}}_{\ell_i}$, where $g_\gamma$ is a holonomy associated to a (rooted) cycle $\gamma$ which does not include the link $\ell_i$ itself.

The parallel transport $g_\gamma$ is a word in the holonomies $g_{\ell_j}$ and their inverses for $j\neq i$, i.e. $g_\gamma=W_\gamma\{g_{\ell_j}\}$. We define $g'_\gamma$ as the same word in the holonomies $g_{\ell'_j}$, i.e. $g'_\gamma=W_\gamma\{g_{\ell'_j}\}$. We can then define the extension of a parallel-transported translation operator as
\be\label{extensionPF}
\mathcal{E}_{\Delta,\Delta'}\left(R^{g_\gamma\sigma g^{-1}_\gamma}_{\ell_i}\right)
\coloneqq\prod_{\ell'\in\mathbf{P}_\text{L}^{-1}(\ell_i)}R^{g'_\gamma\sigma(g_\gamma')^{-1}}_{\ell'}\times\prod_{\ell'\in\mathbf{P}_\text{L}^{-1}(\ell_i^{-1})}L^{g'_\gamma \sigma(g_\gamma')^{-1}}_{\ell'}.
\ee
Note that $g'_\gamma$ does not include any leaves in $\mathcal{S}(\ell_i)\coloneqq\mathbf{P}^{-1}_\text{L}(\ell_i)\cup\mathbf{P}^{-1}_\text{L}(\ell^{-1}_i)$, and that therefore there are no ordering issues. 

Let us now discuss the interpretation of the parallel transports $g'_\gamma$ in the finer triangulation. The holonomy $g_\gamma$ describes a priori a rooted cycle in the coarse triangulation. But it can also be interpreted as giving a parallel transport, alternative to the one provided by the tree, for the flux based at the source node $l_i(0)$ of $l_i$. Let us write $g_\gamma$ as $g_\gamma=W_\gamma\{g_{\ell_j}\}$. Then, this alternative path is going from the root $r$ to the source vertex $\ell_i(0)$ of $\ell_i$, and is given by   
\be 
t_{r\ell_i(0)}\circ W_\gamma\big\{t_{r\ell_j(1)}^{-1}\circ l_j\circ t_{r\ell_j(0)}\big\}.
\ee
Here $t_{r\ell_j(0)}$ is the path along the tree from the root $r$ to the source node of $\ell_j$, and $t_{r\ell_j(1)}$ is the path along the tree from the root $r$ to the target node of $\ell_j$.

In the refined triangulation, $g'_\gamma$ provides in the same way alternative paths for the parallel transports of the fluxes $\ell'\in\mathcal{S}(\ell_i)$. For a fixed $\ell' \in   \mathbf{P}^{-1}_\text{L}(\ell_i)$, the path starts at the root $r'$ and ends at $\ell'(0)$. It is given by 
\be
t_{r\ell'(0)}\circ W_\gamma\big\{t_{r\ell'_j(1)}^{-1}\circ l'_j\circ t_{r\ell'_j(0)}\big\}.
\ee
Likewise, if $\ell'\in\mathbf{P}^{-1}_\text{L}(\ell^{-1}_i)$, the path ends at $\ell'(1)$, and is given by 
\ba
t_{r\ell'(1)}\circ W_\gamma\big\{t_{r\ell'_j(1)}^{-1}\circ l'_j\circ t_{r\ell'_j(0)}\big\}.
\ea

Again, one can also choose alternative extensions for the parallel-transported translation operators $R^{g_\gamma\sigma g^{-1}_\gamma}_{\ell_i}$, in particular by changing the parallel transports $g'_\gamma$. For each $\ell'\in\mathcal{S}(\ell_i)$, one can replace $g'_\gamma$ with a ($\ell'$-dependent) transport along an homotopy-equivalent (in $\Sigma\backslash\Delta_{(d-2)}$) path to $\gamma$. As before, in order to avoid ordering issues, one can demand that these new paths should not include any $\ell'\in\mathcal{S}(\ell_i)$. This condition can be weakened in a similar way as discussed previously for the translation operators.

\section{Continuum operators}
\label{contop}

\noindent In the previous section, we have discussed extensions of operators on a given triangulation $\Delta$ to a refined triangulation $\Delta'$. This allows to define a consistent family of operators given on a sequence $\Delta\prec\Delta'\prec\Delta''\prec\dots$ of finer and finer triangulations. By taking the operator on a very fine triangulation $\Delta^{(n)}$ in this sequence, it is also possible to extend this consistent family to all triangulations coarser than $\Delta^{(n)}$. For this, we just have to first refine states from the coarser triangulation to $\Delta^{(n)}$, and then apply the operator $\mathcal{O}_{\Delta^{(n)}}$ to such states. This can be repeated with an even finer triangulation $\Delta^{(n')}\succ\Delta^{(n)}$, which then gives a consistent definition on all triangulations which are coarser than $\Delta^{(n')}$ (agreeing with the previous definition on triangulations where it has already been defined). However, there is no guarantee that starting with an operator $\mathcal{O}_\Delta$ and then extending this operator to $\Delta'$ and $\Delta''$ independently will give a definition which agrees on a triangulation that can arise as coarse-graining of $\Delta'$ and $\Delta''$ respectively.

The challenge is therefore to extend a given observable $\mathcal{O}_\Delta$ to all triangulations at once in a consistent way. Let us first discuss one possibility which does lead to a consistent family of observables, but however \textit{not} to a continuum observable in the usual sense (although it can be defined on the inductive limit Hilbert space $\mathcal{H}_\infty$). Given $\mathcal{O}_{\Delta_0}$ on $\mathcal{H}_{\Delta_0}$ we define extensions to any finer $\mathcal{H}_{\Delta'}$ by
\be\label{nullextension}
(\mathcal{O}_{\Delta_0})_{\Delta'}
\coloneqq(\mathcal{O}_{\Delta_0})'\circ\prod_I\delta(\mathcal{C}_I,\openone),
\ee
where $\mathcal{C}_I$ are the (coarse-graining) constraints describing states in $\mathcal{H}_{\Delta'}$ which arise as refinements from $\mathcal{H}_{\Delta_0}$. Here $(\mathcal{O}_{\Delta_0})'$ is any of the allowed extensions of $\mathcal{O}_{\Delta_0}$ to $\Delta'$. It does not matter which extension one chooses, since the operator  $(\mathcal{O}_{\Delta_0})_{\Delta'}$ is only non-vanishing on states which arise as refinements of $\Delta_0$. The actions of different extensions $(\mathcal{O}_{\Delta_0})'$ do agree on such states.

We can also define $(\mathcal{O}_{\Delta_0})_{\Delta}$ for any $\Delta$ which is not a refinement of $\Delta_0$, by first refining a state in $\mathcal{H}_\Delta$ to a common refinement of $\Delta_0$ and $\Delta$, and then applying the prescription \eqref{nullextension}.

The family $(\mathcal{O}_{\Delta_0})_{\Delta'}$ of observables constructed in this way satisfies the cylindrical-consistency conditions \eqref{ccO}. It therefore defines an operator on the inductive limit Hilbert space $\mathcal{H}_\infty$. The observables do however vanish on any state which is not cylindrical over $\Delta_0$ (i.e. which is not a refinement of a state in $\mathcal{H}_{\Delta_0}$). Such operators do therefore not correspond to quantizations\footnote{Note that the problem of defining observables at once for all triangulations is not a problem of quantization as such, since it arises already at the classical formulation of the continuum phase space as a (modified) projective limit, as explained in \cite{fluxC}.} of ``continuum'' flux and holonomy functions. The classical version they correspond to are rather such functions multiplied with a (infinite) product of delta functions imposing that curvature is vanishing almost everywhere.

Let us emphasize again that, nevertheless, the observables are well defined. Likewise, their spectrum has a well defined meaning. This holds in particular with respect to the interpretation of the inductive Hilbert space construction as corresponding classically to a symplectic reduction \cite{FGZ,fluxC}. That is, the spectrum of $\mathcal{O}_\Delta$ coincides with the spectrum of $(\mathcal{O}_\Delta)'$ restricted to the subspace of states which are cylindrical over $\Delta$. The spectrum of $(\mathcal{O}_{\Delta_0})_{\Delta'}$ is given by $\text{Spec}(\mathcal{O}_{\Delta_0})\cup\{0\}$.
 
This leaves us with the question of whether one can construct quantum observables on $\mathcal{H}_\infty$ which would actually correspond to quantizations of continuum versions of flux and holonomy observables. Let us sketch such a construction here, starting with an holonomy operator.

\subsection{Holonomy operators}

\noindent To discuss the holonomy operator, we will assume that the path $\gamma$ underlying the holonomy is piecewise-geodesic in the auxiliary metric on $\Sigma$. The holonomy operator is unambiguously defined on a triangulation $\Delta$ if the underlying path is a path in the inside of the manifold $\Sigma\backslash\Delta_{(d-2)}$, i.e. if it avoids the defects of this manifold. We call the set of triangulations $\Delta$ on which this is the case $\text{Triang}(\gamma)$. This is itself a partially-ordered directed set, and the corresponding inductive limit Hilbert space gives a subspace $\mathcal{H}_\infty(\gamma)$ of the full inductive limit Hilbert space $\mathcal{H}_\infty$. The holonomy operator can unambiguously be defined on this subspace.
 
In the case in which the path $\gamma$ does hit the boundary of $\Sigma\backslash\Delta_{(d-2)}$ one needs to choose a regularization of the operator (or rather of the underlying path). Any such choice of regularization corresponds to an extension of an operator from $\mathcal{H}(\gamma)$ to a bigger subspace. Such extensions of a (bounded) operator from $\mathcal{H}_\infty(\gamma)$ to a (bounded) operator $\mathcal{H}_\infty$ always exist, but they are in general not unique.

\subsection{Translation operators}

\noindent For the translation operator (or the parallel-transported translation operator), the main issue is again to unambiguously define the paths underlying the parallel transports of the translation operators associated to the finer links.
 
In particular, for the notion of integrated flux over a surface $S$ (in $d=3$ spatial dimensions), we have to specify a parallel transport for all points on this surface to, e.g., a surface root. The complicated case is here $d=3$. Indeed, for $d=2$ there is a canonical construction for the parallel transport of the fluxes, which are integrated along a one-dimensional ``shadow graph'' as explained in \cite{fluxC}. For the following discussion, we therefore restrict ourselves to $d=3$, and furthermore assume that the auxiliary metric is flat, so that the notion of piecewise-geodesic reduces to that of piecewise-linear.
 
Let us consider the integrated flux observables associated to a surface $S$, which is itself piecewise-linear and more precisely comes from the gluing of (flat) triangular surfaces $t$. The continuum parallel transport for this surface is specified as follows. First, this parallel transport is required to take place in the surface $S_\epsilon$ which is infinitesimally shifted below $S$ (with respect to the orientation of $S$). Second, we choose in each triangle $t$ of $S$ an inner reference point $r(t)$, referred-to as a $t$-root, to which all the inside points of $t$ are parallel-transported along geodesics (which are here straight lines). Finally, we choose a tree, referred-to as an $S$-tree, in the graph dual to the triangulation of $S$. The root of this $S$-tree should agree with that of the $t$-roots $r(t_0)$. Furthermore, the tree should connect the remaining $t$-roots $r(t)$ to $r(t_0)$ as follows: the nodes $r(t)$ of two neighboring triangles are connected by a geodesic restricted to $S_\epsilon$, i.e. a possibly kinked line in the $d=3$ flat space, and a straight line if considered from the induced (flat) metric of the two triangles.
 
This description defines the parallel transport for the continuum observable. We now have to construct a projection of this parallel transport to a given triangulation $\Delta$. This triangulation should be sufficiently fine in order to support the surface $S$, i.e. the triangular surfaces $t$ are unions of two-dimensional simplices $\sigma^2$ of $\Delta$. 

Consider the cut of the surface $S_\epsilon$ with the triangulation $\Delta$. By shifting $S$ infinitesimally to $S_\epsilon$, we avoid the potential curvature singularities (which in $d=3$ are along line defects) along the edges of the triangulation $\Delta$ (remember that we also assume that the edges of the triangulation are geodesics, i.e. straight lines, and that the triangles of the triangulation are minimal surfaces, i.e. flat). The edges of $S_\epsilon$ might now be ``thickened'', since it could happen that more than two (downside) tetrahedra hinge on a given edge of $S$. We can however ignore this thickening since the path of the parallel transport across these additional tetrahedra is unambiguous.

\begin{center}
\begin{figure}[h]
\includegraphics[scale=0.6]{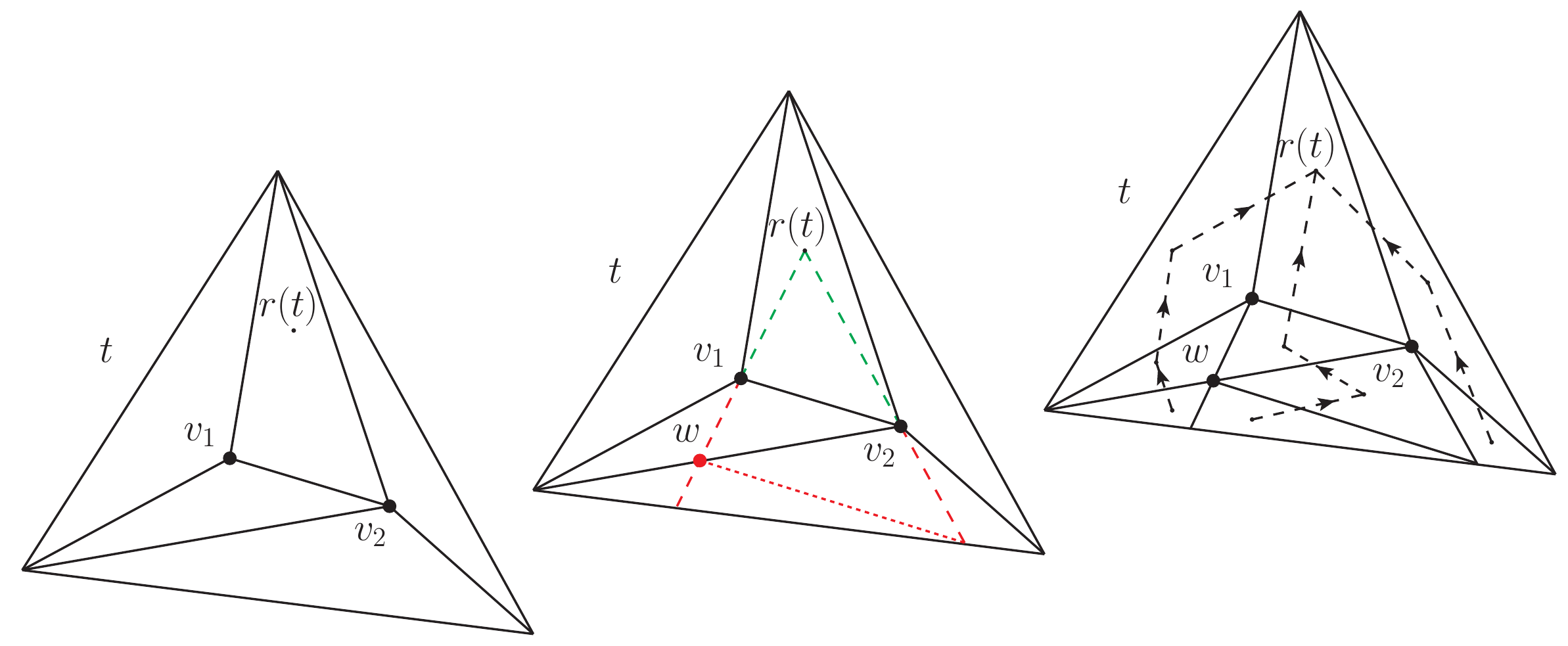}
\caption{The left panel shows the triangle $t$ and its induced subdivision by a triangulation $\Delta$. The vertices $v_1$ and $v_2$ lead to a further subdivision into homotopy-equivalent regions with respect to the transport to $r(t)$, which is shown in the middle panel. This subdivision is completed to a triangulation of $t$, and therefore the triangulation $\Delta$ of $\Sigma$ needs to be refined to a triangulation $\Delta'$ such that the induced triangulation for $t$ coincides with the one in the middle panel. The right panel shows the (surface) tree for the parallel transport in the triangulation $\Delta'$.}
\label{fig:subdivision}
\end{figure}
\end{center}

What needs to be clarified is the path for the parallel transport across the two-dimensional triangulation of $S$ which is induced by $\Delta$. This induced triangulation $\Delta^2$ of $S$ will be a refinement of the original triangulation of $S$ into triangles $t$, as shown on figure \ref{fig:subdivision}. We have to specify the parallel transport paths on $S\backslash V(\Delta^2)$, where $V(\Delta^2)$ denotes the vertices of $\Delta^2$. We assume that the triangulation $\Delta$ is such that none of the $t$-roots $r(t)$ coincide with a vertex of the induced triangulation or lies on an edge of this induced triangulation. We furthermore assume that the vertices of $V(\Delta^2)$ do not touch any of the paths along the $S$-tree. Therefore, we are just left with the task of specifying the tree for the parallel transport inside a given triangle $t$.
 
We construct the parallel transport on $t\backslash V(\Delta^2)$ as follows. Since the $t$-root $r(t)$ lies in the inside of one of the induced triangles $\sigma^2$, the parallel transport for a point $p$ of this triangle $\sigma^2_0$ to $r(t)$ is therefore uniquely specified along the geodesic from $p$ to $r(t)$. Let us consider some other induced triangle $\sigma^2_i$ of $t$. Then we can have the following two cases:
\begin{enumerate}
\item The parallel transports along straight lines of all the points in $\sigma^2_i$ are homotopy-equivalent in $S\backslash V(\Delta^2)$. This means that for any two points $p_1$ and $p_2$ lying inside this induced triangle, the piecewise-geodesic path from $r(t)$ to $p_1$, then to $p_2$, and back to $r(t)$, is homotopy-equivalent to the trivial cycle. We can then choose any of the inside points $p$, and adopt as a parallel transport for all points of this triangle a representative path which crosses through all the triangles which are crossed by the path from $p$ to $r(t)$.
\item There are points $p_1$ and $p_2$ of the inside of the induced triangle for which the piecewise geodesic path from $r(t)$ to $p_1$, then to $p_2$, and back to $r(t)$, is not homotopy-equivalent (in $t\backslash V(\Delta^2)$) to the trivial cycle. In this case, we have to subdivide the triangulation $\Delta$ further, until this situation eventually does not appear (now with respect to $t\backslash V\big((\Delta')^2\big)$).
\end{enumerate}

To see that this procedure does indeed stabilize after a finite number of steps (and does not produce further and further subdivisions), consider the example of figure \ref{fig:subdivision}. There, we construct the subdivision of $t$ into regions which are homotopy-equivalently transported to $r(t)$. To this end, we need to draw straight lines (appearing as dashed lines in the middle panel of figure \ref{fig:subdivision}) from the $t$-root $r(t)$ to all the vertices $V(\Delta^2)$, and continue these lines until they hit the boundary of $t$. We can ignore the parts of the lines which lie in $\sigma^2_0$ (which are in dashed green in figure \ref{fig:subdivision}). The remaining parts of the lines (in dashed red) will subdivide the other triangles $\sigma^2_i$ into homotopy-equivalent regions. We then need to subdivide the triangulation further, until these regions are unions of triangles $(\sigma')^2$ of the refined triangulation. This introduces the finely dashed red line in the middle panel of figure \ref{fig:subdivision}. However, note that the lines emanating from $r(t)$ do not cross themselves, and therefore only form new vertices with the edges which are already subdividing the triangle $t$ (this is the vertex $w$ in the middle panel). These new vertices can in principle refine the notion of homotopy-equivalent regions, which would then require an iteration of the procedure. All the new vertices are however exactly on the boundary of the homotopy-equivalent regions which have just been defined. Therefore, an iteration of this procedure is not necessary. We now have a triangulation of $t$ which is such that any two points in the inside of any given triangle are homotopy-equivalent as far as the transport to $r(t)$ is concerned. This allows to define a sub-tree, which specifies the parallel transport for the triangle $t$ itself (see the right panel of figure \ref{fig:subdivision}). As a last step, one needs to find a refined triangulation $\Delta'$ of $\Sigma$ which induces a triangulation of $t$ that coincides with the one we have just constructed. 

In summary, if the above assumptions are satisfied, we will eventually be able to define the parallel transport on a possibly refined triangulation $\Delta'$, which is required in order to specify the translation operator (on $\mathcal{H}_{\Delta'}$). We can also apply this translation operator to states in $\mathcal{H}_\Delta$ since we just need to refine these states to $\Delta'$.

The discussion on alternative extensions to $\Delta'$ for a given translation operator defined on a triangulation $\Delta$ in section \ref{extflux} included in particular the change of parallel transport. Therefore, we can adjust the parallel transport of a given extension to $\Delta'$ so that it matches the one given by the projection of the continuum translation operator.

Consider the (integrated) flux associated to a fixed choice of a triangulated surface $S$, a choice of $t$-roots, and a choice of $S$-surface tree. To define the corresponding translation operator on the continuum Hilbert space, we first consider a partially-ordered subset of triangulations such that the assumptions regarding the reconstruction of an unambiguous parallel transport are satisfied for each of the triangulations in this subset. We can then argue in a similar manner as for the holonomy operators, and define an integrated translation operator on the corresponding subspace of $\mathcal{H}_\infty$. Extensions of this operator always exist, but are again not unique. 

We therefore obtain continuum operators which are quantizing the holonomies and (the exponentiated flow of the) fluxes. This allows to consider the (Weyl form of) the holonomy-flux algebra, or more precisely its commutation relations. Due to the cylindrical-consistency of the operators, one can evaluate the commutators between two operators on a triangulation $\Delta$ on which both operators can be unambiguously defined. Due to our quantization prescription, one then obtains the expected result for the commutators between holonomies and translation operators.

In \cite{fluxC}, we have also evaluated the Poisson algebra of the fluxes explicitly. Note that this cannot be performed in the AL framework, since there the Poisson bracket of two fluxes might lead to a more singular object, e.g. a flux only smeared over a one-dimensional path in $d=3$ spatial dimensions. In contrast, in \cite{fluxC} one obtains again (parallel-transported) fluxes smeared over $(d-1)$-dimensional surfaces as a result of the commutator between two fluxes. The reason is that the main contribution to the commutators arises from the parallel transport and the basic fluxes. Therefore, all the parallel transport holonomies for a certain ($(d-1)$-dimensional) part of the flux surface contribute to the commutator. The result of the commutator is therefore a flux which is smeared over this part of the surface (see \cite{fluxC} for details). The complete consideration of these commutators requires to go through many separate cases, and we therefore leave the consideration of the quantum commutators for future work.

\section{Discussion}
\label{sec:discussion}

\noindent We have constructed in this paper a new realization of quantum geometry. Our construction being based on the holonomy and flux variables of loop quantum gravity (LQG), it shares some features with the already existing Ashtekar--Lewandowski (AL) representation, but, most importantly, exhibits crucial differences.

First of all, both the AL and the BF representations provide a concrete realization of the idea of seeing (quantum) gravity as a topological field theory with defects. However, while in the AL representation the defects are the carriers of the geometrical excitations themselves, which leads to complications when trying to construct semi-classical smooth geometries, in the BF representation the situation is reversed. Indeed, as we have seen, the BF representation is based on a vacuum peaked on locally and globally flat connections, and excitations on top of this vacuum are carried by defects encoding non-trivial local curvature (while the conjugated intrinsic geometry is maximally uncertain).

Second, the construction of the BF representation relies on the triangulation of the spatial manifold itself, and does not consider the (arbitrary) dual graphs as fundamental structures, as it is the case in the AL representation. Because of this, the notion of refinement and embedding of dual graphs in the AL representation is replaced here by the refinement of triangulations and the associated notion of embedding of coarser triangulations into finer ones. This notion of embedding is crucial for the construction of a continuum inductive limit Hilbert space.

As recalled in the introduction, a fundamental result of LQG, which is based on the AL representation, is the existence of a continuum inductive limit Hilbert space which carries a unitary representation of the holonomy-flux algebra and of the action of spatial diffeomorphisms. We have shown here that this result can be extended to the BF representation as well, and that it is possible to construct an inductive limit Hilbert space in which the Hilbert spaces associated to triangulations carrying a finite number of curvature excitations are embedded.

The existence of this continuum Hilbert space relies on several key technical points. The most important one is that we need to work with exponentiated (i.e. compactified) fluxes, which is in turn equivalent to considering a discrete topology on the group labelling the curvature excitations. This is the reason why the BF representation evades the F--LOST uniqueness theorem \cite{fleischhack,LOST}. Indeed, this theorem states that the AL representation is unique given a number of assumptions, and one of these is that the fluxes have to exist as well-defined operators.

The introduction of a discrete topology on the group is one of the key features which leads to different properties for the geometric observables in the BF representation as compared to the AL representation. In particular, the spectrum of the area operator, discussed in section \ref{sec:spectrumAr}, is bounded in the BF representation. For the structure group $\U(1)$, the spectrum can be either discrete or continuous, depending on a parameter $\mu$ appearing in the definition of the area operator. We expect a similar result for the (fully gauge-invariant) $\SU(2)$ area operator\footnote{The continuity of the spectrum can arise through the winding of the (generalized) eigenvalues  associated to infinitely many spin labels in a bounded interval. This does not occur if one replaces the Lie group with a quantum group at root of unity, although in this case the description of the defects becomes technically more involved (see the comment below). On the other hand, a continuum spectrum arising from a (Bohr) compactification offers also advantages, as exemplified in relation to chaotic dynamics and constrained quantization in \cite{chaostoappear}.}. Furthermore, there are different possibilities to implement the presence of the Barbero--Immirzi parameter into the area operator, and it would be very interesting to see whether this parameter will continue to appear in the black hole entropy state counting, for which the area spectrum plays a crucial role \cite{AshKras}. The fact that the spectra of the operators can be different in different representations emphasizes also the point raised in \cite{discretespectra?}, namely that the spectra of physical observables on the physical Hilbert space might be quite unrelated to the spectra of (kinematical) observables on kinematical Hilbert spaces.

Further key differences between the AL and BF representations are the different requirements for the behavior of operators under refinement. In the case of the BF representation, we can introduce for the geometric operators a measurement scale in the form of a subdivision of the region to be measured into a triangulation. This triangulation is attached to the operator and not the the state. Therefore, the area of each triangle in such a triangulation is bounded. This still allows to achieve large areas by sufficiently subdividing a region into pieces.

It will be important to investigate how various results relying on the AL representation will change if one works with the BF representation instead. For instance, the results of the black hole entropy calculation \cite{BHentropyLQG,AshKras} might change if one uses the BF representation. In this direction, the work \cite{hannoBH} is interesting since it uses a kind of mixture between the BF and AL representations in order to accommodate the isolated horizon constraints describing a certain family of black holes \cite{ihorizon}.

The advantages of considering this new BF representation are manifold, and there are lots of interesting and important directions to be explored, as explained in the following points:
\begin{itemize}
\item The BF representation allows to have a nicer geometric interpretation since the states are not based anymore on degenerate spatial geometrical excitations. One should explore this interpretation in particular in $(3+1)$ spacetime dimensions, and its relation to simplicial geometry. Since the classical phase space on a fixed triangulation agrees with that of loop quantum gravity on a fixed (dual) graph, one will find that the space of geometries, if interpreted in a simplicial manner, does not agree with Regge geometries \cite{dittrichryan1,dittrichryan2,areaangle}, but rather defines twisted geometries \cite{bonzomarea,twisted}. However, the basis states are in fact not semi-classical, but rather squeezed states. In particular, the spatial geometrical operators have maximal uncertainties, whereas the curvature operators are maximally peaked. An interesting question is therefore to what extent the connection determines the four-dimensional geometry. Note that the construction of such states with curvature line defects was also considered in \cite{thooftstudent}.

\item The BF representation forces us to work with the gauge-covariant fluxes, i.e. the fluxes with parallel transport. This is intimately related to the behavior of the fluxes under refining and coarse-graining. It furthermore explains the effect of ``curvature-induced torsion'' \cite{eteradeformed,fluxC}, which can be (easily) understood as the fact that a curved triangle or curved tetrahedron does not satisfy the so-called closure constraints. This effect leads to a difficulty for coarse-graining, in the sense of mapping spin networks to coarser spin networks. As noted in \cite{eteradeformed}, the coarse spin networks do not necessarily need to satisfy the coupling rules. However, this issue is largely connected to not providing a clean splitting of the observable algebra\footnote{An alternative is to ``enlarge'' the spin network states, as proposed in \cite{eteradeformed}.}, as for instance used in \cite{lanery4} for the case of the standard non-parallel-transported fluxes. Here, we provided such a splitting in section \ref{splitting algebra}, which allows for a clear geometric interpretation of coarse-graining in terms of which observables are kept and which are averaged over. This also answers the question raised in \cite{howmany} of how may quanta there are in a quantum spacetime: the phase space factor associated to the observables averaged over counts the number of quanta.

\item In $(2+1)$ spacetime dimensions, the BF representation constructed here shares many properties with the so-called combinatorial quantization scheme \cite{Fock:1998nu,Alekseev:1994pa,Alekseev:1994au,Alekseev:1995rn,Meusburger:2003ta,Meusburger:2003hc,MeusburgerNoui}, which is based on the quantization of the space of flat connections on a manifold with defects. In this combinatorial approach, the number of defects is fixed, so there is no need to invoke a continuum limit and therefore to use a Bohr compactification like we do here. Nevertheless, it would be interesting to construct an explicit maps between these approaches. Reference \cite{MeusburgerNoui2} already provides a map between the observable algebras.

\item Since the roles of the holonomies and the fluxes are inverted as compared to the AL representation, one can now work with the fluxes themselves, and therefore describe the coarse-graining of geometrical quantities. This would be a completion of the proposal \cite{aristideandco,guedes} to use a (non-commutative) flux representation for loop quantum gravity. In fact, in appendix \ref{appendix:spin}, we discuss the spin representation and a non-commutative product (which is basically the matrix product) based on this representation. One needs this product in order to express the inner product of two states as a functional applied to the (non-commutative) product of these states.

In this respect, it is also interesting to explore the full algebra of the exponentiated flux operators. In contrast, the full commutator algebra of the flux operators cannot be described within the AL representation since it would lead to fluxes smeared over singular surfaces in the case where the flux surfaces cut each other. It has been shown in \cite{fluxC} that in the case of the BF representation the classical Poisson algebra of the fluxes is well-defined and does not lead to such an appearance of singular surfaces.

\item In this work, we have left largely open the question of how to impose the spatial diffeomorphism and the Hamiltonian constraints. Let us first comment on the spatial diffeomorphism constraints. Here, we expect that spatial diffeomorphisms will act by moving the defects around, i.e. by changing the embedding of the triangulation network which supports the defects. This is basically the same action as in the AL representation, and one would therefore use, as in the AL representation, a group averaging procedure in order to define the (spatially) diffeomorphism invariant Hilbert space.

In \cite{fluxC}, we have provided a generator (on phase space) for this action in the case of $(2+1)$ spacetime dimensions. As part of the task of understanding the geometries encoded in the $(3+1)$-dimensional case, it would be important to investigate also this case. The question there is whether and how the diffeomorphism symmetry coincides with the BF translation symmetry. This is related to a proposal by Zapata \cite{zapataBF} to construct a theory of topological gravity as a BF theory with certain broken symmetries.

\item Let us now turn to the Hamiltonian constraints. We do believe that Hamiltonian constraint operators can in principle be constructed, since the main regularization mechanism pointed out in \cite{thiemannH1} should also hold in our case. However, we expect that the problems with the constraint algebra \cite{Lewandowskietal} will persist. This is ultimately related to the problem of preserving full diffeomorphism symmetry if lattices are introduced, even if this happens only on an auxiliary level \cite{bddiffeo,bahrdittrich09a,bdreview12}. An advantage of the BF representation is however the nicer geometric interpretation, which can facilitate the discussion of these issues. The work \cite{wolfgangH,wolfgang} is also aimed at understanding the dynamics of spin foam gravity as a continuum theory, starting from BF theory.

An alternative to directly imposing the Hamiltonian constraints is to use a discrete time dynamics \cite{consistentdiscr}, and then to consider the continuum limit. This would in fact fully follow the philosophy of approximating the dynamics by using defects in a Regge-like manner \cite{regge,thooft}. A framework for describing a simplicial canonical dynamics has been described in \cite{bdhoehn2,bdhoehn3}. The question of how to reconstruct the continuum limit has been considered in e.g. \cite{timeevol,uniform,cylconsis,phoehn2,holonmysf,bahr14,bd14}. In this continuum limit, one can also hope to restore diffeomorphism symmetry as exemplified in \cite{improved,bahrdittrichsteinhaus,bahrdittrich09b}.
\end{itemize}

The above points concern possible future work directions related to the BF representation. In addition, we believe that the techniques developed in this paper will also be helpful to construct further realizations of quantum geometry. Any topological field theory whose defects support the holonomy-flux algebra (or an alternative observable algebra) can serve as a starting point for such a construction, as explained in \cite{timeevol}. A main question for future research is therefore whether there exists any such topological field theories in the case of four spacetime dimensions. Positive indications have been found in \cite{merce,intertwiner,qgroupspinnet} by studying the coarse-graining of (simplified) spin foam models, and work is in progress to find new topological field theories starting from spin foam models \cite{decoratedtnw,clement}. Again, there are a number of further directions to explore, including the following ones:

\begin{itemize}
\item A long standing problem has been that of constructing a Hilbert space based on the structure group $\SL(2,\mathbb{C})$ within the AL representation. The main difficulty in trying to do so is the non-compactness of the group. Even if one finds a way to appropriately compactify the group, there is the problem of how to construct an invariant measure. In the case of the BF representation, we need to compactify instead the dual of the group, which leads to a discrete topology on the group itself. Therefore, the advantage might be that there is an invariant discrete measure on the group which might allow to construct a BF-like representation for a non-compact gauge group.

\item An important question is whether there is a way to avoid the requirement of compactifying the configuration space (which, as we have discussed, can be understood from the necessity to define defects as stable physical entities), possibly along the lines of \cite{okolov}. In this vein, it would be interesting to understand the general properties of inductive limit Hilbert spaces and their limitations, and in particular to revisit the F--LOST uniqueness theorem \cite{fleischhack,LOST}. In particular, the question is what kind of representations are allowed if one relaxes the assumption of the existence of (non-exponentiated) flux operators?

\item One can also consider cases in which the dual of the group is already compact. Apart from finite groups, this is the case of quantum groups at root of unity. These describe (at least in three spacetime dimensions, since the results are only conjectured for the four-dimensional case) quantum gravity with a positive cosmological constant (and a Euclidean signature). There is by now a large body of work concerned with the issue of implementing the cosmological constant or quantum groups \cite{aldo,lee,girelli1,girelli2,pranzetti,rovvid}. The introduction of a quantum group changes the almost everywhere flat connection to an almost everywhere homogeneously-curved connection. A corresponding description in terms of simplicial geometry is given in \cite{newregge}.

When replacing $\SU(2)$ with its quantum deformation at root of unity, the first question is to understand the structure of the defects. In the case of $(2+1)$ dimensions, the defects have been understood to be given (in the context of higher categories) by the Drinfeld center associated to the topological theory (here the Tuare--Viro \cite{TV} topological quantum field theory), as explained for instance in \cite{balsam}. Therefore, the construction of a Tuarev--Viro-based representation can be achieved by combining the methods of \cite{balsam} and the methods developed in this work \cite{bdtoappear}. For the four-dimensional case, recent developments have been achieved in \cite{barrettmeus}.

Using a quantum group at root of unity will change the results for the spectrum of the area operator. The compactification of the spectrum for $\SU(2)$ means that the eigenvalues have to wind around in the interval allowed by the bound. This can lead to a continuous spectrum in the case of Lie groups. In the case of a quantum group at root of unity, the bound is on the (admissible) spins themselves, which means that the number of eigenvalues is itself actually bounded. This would prevent the spectrum from being continuous.
\end{itemize}

In summary, the existence of this new representation should not come as a surprise, and in fact indicates the presence of unsuspected rich phase structures in discrete approaches to quantum gravity. In ordinary quantum field theory, the existence of different vacua is tight to the description of condensation mechanisms, along with phase transitions and symmetry breakings. Different physical situations might be easier to describe within different representations based on different vacua. Recent results \cite{eckert,qgroupspinnet,clement} indicate that spin foam models can have non-trivial phase diagrams, with phases corresponding to particular topological field theories. Given such a topological field theory, one can describe excitations on top of a vacuum, and then dynamics for these excitations. In this way, it can be expected that each phase will correspond to a different realization of quantum geometry. This will not only open up a rich field of research, but also enrich our methods for constructing and understanding quantum geometries.

\section*{Acknowledgements}

\noindent We would like to thank Abhay Ashtekar, Klaus Fredenhagen, and Aldo Riello for helpful discussions. This research was supported by Perimeter Institute for Theoretical Physics. Research at Perimeter Institute is supported by the Government of Canada through Industry Canada and by the Province of Ontario through the Ministry of Research and Innovation. BB was funded by project 4966/1-1 of the German Research Foundation (DFG). MG is supported by the NSF Grant PHY-1205388 and the Eberly research funds of The Pennsylvania State University.

\appendix

\section*{Appendices}

\section{First fundamental group of manifolds}

\subsection{Manifolds without defects}

\label{appendix:pi1}

\noindent In this appendix, we briefly recall how the first fundamental group $\pi_1(\Sigma)$ can be obtained in the case of two- and three-dimensional manifolds without defects.

First, if $\Sigma$ is a closed and orientable Riemann surface of genus $g$, its fundamental group $\pi_1(\Sigma)$ is given by the free group on $2g$ generators $a_1,b_1,\dots,a_g,b_g$, divided by a normal subgroup generated by the single relator $R_1=[a_1,b_1]\dots[a_g,b_g]$, where the commutator is $[a,b]=aba^{-1}b^{-1}$. This can be shown by using the Seifert--van Kampen theorem to write $\pi_1(\Sigma)\simeq\pi_1\big(\bigvee_{i=1}^{2g}\mathbb{S}^1_i\big)/N$, where $\bigvee_{i=1}^{2g}\mathbb{S}^1_i$ is the bouquet obtained from the wedge sum of $2g$ circles, and $N$ is the normal subgroup generated by $R_1$. Since this relator corresponds to a null-homotopic loop, we can write the presentation
\be\label{2dpresentation}
\pi_1(\Sigma)
=\big\langle a_1,b_1,\dots,a_g,b_g\big|[a_1,b_1]\dots[a_g,b_g]=\openone\big\rangle.
\ee
Notice that this first fundamental group, although always being finitely presented, is not free for $g\geq2$.

For a three-dimensional compact, connected, and orientable manifolds $\Sigma$, the first fundamental group can also be finitely presented. This is most easily seen by using the fact that any such manifold admits a CW complex. Indeed, if we denote by $\Sigma_2$ the two-skeleton of the CW complex of $\Sigma$, then we have that $\pi_1(\Sigma)\simeq\pi_1(\Sigma_2)$, which can in turn be computed using the Seifert--van Kampen theorem and the fact that $\Sigma_2$ is the union of the one-skeleton $\Sigma_1$ with all the two-cells of the CW complex. More precisely, $\pi_1(\Sigma_1)$ is always given by a free group $\la a_1,\dots,a_{n_1}\ra$ on $n_1$ generators, and the presentation of the first fundamental group of $\Sigma$ is given by
\be
\pi_1(\Sigma)
=\big\langle a_1,\dots,a_{n_1}\big|R_1,\dots,R_{n_2}\big\rangle,
\ee
where the relator $R_i$ is the homotopy class in $\pi_1(\Sigma_1)$ of the boundary of the $i$-th two-cell (with $i\in\llbracket1,n_2\rrbracket$), and therefore a word in the generators. This can equivalently be described in terms of the so-called spine of the manifold $\Sigma$, which is a two-complex consisting of a single vertex, one-cells and two-cells. If the manifold $\Sigma$ is closed, its spine can be obtained by collapsing it after first removing an open three-ball. In this language, the generators are in one-to-one correspondence with the one-cells of the spine, the relators are in one-to-one correspondence with the two-cells, and the explicit expression for a relator $R_i$ describes how the $i$-th two-cell is attached to the one-skeleton.

\subsection{Manifolds with defects}
\label{appendix:pi1defects}

\noindent The case of the defected manifold $\overline{\Sigma}$ is simpler to deal with than the case of a manifold without defects. This is due to the fact that the graph $\Gamma$, which is the one-skeleton of the simplicial complex dual to the triangulation $\Delta$ of $\Sigma$, is a deformation-retract of $\overline{\Sigma}$. As a consequence, their first fundamental groups are isomorphic, i.e. $\pi_1\big(\overline{\Sigma}\big)\simeq\pi_1(\Gamma)$, which is in turn immediate to obtain. Indeed, it is well-known that the first fundamental group of a graph is free and generated by $|l|-|n|+1$ generators corresponding to its fundamental cycles \cite{hatcher}. By using a maximal spanning tree to characterize the fundamental cycles of the graph, we can therefore say that the first fundamental group of $\Sigma\backslash\Delta_{(d-2)}$ is freely generated by generators associated to the leaves of the tree.

For the sake of completeness, let us comment on the counting of the number of generators for the first fundamental group in the case of a punctured Riemann surface of arbitrary genus. Any such surface with $p$ (thickened) punctures can be decomposed as a connected sum $\overline{\Sigma}=A\text{\small{$\#$}}B$, where $A$ has genus $g-1$ and one puncture, while $B$ has genus one and $p+1$ punctures\footnote{This is possible since moving the punctures on the surface does not change its first fundamental group.}. In this decomposition, the surfaces $A$ and $B$ are glued along one common puncture (so that the total number of punctures is indeed $p$), and $A\cap B=\mathbb{S}^1$. By the Seifer--van Kampen theorem, we have that the first fundamental group of $\overline{\Sigma}$ is given by the amalgamated free product
\be\label{freeproduct}
\pi_1\big(\overline{\Sigma}\big)\simeq\pi_1(A)*_{\pi_1(A\cap B)}\pi_1(B)
=\pi_1(A)*_{\pi_1(\mathbb{S}^1)}\pi_1(B).
\ee
Each factor in this expression can easily be computed. First, for a surface of genus $g$ with a single puncture, the first fundamental group is isomorphic to that of a wedge sum of $2g$ circles, so we can write that $\pi_1(A)\simeq\pi_1\big(\bigvee_{i=1}^{2(g-1)}\mathbb{S}^1_i\big)$, which is the free group $F_{2(g-1)}$ on $2(g-1)$ generators. Second, for a surface of genus one with $p$ punctures, the first fundamental group is isomorphic to that of a wedge sum of $p+1$ circles. This implies that $\pi_1(B)\simeq\pi_1\big(\bigvee_{i=1}^{p+2}\mathbb{S}^1_i\big)$, which is the free group on $F_{p+2}$ on $p+2$ generators. Finally, we have that $\pi_1(\mathbb{S}^1)=F_1$, which is the free group on a single generator. Gathering these results, we get that the amalgamated free product \eqref{freeproduct} defining $\pi_1\big(\overline{\Sigma}\big)$ is generated by the generators of $\pi_1(A)$ and $\pi_1(B)$, with one generator removed due to $\pi_1(\mathbb{S}^1)$. Therefore, this leads to a free group on $2(g-1)+(p+2)-1=2g+p-1$ generators. This result is consistent with the usual presentation which generalizes \eqref{2dpresentation} to the case of punctured surfaces and takes the form
\be\label{punctured presentation}
\pi_1\big(\overline{\Sigma}\big)
=\big\langle a_1,b_1,\dots,a_g,b_g,g_1,\dots,g_p\big|[a_1,b_1]\dots[a_g,b_g]=g_1\dots g_p\big\rangle.
\ee
This presentation is written in terms of $2g$ generators corresponding to the non-contractible paths in $\overline{\Sigma}$, and the $p$ generators corresponding to the paths around the punctures. Now, this presentation can be reduced by a so-called Nielsen transformation which, since there is only one relation in the presentation \eqref{punctured presentation}, indeed leads to the free group on $2g+p-1$ generator.

Now, since for triangulated surfaces the Euler characteristic is given by $\chi=2-2g=|n|-|l|+|f|$, and since the punctures are ``piercing'' the faces (which means that $p=|f|$), we get that $2g+p-1=|l|-|n|+1$, which indeed corresponds to the number of leaves in a maximal spanning tree.

\section{Reconstruction of branch fluxes}
\label{appendix:branchX}

\noindent We explain in this appendix how to reconstruct the fluxes (and operators) associated to the branches from the ones associated to the leaves. To this end we have to employ  the Gau\ss~constraints\footnote{Recall that a maximal spanning tree has $|b|=|n|-1$ branches and $|\ell|=|l|-|b|$ leaves. As far as the counting is concerned, it is therefore possible to reconstruct all the fluxes associated with the links $l$ from the knowledge of the fluxes associated to the leaves $\ell$ and $|n|-1$ Gau\ss~ constraints.}. Recall that the Gau\ss~constraint for a node $n$  takes the form
\be\label{gauss1}
\mathcal{G}_n
=\sum_{l|l(0)=n}\bX_l+\sum_{l|l(1)=n}\bX_{l^{-1}}
\stackrel{!}{=}0,
\ee
where the rooted fluxes are given by
\be
\bX_l=g^{-1}_{rl(0)}X_lg_{rl(0)},\q
\bX_{l^{-1}}=g^{-1}_{rl^{-1}(0)}X_{l^{-1}}g_{rl^{-1}(0)}=-g^{-1}_{rl^{-1}(0)}h_lX_lh_l^{-1}g_{rl^{-1}(0)}.
\ee
The last definition implies that for branches $l=b$ we simply have $\bX_{b^{-1}}=-\bX_b$, while for leaves $l=\ell$ we have $\bX_{\ell^{-1}}=-g_\ell\bX_\ell g_\ell^{-1}$, where $g_\ell=g_{r\ell(1)}h_\ell g_{r\ell(0)}$.

Given a choice of tree, one can successively solve for all the fluxes by starting from the end nodes of the tree, where the Gau\ss~constraints each involve only one flux associated to a branch. By solving for these fluxes associated to the end nodes of the tree, we can go to the nodes which are only one step away from the end nodes along the tree.  Again, for every node there is only one unknown branch flux. One can then iterate this procedure and solve all the Gau\ss~constraints except for the one at the root. The fluxes associated to branches are then given by linear combinations of the fluxes associated to the leaves and to the inverse leaves.

To give the result of this procedure in a compact form, let us denote by $\mathcal{N}_{\partial\mathcal{T}>b(1)}$ the set of end nodes of the tree which lie after the node $b(1)$ when following the orientation of the tree. For each end node $n_\text{e}$ in $\mathcal{N}_{\partial\mathcal{T}>b(1)}$ there is a unique path $\gamma_{b(1)n_\text{e}}$ going from the target node $b(1)$ of the branch to the end node $n_\text{e}$. Considering the set of all such paths for a given $b$, we can then denote by $\mathcal{L}_\text{s}(b)$ the set of leaves which have their source node $\ell(0)$ along the paths $\gamma_{b(1)n_\text{e}}$, and by $\mathcal{L}_\text{t}(b)$ the set of leaves which have their target node $\ell(1)$ along the paths $\gamma_{b(1)n_\text{e}}$. As suggested by the notation, a choice of branch $b$ automatically determines the sets $\mathcal{L}_\text{s}(b)$ and $\mathcal{L}_\text{t}(b)$ (which can have a non-empty intersection). The flux is then given by
\be
\bX_b
=\sum_{\ell\in\mathcal{L}_\text{s}(b)}\bX_\ell+\sum_{\ell\in\mathcal{L}_\text{t}(b)}\bX_{\ell^{-1}}.
\ee
In this formula, each leaf $\ell$ can appear at most once with the orientation $\ell$ and at most once with the orientation $\ell^{-1}$. The right translations corresponding to a branch can therefore be defined as
\be
R_b^\alpha
=\prod_{\ell\in\mathcal{L}_\text{s}(b)}R_\ell^\alpha\prod_{\ell\in\mathcal{L}_\text{t}(b)}L_\ell^\alpha.
\ee

\section{The spin representation and measure functional}
\label{appendix:spin}

\noindent In most of this article we have worked in the holonomy representation for the wave functions. In this appendix we discuss a group Fourier transform with respect to a discrete measure on the group, which defines a spin representation.

\subsection{The group U(1)}

\noindent The discrete group $\U(1)$ can be understood as arising from a Bohr compactification of its dual group $\mathbb{Z}$. Consider the space of almost periodic functions $\psi(k)$ on $\mathbb{Z}$. Almost periodic functions are constructed as finite combinations of functions from the set $\{e^{-\i k\alpha}\,|\,\alpha\in[0,2\pi)\}$. Since there exists an inner product, we also take the norm completion of this space.

This Bohr compactification inner product is given by
\be
\langle\psi_1|\psi_2\rangle
=\lim_{T\rightarrow\infty}\f{1}{(2T+1)}\sum_{|k|\leq T}\overline{\psi_1(k)}\psi_2(k),
\ee
and it turns the functions $\psi_\alpha(k)=e^{-\i k\alpha}$ into an (over-countable) orthonormal basis. This basis can be used to transform states to the ``holonomy representation'', which is here in terms of an angle $\theta\in[0,2\pi)$. This transformation is given by
\be\label{trafo1}
\psi(\theta)
=\langle\psi_\theta|\psi\rangle
=\lim_{T\rightarrow\infty}\f{1}{(2T+1)}\sum_{k\leq T}e^{\i k\theta}\psi(k),
\ee
from which we recover the following Kronecker delta form of the basis:
\be
\psi_\alpha(\theta)
=\delta(\alpha,\theta),
\ee
where $\delta(\alpha,\beta)=1$ if $\alpha=\beta$ mod $2\pi$ and is vanishing otherwise. The inverse transformation to \eqref{trafo1} is given by
\be
\psi(k)
=\int\de\mu_\text{d}(\theta)\,e^{-\i k\theta}\psi(\theta),
\ee
with a discrete measure $\mu_\text{d}$. The translation operator
\be
R^\phi\psi(\theta)
\coloneqq\psi(\theta+\phi)
\ee
acts on a basis vector as $R^\phi\psi_\alpha=\psi_{\alpha-\phi}$, and therefore in the $k$-representation as multiplication
\be
R^\phi\psi(k)
=e^{\i\phi k}\psi(k).
\ee

We can define a measure
\be
\mu_\text{Bohr/BF}(\psi_\alpha)
=\delta(\alpha,0)
\ee
for the basis functions $\psi_\alpha(k)=e^{-\i k\alpha}$, from which one can recover the inner product in the $k$-representation as
\be
\langle\psi_1|\psi_2\rangle
=\mu_\text{d}\big(\overline{\psi_1}\cdot_k\psi_2\big).
\ee
This measure is the analogue (or rather the dual) of the Haar measure (which underlies the Ashtekar--Lewandowski measure), which in the context of $\U(1)$ would be defined as
\be
\mu_\text{Haar/AL}(\psi_k)
=\delta(k,0)
\ee
for basis functions $\psi_k(\theta)=e^{\i k\theta}$ in the $\theta$-representation.

\subsection{The group SU(2)}
\label{spinrepsu2}

\noindent In order to define a spin representation for $\SU(2)$, we proceed as in the case of $\U(1)$. Consider a suitable space of functions on the set
\be
\big\{j,M,N\,\big|\,j\in\mathbb{N}/2,\ M,N=-j,-j+1,\dots,j-1,j\big\}.
\ee
By analogy with the case of $\U(1)$, we define as ``suitable'' functions the functions which arise as finite linear combinations of functions from the set
\be
\big\{\sqrt{d_j}\,D^j_{MN}(\alpha)\,\big|\,\alpha\in\SU(2)\big\},
\ee
where $D^j_{MN}(\cdot)$ are the spin-$j$ representation matrices.

We write these functions into the matrix form $\psi_{MN}(j)=\psi(j,M,N)$, and equip the space of functions with the inner product
\be\label{C12}
\langle\psi_1|\psi_2\rangle
=\lim_{\Lambda\rightarrow\infty}\f{1}{\mathcal{N}(\Lambda)}\sum_{j=0}^\Lambda\tr\big((\psi_1^\dagger(j)\psi_2(j)\big),
\ee
where $\big(\psi_{MN}(j)\big)^\dagger=\overline{\psi_{NM}(j)}$. In order to define this inner product, we have introduced the regularization factor
\be\label{nfactor}
\mathcal{N}(\Lambda)
=\sum_{j=0}^\Lambda d_j^2
=\f{1}{6}(4\Lambda+3)(2\Lambda+2)(2\Lambda+1),
\ee
where $d_j=2j+1$. In this inner product, the following functions are orthonormal:
\be
\psi_\alpha(j)_{MN}
=\langle j,M,N|\psi_\alpha\rangle
=\sqrt{d_j}\big(D^j_{MN}(\alpha)\big)^\dagger,
\ee
and with $\alpha\in\SU(2)$ constitute a (over-countable) basis.

We can now define the transformation to the group representation as
\be\label{C15}
\psi(h)
=\langle\psi_h|\psi\rangle
=\lim_{\Lambda\rightarrow\infty}\f{1}{\mathcal{N}(\Lambda)}\sum_{j=0}^\Lambda\sqrt{d_j}\,\tr\big(D^j(h)\psi(j)\big),
\ee
from which we therefore obtain
\be
\psi_\alpha(h)
=\lim_{\Lambda\rightarrow\infty}\f{1}{\mathcal{N}(\Lambda)}\sum_{j=0}^\Lambda d_j\tr\big(D^j(h\alpha^{-1})\big)
=\delta(\alpha,h),
\ee
with $\delta(alpha,h)=1$ iff $\alpha=h$ and vanishing otherwise. The inverse transformation to \eqref{C15} is given by
\be
\psi(j)
=\int\de\mu_\text{d}(h)\,\sqrt{d_j}\big(D^j(h)\big)^\dagger\psi(h),
\ee 
with the discrete measure $\mu_\text{d}$ on the group. The translation operator
\be
R^\sigma\psi(h)
=\psi(h\sigma)
\ee
acts in the spin representation as
\be
(R^\sigma\psi)(j)
=\big(D^j(\sigma)\big)\cdot\psi(j).
\ee

We can now also attempt to define a measure in the spin representation space which would recover the inner product. However, this can only be accomplished with a non-commutative product. We therefore define a measure 
\be\label{measureBF}
\mu_\text{BF}(\psi_\alpha)
=\delta(\alpha,\openone),
\ee
and introduce a non-commutative (matrix) product for the basis states via
\be
(\psi_\alpha\star\psi_\beta)(j)
=\psi_{\alpha\beta}(j),
\ee
which we extend by bi-linearity. We can then express the inner product \eqref{C12} as
\be
\langle\psi_1|\psi_2\rangle
=\mu_\text{BF}(\psi_1^\dagger\star\psi_2).
\ee

\subsection{Intertwiner and spin representation}
\label{intertwinerspin}

\noindent The spin representation discussed in the previous section can easily be extended from functions on $G$ to functions on $G^{|\ell|}$. For this, we consider states of the form
\be
\psi\{j\}_{\{MN\}}
=\sum_{\{\alpha \}}c_{\{\alpha\}}\left(\prod_\ell\sqrt{d_{j_\ell}}\,D^{j_\ell}(\alpha_\ell^{-1})_{M_\ell N_\ell}\right),
\ee
where we have used the basis states
\be
\psi_\alpha\{j\}_{\{MN\}}
=\left(\prod_\ell\sqrt{d_{j_\ell}}\,D^{j_\ell}(\alpha_\ell^{-1})_{M_\ell N_\ell}\right)
\ee
which are orthonormal in the inner product
\be\label{C25}
\langle\psi_1|\psi_2\rangle
=\lim_{\Lambda\rightarrow\infty}\f{1}{(\mathcal{N}(\Lambda))^{|\ell|}}\sum_{\{j\}}^\Lambda\tr\big(\psi_1^\dagger\{j\}\psi_2\{j\}\big).
\ee
Here $\tr(\psi_1^\dagger\psi_2)$ is a shorthand notation for
\be
\tr(\psi_1^\dagger \psi_2)
=\sum_{M_1,N_1,\dots,M_{|\ell|},N_{|\ell|}}(\overline{\psi_1})_{M_1N_1\dots M_{|\ell|}N_{|\ell|}}(\psi_2)_{M_1N_1\dots M_{|\ell|}N_{|\ell|}}.
\ee

We have therefore expressed the inner product \eqref{C25} as a summation over $j$ as well as on the magnetic indices $M_1,N_1,\dots,M_{|\ell|},N_{|\ell|}$. These latter transform under the adjoint action, namely in a representation
\be
V_{j_1}\otimes V_{j_1}^*\otimes\dots\otimes V_{ j_{|\ell|}}\otimes V_{j_{|\ell|}}^*.
\ee
Let us now choose a recoupling scheme and a unitary intertwining map
\be
U^{\{j\}}_{\mathcal{I},J,L;\{M,N\}}:V_{j_1}\otimes V_{j_1}^*\otimes\dots\otimes V_{ j_{|\ell|}}\otimes V_{j_{|\ell|}}^*\ \longrightarrow\ \bigoplus_{\mathcal{I},J}V_J,
\ee
where $J$ is labelling the resulting representation, $L$ is the associated magnetic index, and $\mathcal{I}$ summarizes the intertwiner (i.e. intermediate spin) labels. Such an intertwining map can be built from $(3jm)$ symbols, which couple two representations $j_1$ and $j_2$ to $j_3$. In this case (and with certain phase conventions), the matrix elements of $U$ are real. The condition of unitarity then means that
\be
\sum U^{\{j\}}_{\mathcal{I},J,L;\{M,N\}}U^{\{j\}}_{\mathcal{I}',J',L';\{M,N\}}
=\delta_{\mathcal{I},\mathcal{I}'}\delta_{J,J'}\delta_{L,L'},
\ee
and
\be\label{c30}
\sum_{\mathcal{I},J,L}U^{\{j\}}_{\mathcal{I},J,L;\{M,N\}}U^{\{j\}}_{\mathcal{I},J,L;\{M',N'\}}
=\delta_{\{M\},\{M'\}}\delta_{\{N\},\{N'\}}.
\ee

We can then change our spin representation to
\be
\psi(\{j\},\mathcal{I},J)_L
\coloneqq\sum_{\{MN\}}U^{\{j\}}_{\mathcal{I},J,L;\{M,N\}}\psi\{j\}_{\{MN\}},
\ee
and also rewrite the inner product \eqref{C25} as a sum over the $\{j\},\mathcal{I},J,L$ indices.

To obtain the (completely) gauge-invariant part of a wave function $\psi(\{j\},\mathcal{I},J)_L$, one would set $J=0$ (and hence $L=0$). Note however that a wave function with (Kronecker) $\delta_{J,0}$ behavior would have vanishing norm in the (rewritten) inner product \eqref{C25}. This is due to the scaling factor $\mathcal{N}(\Lambda)^{|\ell|}$ in \eqref{C25}). However, it is not possible to change this scaling factor such that a sufficiently large class of wave functions have finite norm. The reason for this is that the conjugacy classes $\{\alpha\}$ require different scaling behaviors depending on which type of adjoint orbit they describe.

We can nevertheless apply the rigging map to a basis state in the recoupled spin representation:
\be\label{c32}
\eta\big(\psi_{\{\alpha\}}\big)(\{j\},\mathcal{I},J)_L
=\int_{G/\text{stab}(u)}\de\mu_\text{d}(u)\sum_{\{M,N\}}\left(\prod_\ell\sqrt{d_{j_\ell}}\,\overline{D^{j_\ell}(u\alpha_\ell^{-1}u^{-1})}_{M_\ell N_\ell}\right)U^{\{j\}}_{\mathcal{I},J,L;\{M,N\}}.
\ee
Note that this will not give a regular function in the indices $(J,L)$ since the $J=0$ is given by a (discrete measure) integral over a constant. This integral only makes sense if one applies $\eta\big(\psi_{\{\alpha\}}\big)$ to a state in the kinematical Hilbert space.

Therefore, the physical inner product between two basis states can also be expressed as a sum over $\mathcal{I},J,L$ and $\{j\}$. The sum over $\mathcal{I},J,L$ leads, due to \eqref{c30}, to Kronecker delta symbols in the indices $\{M,N\}$, and the sum over these indices leads to the representation of the (Kronecker) delta in the group elements, i.e.
\be
\eta\big(\psi_{\{\alpha\}}\big)\psi_\beta
&=\int_{G/\text{stab}(u)}\de\mu_\text{d}(u)\prod_\ell\lim_{\Lambda\rightarrow\infty}\f{1}{\mathcal{N}(\Lambda)}\sum_{j=0}^\Lambda d_j\chi_j(u\alpha_\ell u^{-1}\beta_\ell^{-1}) \nn\\
&=\int_{G/\text{stab}(u)}\de\mu_\text{d}(u)\prod_\ell\delta\big(u\alpha_\ell u^{-1},\beta_\ell\big)\nn\\
&=\delta\big([\{\alpha\}],[\{\beta\}]\big).
\ee

\section{Spectrum of the translation operator for irrational angle}
\label{appendix:spectrum}

\noindent In this appendix, we prove that the translation operator $R^\phi$ acting on $L^2\big(\U(1),\de\mu_\text{d}\big)$ has a spectrum given by $\U(1)$ when the angle $\phi$ is irrational. This actually holds when equipping $\U(1)$ with either a continuous or a discrete topology. The difference between these two cases is just that the spectral measure for $L^2\big(\U(1),\de\mu_\text{d}\big)$ is continuous, while for $L^2\big(\U(1),\de\mu_\text{Haar}\big)$ it is discrete.

Let us introduce the ket $|n\rangle\coloneqq\psi_{\alpha-n\phi}$ for $n\in \mathbb{Z}$. Note that this is \emph{not} the usual eigenvector of the momentum operator for eigenvalue $n$. The translation operator $R^\phi$ acts as a right shift, i.e. $R^\phi|n\rangle=|n+1\rangle$. Let $e^{\i\rho}\in\U(1)$ for $\rho\in[0,2\pi)$. We are now going to show that $e^{\i\rho}$ is in the spectrum of $R^\phi$. This is equivalent to saying that the operator $R^\phi-e^{\i\rho}\openone$ does not have a bounded inverse. First, if this operator has no inverse, we have the desired result and there is nothing to show. If it does have an inverse, then we have to show that this inverse is not bounded. For this, it is sufficient to show that there exists a sequence $\{\Psi_N\}_N$ of vectors in $\mathcal{H}=L^2\big(\U(1),\mu_\text{d}\big)$ such that $\|\psi_N\|^2=1$ but $\big\|\big(R^\phi-e^{\i\rho}\openone\big)\psi_N\big\|^2\rightarrow0$. We now define this sequence. It is given by $\psi_N=\sum_{n\in\mathbb{Z}}c_n^{(N)}|n\rangle$, with
\be
c_n^{(N)}
=\left\{\begin{array}{cl}
\displaystyle\f{1}{\sqrt{2N+1}}e^{\i n\rho}&\text{if $|n|\leq N$},\\[10pt]
0&\text{otherwise}.
\end{array}\right.
\ee
This sequence satisfies $\|\Psi_N\|^2=1$, as well as
\be
\big\|\big(R^\phi-e^{\i\rho}\openone\big)\Psi_N\big\|^2
=\f{2}{2N+1}, 
\ee
where the right-hand side converges to zero. This finishes the proof.

In fact, the (formal) limit $\lim_{N\rightarrow\infty}\Psi_{N}$ does not converge in $\mathcal{H}$, but $\sqrt{2N+1}\Psi_N$ converges (in the dual space of a suitable dense subspace) to the generalized eigenvector \eqref{irreigen} corresponding to the point $e^{\i\rho}$ in the spectrum. The whole spectrum is $\U(1)$, and is continuous because the operator does not have any (normalizable) eigenvectors, but only non-normalizable ones.

\section{Refined trees}
\label{appendix:trees}

\subsection{On the notion of refined trees}

\begin{center}
\begin{figure}[h]
\includegraphics[scale=0.7]{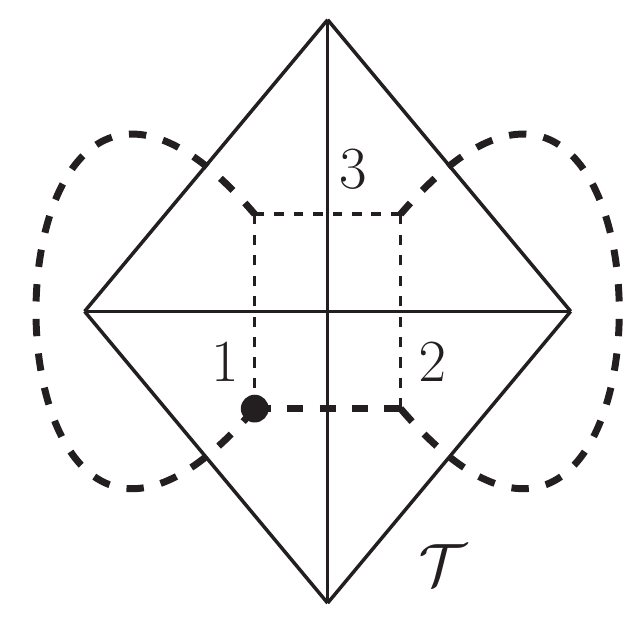}
\caption{Triangulation $\Delta$ with a tree $\mathcal{T}$ (thick dashes) in its dual graph, three leaves (thin dashes) labeling the three fundamental cycles, and a root $r$ (thick node). We have omitted the orientation of the simplices and of the dual graph for the sake of clarity.}
\label{fig:refine-tree-1}
\end{figure}
\end{center}

\begin{center}
\begin{figure}[h]
\includegraphics[scale=0.7]{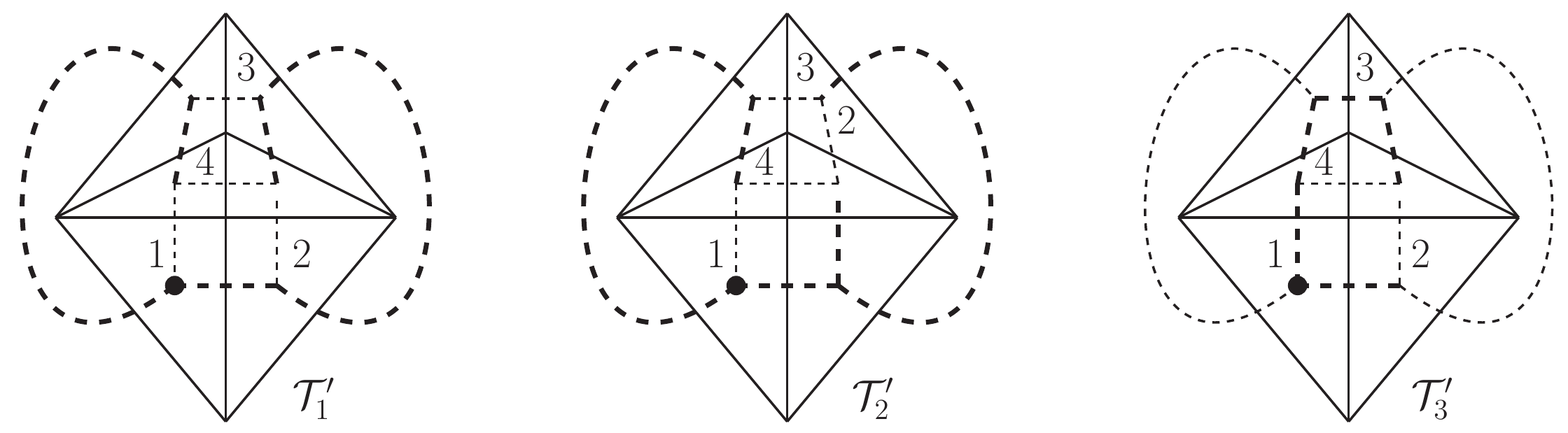}
\caption{Triangulation $\Delta'\succ\Delta$ with three different trees in its dual graph.}
\label{fig:refine-tree-2}
\end{figure}
\end{center}

\noindent In order to understand the subtleties underlying the notion of refined trees, let us focus on the example depicted on figure \ref{fig:refine-tree-1}. There, we have a triangulation $\Delta$ consisting of four triangles, a tree $\mathcal{T}$ in its dual graph, and three leaves, $\ell_1$, $\ell_2$ and $\ell_3$, labeling the fundamental cycles $\gamma_{\ell_1}$, $\gamma_{\ell_2}$ and $\gamma_{\ell_3}$. By subdividing the edge shared by the two upper triangles we get a finer triangulation $\Delta'$, and we have represented on figure \ref{fig:refine-tree-2} three possible trees (there are finitely many other possibilities) in its dual graph $\Gamma'$. For each tree $\mathcal{T}'$ and the corresponding set $\mathcal{L}'$ of leaves, we can build paths $\gamma_{\ell'}$ describing the fundamental cycles by starting at the root, going along the tree until the source of $\ell'$, then going along $\ell'$, and back to the root along the tree. Then, we can project these paths with $\mathbf{P}$, and compare the resulting projections with the fundamental cycles described by $\mathcal{T}$. Let us do this for the three trees $\mathcal{T}'$ represented on figure \ref{fig:refine-tree-2}.
\begin{enumerate}
\item With $\mathcal{T}'_1$, we have that $\mathbf{P}(\gamma_{\ell'_1})=\gamma_{\ell_1}$, $\mathbf{P}(\gamma_{\ell'_2})=\gamma_{\ell_2}$, and $\mathbf{P}(\gamma_{\ell'_3})=\mathbf{P}(\gamma_{\ell'_4})=\gamma_{\ell_3}$. This means that the projections of the fundamental cycles described by the leaves of $\mathcal{T}'_1$ correspond to the fundamental cycles described by the leaves $\mathcal{L}$ of $\mathcal{T}$. In addition to this, one can see that all possible paths in $\mathcal{T}'_1$ are mapped under $\mathbf{P}$ to paths in $\mathcal{T}$.
\item With $\mathcal{T}'_2$, we have that $\mathbf{P}(\gamma_{\ell'_1})=\gamma_{\ell_1}$, $\mathbf{P}(\gamma_{\ell'_2})=\gamma_{\ell_2}$, and $\mathbf{P}(\gamma_{\ell'_3})=\gamma_{\ell_3}$, while $\mathbf{P}(\gamma_{\ell'_4})$ can be described in terms of the cycles $\gamma_{\ell_2}$ and $\gamma_{\ell_3}$ in $\Gamma$. However, at the difference with the tree $\mathcal{T}'_1$, one can see that the tree $\mathcal{T}'_2$ has paths which under $\mathbf{P}$ are not mapped to paths in $\mathcal{T}$.
\item With $\mathcal{T}'_3$, there are paths which under $\mathbf{P}$ are not mapped to paths in $\mathcal{T}$, and the projections of the fundamental cycles do not correspond to the fundamental cycles determined by $\mathcal{T}$. Indeed, although we have that $\mathbf{P}(\gamma_{\ell'_1})=\gamma_{\ell_1}$, there is no leaf $\ell'$ in $\mathcal{T}'_3$ such that $\mathbf{P}(\gamma_{\ell'})=\gamma_{\ell_2}$ or $\mathbf{P}(\gamma_{\ell'})=\gamma_{\ell_3}$.
\end{enumerate}

With this example, we see that the different possible trees which can be chosen in the refined graph $\Gamma'$ do not characterize in an equivalent manner under $\mathbf{P}$ the paths and fundamental cycles described by the coarser tree $\mathcal{T}$.

\subsection{Proof that all refined trees arise from the construction procedure}
\label{app:Construction}

\begin{Lemma}
Let $\Delta\prec\Delta'$, and $\mathcal{T}$ be a maximal tree with a refined tree $\mathcal{T}'$. Then $\mathcal{T}'$ is a result of the construction procedure described in section \ref{refinementtrees}.
\end{Lemma}

In order to prove this result, we first look at a single $d$-dimensional simplex $\sigma_\Delta^d\subset\Delta$. Consider the edges of $\mathcal{T}'$ which start and end in this simplex, i.e. which start and end at vertices dual to $d$-dimensional simplices $\sigma_{\Delta'}^d\subset\mathbf{P}^{-1}(\sigma_\Delta^d)$. This collection gives a sub-tree of $\mathcal{T}'$, which is connected due to property $(ii)$ of a refined tree.

Next, consider a $(d-1)$-dimensional simplex $\sigma_\Delta^{d-1}$ dual to a branch in $\mathcal{T}$. Because of $(i)$, $\mathbf{P}^{-1}(\sigma_\Delta^{d-1})$ contains at least one $(d-1)$-dimensional simplex dual to a branch in $\mathcal{T}'$. If it did contain more than one, one could construct a closed loop in $\mathcal{T}'$ because of property $(i)$. This means that $\mathbf{P}^{-1}(\sigma_\Delta^{d-1})$ contains precisely one branch in $\mathcal{T}'$.

All that remains to show is that the branches which we have singled out so far are the only ones in $\mathcal{T}'$. But the only edges which we have not considered yet are the ones which are mapped under $\mathbf{P}$ to a leaf in $\mathcal{T}$. So none of these edges can be a branch, because of property $(i)$, and we are therefore done. \qed

\subsection{Proof of the transitivity of refined trees}
\label{proof:Ttransitivity}

\noindent
\textbf{Lemma \ref{lemma:Ttransitivity}} (Transitivity of the refined trees)\textbf{.}
\textit{Let us consider three triangulations $\Delta\prec\Delta'\prec\Delta''$. Let $\mathcal{T}'$ be a refined tree for $\mathcal{T}$, and $\mathcal{T}''$ be a refined tree for $\mathcal{T}'$. Then we have that $\mathcal{T}''$ is also a refined tree for $\mathcal{T}$.}\\

In order to prove this lemma, we need to show that the tree $\mathcal{T}''$ satisfies the properties $(i)$ and $(ii)$ of definition \ref{def:refine-tree} with respect to the tree $\mathcal{T}$.

Property $(i)$ follows from result \eqref{Ptransitivity} of lemma \ref{lemma:Ptransitivity}, and from the fact that the trees $\mathcal{T}'$ and $\mathcal{T}''$ are refinements of $\mathcal{T}$ and $\mathcal{T}'$ respectively. Indeed, since $\mathbf{P}_{\Delta',\Delta}(\mathcal{T}')=\mathcal{T}$ and $\mathbf{P}_{\Delta'',\Delta'}(\mathcal{T}'')=\mathcal{T}'$, by virtue of \eqref{Ptransitivity} we also have that $\mathbf{P}_{\Delta'',\Delta}(\mathcal{T}'')=\mathcal{T}$.


For property $(ii)$, we have to show that any two nodes $n''_1$ and $n''_2$, such that the dual simplices $\sigma_{\Delta''}^d(n_1'')$ and $\sigma_{\Delta''}^d(n''_2)$ are in $\mathbf{P}^{-1}_{\Delta'',\Delta}(\sigma^d_\Delta)$ for a fixed $\sigma^d_\Delta$, can be connected by a path in $\mathcal{T}''$ without crossing the boundary $\partial \big(\mathbf{P}^{-1}_{\Delta'',\Delta}(\sigma^d_\Delta)\big)=\mathbf{P}^{-1}_{\Delta'',\Delta}(\partial\sigma^d_\Delta)$. Let $(\sigma^d_{\Delta'})_1$ be the simplex in $\Delta'$ containing the simplex $\sigma_{\Delta''}^d(n_1'')$ dual to $n_1''$, i.e. such that $\sigma_{\Delta''}^d(n_1'')\in\mathbf{P}^{-1}_{\Delta',\Delta}\big((\sigma^d_{\Delta'})_1\big)$, and likewise for $(\sigma^d_{\Delta'})_2$. We have $(\sigma^d_{\Delta'})_i\in\mathbf{P}^{-1}_{\Delta',\Delta}(\sigma^d_\Delta)$ for $i=1,2$. Then the following two situations can occur:
\begin{enumerate}
\item If $(\sigma^d_{\Delta'})_1=(\sigma^d_{\Delta'})_2$, the tree path from $n''_1$ to $n''_2$  is ``inside'' $\mathbf{P}^{-1}_{\Delta',\Delta}\big((\sigma^d_{\Delta'})_1\big)=\mathbf{P}^{-1}_{\Delta',\Delta}\big((\sigma^d_{\Delta'})_2\big)$ by virtue of property $(ii)$ with respect to the pair $(\mathcal{T}',\mathcal{T}'')$ of trees. Therefore, this path does also not cross the boundary of $\mathbf{P}^{-1}_{\Delta'',\Delta}(\sigma^d_\Delta)$, which shows that $(ii)$ holds in this case also for the pair $(\mathcal{T},\mathcal{T}'')$.
\item If $(\sigma^d_{\Delta'})_1\neq(\sigma^d_{\Delta'})_2$, the path from $n''_1$ to $n''_2$  along $\mathcal{T}''$ projects under $\mathbf{P}_{\Delta'',\Delta'}$ to a path along the tree $\mathcal{T}'$, which is from the node dual to $(\sigma^d_{\Delta'})_1$ to the node dual to $(\sigma^d_{\Delta'})_2$. By virtue of property $(ii)$ for the pair $(\mathcal{T},\mathcal{T}')$ of trees, the path does not cross the boundary of $\mathbf{P}^{-1}_{\Delta',\Delta}(\sigma^d_\Delta)$. Together with the first point above, this shows that also the original path from $n''_1$ to $n''_2$ along $\mathcal{T}''$ does not cross the boundary of $\mathbf{P}^{-1}_{\Delta'',\Delta}(\partial\sigma^d_\Delta)$.
\end{enumerate}
Therefore, $(ii)$ holds for the pair $(\mathcal{T},\mathcal{T}'')$ in general, which finishes the proof. \qed

\section{Splitting of the observable algebra}
\label{proof:separation}

\noindent
\textbf{Lemma \ref{lemma:separation}.}
\textit{The coarse observables \eqref{HFcoarse} and the fine observables \eqref{HFfine} form two mutually-commuting subalgebras of the rooted holonomy-flux algebra $\mathfrak{A}^r$, which we denote respectively by $\mathfrak{A}^r_\mathrm{c}$ and $\mathfrak{A}^r_\mathrm{f}$.}\\

The proof of lemma \ref{lemma:separation} can be obtained by a direct calculation. First of all, it is clear that $\mathfrak{A}^r_\mathrm{c}$ and $\mathfrak{A}^r_\mathrm{f}$ each form a subalgebra which reproduces the Poisson structure on $T^*G$. The nontrivial statement is to show that these subalgebras are mutually commuting.

Since holonomies are always Poisson-commuting, the holonomies in $\mathfrak{A}^r_\mathrm{c}$ have vanishing Poisson brackets with the holonomies in $\mathfrak{A}^r_\mathrm{f}$. In addition to this, since the leaves $\ell'_i$ and $\ell'_I$ are always distinct, the holonomies in $\mathfrak{A}^r_\mathrm{c}$ commute with all the fluxes in $\mathfrak{A}^r_\mathrm{f}$.

Let us now turn to the commutation relations between the fluxes in $\mathfrak{A}^r_\mathrm{c}$ and the holonomies in $\mathfrak{A}^r_\mathrm{f}$. In \eqref{Hfine1}, in addition to the leaf $\ell'_i$ which always exists in $\mathbf{P}_\text{L}^{-1}(\ell_i)$, we have a leaf $\ell'_I$ in the same pre-image. Therefore, for a given $\ell_i$ the Poisson bracket between $\bX'_{\ell_i}$ and $\mathcal{C}_I$ contains two contributions, which are in turn vanishing since
\be
\lb(\bX'_{\ell_i})^k,\mathcal{C}_I\rb
=\lb\bX^k_{\ell'_i}+\bX^k_{\ell'_I},g_{\ell'_i}g_{\ell'_I}^{-1}\rb
=g_{\ell'_i}\tau^kg_{\ell'_I}^{-1}-g_{\ell'_i}\tau^kg_{\ell'_I}^{-1}
=0.
\ee
Similarly, for \eqref{Hfine2} we have that
\be
\lb(\bX'_{\ell_i})^k,\mathcal{C}_I\rb
=\lb\bX^k_{\ell'_i}+\bX^k_{(\ell'_I)^{-1}},g_{\ell'_i}g_{\ell'_I}\rb
=g_{\ell'_i}\tau^kg_{\ell'_I}-g_{\ell'_i}\tau^kg_{\ell'_I}
=0.
\ee
For \eqref{Hfine3}, since the leaf $\ell'_I$ does not belong to the pre-image of any leaf $\ell_i$, it does not appear in the coarse observable $\bX'_{\ell_i}$ and therefore the Poisson brackets $\lb\bX'_{\ell_i},g_{\ell'_I}^{-1}\rb$ are identically vanishing. This shows that the fluxes in $\mathfrak{A}^r_\mathrm{c}$ commute with the holonomies in $\mathfrak{A}^r_\mathrm{f}$.

Finally, let us look at the commutation relations between the fluxes in $\mathfrak{A}^r_\mathrm{c}$ and the fluxes in $\mathfrak{A}^r_\mathrm{f}$. For \eqref{Hfine1}, since $\ell'_I\in\mathbf{P}_\text{L}^{-1}(\ell_i)$, the unique contribution coming from this finer leaf to \eqref{fluxrefined} will be a flux $\bX_{\ell'_I}$. However, because of the reversed orientation of the leaf in \eqref{Hfine1}, the Poisson brackets are vanishing since $\lb\bX_{\ell'_I},\bX_{(\ell'_I)^{-1}}\rb=0$. A similar reasoning applies to \eqref{Hfine2}. For \eqref{Hfine3}, since $\ell'_I\in\mathcal{L}'_0$, the flux $\bX_{(\ell'_I)^{-1}}$ never appears in \eqref{fluxrefined} and the Poisson brackets are therefore identically vanishing.
\qed


\begin{thebibliography}{99}

\bibitem{regge} T. Regge,
``General Relativity Without Coordinates'',
Nuovo Cim. \textbf{19}, 558 (1961).

\bibitem{bahrdittrich09a} B. Bahr and B. Dittrich,
``(Broken) gauge symmetries and constraints in Regge calculus'',
Class. Quant. Grav. \textbf{26} 225011 (2009), \texttt{arXiv:0905.1670 [gr-qc]}.

\bibitem{dittrichsteinhaus} B. Dittrich and S. Steinhaus,
``Path integral measure and triangulation independence in discrete gravity'',
Phys. Rev. D \textbf{85} 044032 (2012), \texttt{arXiv:1110.6866 [gr-qc]}.

\bibitem{dittrichkaminskisteinhaus} B. Dittrich, W. Kaminski and S. Steinhaus,
``Discretization independence implies non-locality in 4D discrete quantum gravity,''
Class. Quant. Grav. \textbf{31} 245009 (2014), \texttt{arXiv:1404.5288[gr-qc]}.

\bibitem{lqg1} C. Rovelli,
\textit{Quantum Gravity},
(Cambridge University Press, Cambridge, 2004).

\bibitem{lqg2} A. Ashtekar and J. Lewandowski,
``Background independent quantum gravity: a status report'',
Class. Quant. Grav. \textbf{21} R53 (2004), \texttt{arXiv:gr-qc/0404018}.

\bibitem{lqg3} T. Thiemann,
\textit{Introduction to Modern Canonical Quantum General Relativity},
(Cambridge University Press, Cambridge, 2007).

\bibitem{ali1} A. Ashtekar and C. J. Isham,
``Representations of the holonomy algebras of gravity and non-Abelian gauge theories'',
Class. Quantum Grav. \textbf{9} 1433 (1992), \texttt{arXiv:hep-th/9202053}.

\bibitem{ali2} A. Ashtekar and J. Lewandowski,
``Representation theory of analytic holonomy C* algebras'',
in J. Baez, ed., \textit{Knots and Quantum Gravity},
(Oxford University Press, 1994), \texttt{arXiv:gr-qc/9311010}.

\bibitem{ali3} A. Ashtekar and J. Lewandowski,
``Projective techniques and functional integration for gauge theories'',
J. Math. Phys. \textbf{36} (1995) 2170, \texttt{arXiv:gr-qc/9411046}.

\bibitem{ali4} A. Ashtekar and J. Lewandowski,
``Differential geometry on the space of connections via graphs and projective limits'',
J. Geom. Phys. \textbf{17} (1995) 191, \texttt{arXiv:hep-th/9412073}.

\bibitem{barbero} J. F. Barbero,
``Real Ashtekar variables for Lorentzian signature spacetimes'',
Phys. Rev. \textbf{D 51} 5507 (1995).

\bibitem{immirzi} G. Immirzi,
``Real and complex connections for canonical gravity'',
Class. Quant. Grav. \textbf{14} L177 (1997), \texttt{arXiv:gr-qc/9612030}.

\bibitem{ashtekardiffeos} A. Ashtekar,
``Some surprising implications of background independence in canonical quantum gravity'',
Gen. Rel. Grav. \textbf{41} 1927 (2009), \texttt{arXiv:0904.0184 [gr-qc]}.

\bibitem{thiemannH1} T. Thiemann,
``Anomaly-free formulation of nonperturbative, four-dimensional Lorentzian quantum gravity'',
Phys. Lett. B \textbf{380} 257 (1996), \texttt{arXiv:gr-qc/9606088}.

\bibitem{thiemannH2} T. Thiemann,
``Quantum spin dynamics (QSD)'',
Class. Quant. Grav. \textbf{15} 839 (1998), \texttt{arXiv:gr-qc/9606089}.

\bibitem{newvac} B. Dittrich and M. Geiller,
``A new vacuum for loop quantum gravity'',
Class. Quant. Grav. (2014), \texttt{arXiv:1401.6441 [gr-qc]}.

\bibitem{fluxC} B. Dittrich and M. Geiller,
``Flux formulation of loop quantum gravity: Classical framework'',
Class. Quant. Grav. (2015), \texttt{arXiv:1412.3752 [gr-qc]}.

\bibitem{dittrichryan1} B. Dittrich and J. P. Ryan,
``Phase space descriptions for simplicial 4d geometries'',
Class. Quant. Grav. \textbf{28} 065006 (2011), \texttt{arXiv:0807.2806 [gr-qc]}.

\bibitem{dittrichryan2} B. Dittrich and J. P. Ryan,
``Simplicity in simplicial phase space'',
Phys. Rev. D \textbf{82} 064026 (2010), \texttt{arXiv:1006.4295 [gr-qc]}.

\bibitem{twisted} L. Freidel and S. Speziale,
``Twisted geometries: A geometric parametrisation of SU(2) phase space'',
Phys. Rev. D \textbf{82} 084040 (2010), \texttt{arXiv:1001.2748 [gr-qc]}.

\bibitem{FGZ} L. Freidel, M. Geiller and J. Ziprick,
``Continuous formulation of the loop quantum gravity phase space",
Class. Quant. Grav. \textbf{30} 085013 (2013), \texttt{arXiv:1110.4833 [gr-qc]}.

\bibitem{KS1} T. A. Koslowski,
``Dynamical quantum geometry (DQG programme)",
(2007), \texttt{arXiv:0709.3465 [gr-qc]}.

\bibitem{KS2} H. Sahlmann,
``On loop quantum gravity kinematics with non-degenerate spatial background",
Class. Quant. Grav. \textbf{27} 225007 (2010), \texttt{arXiv:1006.0388 [gr-qc]}.

\bibitem{KS3} T. Koslowski and H. Sahlmann,
``Loop quantum gravity vacuum with nondegenerate geometry'',
Sigma \textbf{8} 026 (2012), \texttt{arXiv:1109.4688 [gr-qc]}.

\bibitem{varadarajan1} M. Varadarajan,
``The generator of spatial diffeomorphisms in the Koslowski--Sahlmann representation",
Class. Quant. Grav. \textbf{30} 175017 (2013), \texttt{arXiv:1306.6126 [gr-qc]}.

\bibitem{varadarajan2} M. Varadarajan and M. Campiglia,
``The Koslowski--Sahlmann representation: Gauge and diffeomorphism invariance",
(2013), \texttt{arXiv:1311.6117 [gr-qc]}.

\bibitem{varadarajan3} M. Varadarajan and M. Campiglia,
``The Koslowski--Sahlmann representation: Quantum configuration space",
(2014), \texttt{arXiv:1406.0579 [gr-qc]}.

\bibitem{thooft} G. 't Hooft,
``A locally finite model for gravity'',
Found. Phys. \textbf{38} 733--757 (2008), \texttt{arXiv:0804.0328 [gr-qc]}.

\bibitem{pullin} R. Gambini, J. Griego and J. Pullin,
``Chern--Simons states in spin network quantum gravity'',
Phys. Lett. B \textbf{413} (1997) 260, \texttt{arXiv:gr-qc/9703042}.

\bibitem{lewandowski2surface} M. Bobienski, J. Lewandowski and M. Mroczek,
``A two surface quantization of the Lorentzian gravity'',
(2001), \texttt{arXiv:gr-qc/0101069}.

\bibitem{bianchi} E. Bianchi,
``Loop quantum gravity \`a la Aharonov--Bohm'',
(2009), \texttt{arXiv:0907.4388 [gr-qc]}.

\bibitem{MeusburgerNoui2} C. Meusburger and K. Noui,
``The Hilbert space of 3d gravity: quantum group symmetries and observables'',
Adv. Theor. Math. Phys. \textbf{14}, 6 (2010) 1651--1716, \texttt{arXiv:0809.2875 [gr-qc]}.

\bibitem{aldo} H. M. Haggard, M. Han, W. Kami\'nski and A. Riello,
``SL(2,C) Chern--Simons Theory, a non-planar graph operator, and 4D loop quantum gravity with a cosmological constant: semi-classical geometry'',
(2014), \texttt{arXiv:1412.7546 [hep-th]}.

\bibitem{aldo2} H. M. Haggard, M. Han and A. Riello,
``Encoding curved tetrahedra in face holonomies: a phase space of shapes from group-valued moment maps''
(2015), \texttt{arXiv:1506.03053 [math-ph]}.

\bibitem{Fock:1998nu} V. V. Fock and A. A. Rosly,
``Poisson structure on moduli of flat connections on Riemann surfaces and r-matrix'',
Am. Math. Soc. Transl. \textbf{191}, 67--86 (1999), \texttt{arXiv:math/9802054}.

\bibitem{Alekseev:1994pa} A. Y. Alekseev, H. Grosse and V. Schomerus,
``Combinatorial quantization of the Hamiltonian Chern--Simons theory'',
Commun. Math. Phys. \textbf{172}, 317--358 (1995), \texttt{arXiv:hep-th/9403066}.

\bibitem{Alekseev:1994au} A. Y. Alekseev, H. Grosse and V. Schomerus,
``Combinatorial quantization of the Hamiltonian Chern--Simons theory. 2.'',
Commun. Math. Phys. \textbf{174}, 561--604 (1995), \texttt{arXiv:hep-th/9408097}.

\bibitem{Alekseev:1995rn} A. Y. Alekseev and V. Schomerus,
``Representation theory of Chern--Simons observables'',
(1995), \texttt{arXiv:q-alg/9503016}.

\bibitem{Meusburger:2003ta} C. Meusburger and B. Schroers,
``Poisson structure and symmetry in the Chern--Simons formulation of (2+1)-dimensional gravity'',
Class. Quant. Grav. \textbf{20}, 2193--2234 (2003), \texttt{arXiv:gr-qc/0301108}.

\bibitem{Meusburger:2003hc} C. Meusburger and B. Schroers,
``The Quantization of Poisson structures arising in Chern--Simons theory with gauge group $G\times\mathfrak{g}^*$'',
Adv. Theor. Math. Phys. \textbf{7}, 1003--1043 (2004), \texttt{arXiv:hep-th/0310218}.

\bibitem{MeusburgerNoui} C. Meusburger and K. Noui,
``Combinatorial quantisation of the Euclidean torus universe'',
Nucl. Phys. B \textbf{841}, 463--505 (2010), \texttt{arXiv:1007.4615}.

\bibitem{timeevol} B. Dittrich and S. Steinhaus,
``Time evolution as refining, coarse graining and entangling",
(2013), \texttt{arXiv:1311.7565 [gr-qc]}.

\bibitem{horowitz} G. T. Horowitz,
``Exactly soluble diffeomorphism invariant theories",
Comm. Math. Phys. \textbf{125} 417 (1989).

\bibitem{baez} J. C. Baez,
``Spin foam models",
Class. Quant. Grav. \textbf{15} 1827 (1998), \texttt{arXiv:gr-qc/9709052}.

\bibitem{oriti-thesis} D. Oriti,
``Spin foam models of quantum spacetime",
(2003), \texttt{arXiv:gr-qc/0311066}.

\bibitem{perez-review} A. Perez,
``Spin foam models for quantum gravity",
Class. Quant. Grav. \textbf{20} R43 (2003), \texttt{arXiv:gr-qc/0301113}.

\bibitem{BFYangMills} D. Birmingham, M. Blau, M. Rakowski and G. Thompson,
``Topological field theory'',
Phys. Rep. \textbf{209} (1991) 129--340.

\bibitem{DefectsCM} L. Kong,  X. G. Wen,
``Braided fusion categories, gravitational anomalies, and the mathematical framework for topological orders in any dimensions'',
(2014), \texttt{arXiv:1405.5858 [cond-mat.str-el]}.

\bibitem{okolov} A. Oko\l\'ow,
``Quantization of diffeomorphism invariant theories of connections with a non-compact structure group - an example'',
Commun. Math. Phys. \textbf{289} 335--382 (2009), \texttt{arXiv:gr-qc/0605138}.

\bibitem{AshBojLewan} A. Ashtekar, M. Bojowald and J. Lewandowski,
``Mathematical structure of loop quantum cosmology'',
Adv. Theor. Math. Phys. \textbf{7} 233--268 (2003), \texttt{arXiv:gr-qc/0304074}.

\bibitem{LOST} J. Lewandowski, A. Oko\l\'ow, H. Sahlmann and T. Thiemann,
``Uniqueness of diffeomorphism-invariant states on holonomy-flux algebras'',
Commun. Math. Phys. \textbf{267} 703 (2006), \texttt{arXiv:gr-qc/0504147}.

\bibitem{fleischhack} C. Fleischhack,
``Representations of the Weyl algebra in quantum geometry'',
Commun. Math. Phys. \textbf{285} 67 (2009), \texttt{arXiv:math-ph/0407006}.

\bibitem{balsam} A. Kirillov Jr. and B. Balsam,
``Turaev--Viro invariants as an extended TQFT'',
(2010), \texttt{arXiv:1004.1533 [math.GT]}.

\bibitem{TV} V. Turaev and O. Viro,
``State sum invariants of 3 manifolds and quantum 6j symbols'',
Topology \textbf{31} 865 (1992).

\bibitem{bdtoappear} B. Dittrich,
``Yet another vacuum for loop quantum gravity",
to appear.

\bibitem{atiyah} M. Atiyah and R. Bott,
``The Yang--Mills equations over Riemann surfaces'',
Philos. Trans. Roy. Soc. London Ser. A \textbf{308} 523--615 (1982).

\bibitem{goldman} W. Goldman,
``The symplectic nature of fundamental groups of surfaces'',
Adv. Math. \textbf{54} 200--225 (1984).

\bibitem{hatcher} A. Hatcher,
\textit{Algebraic Topology},
(Cambridge University Press, Cambridge, 2001).

\bibitem{benjaminBF} B. Bahr and C. Fleischhack,
to appear.

\bibitem{bonzom-smerlak1} V. Bonzom and M. Smarlak,
``Bubble divergences from twisted cohomology'',
Commun. Math. Phys. \textbf{312} 2 399--426 (2012), \texttt{arXiv:1008.1476 [math-ph]}.

\bibitem{bonzom-smerlak2} V. Bonzom and M. Smarlak,
``Bubble divergences: sorting out topology from cell structure'',
Annales Henri Poincar\'e \textbf{13} (2012) 185--208, \texttt{arXiv:1103.3961 [gr-qc]}.

\bibitem{heatkernel} B. C. Hall, 
``Harmonic analysis with respect to heat kernel measure",
Bull. (N.S.) Amer. Math. Soc. \textbf{38} (2001) 43, \texttt{arXiv:quant-ph/0006037}.

\bibitem{bahrthiemannK} B. Bahr and T. Thiemann,
``Gauge-invariant coherent states for loop quantum gravity II: Non-abelian gauge groups'',
Class. Quant. Grav. \textbf{26} 045012 (2009), \texttt{arXiv:0709.4636 [gr-qc]}.

\bibitem{RAQ}  A. Ashtekar and R. S. Tate,
``An algebraic extension of Dirac quantization: Examples'', 
J. Math. Phys. \textbf{35} (1994) 6434.

\bibitem{1308.5648} J. Diaz-Polo and I. Garay,
``semi-classical states in quantum gravity: Curvature associated to a Voronoi graph'',
Class. Quant. Grav. \textbf{31} 085018 (2014), \texttt{arXiv:1308.5648 [gr-qc]}.

\bibitem{gr-qc/0603008} E. R. Livine and D. R. Terno,
``Reconstructing quantum geometry from quantum information: Area renormalisation, coarse-graining and entanglement on spin networks'',
(2006), \texttt{arXiv:gr-qc/0603008}.

\bibitem{RovelliSmolin} C. Rovelli and L. Smolin,
``Discreteness of area and volume in quantum gravity'',
Nucl. Phys. \textbf{B 442} 593 (1995), \texttt{arXiv:gr-qc/9411005}.

\bibitem{AshtekarArea} A. Ashtekar and J. Lewandowski,
``Quantum theory of geometry I: Area operators'',
Class. Quant. Grav. \textbf{14} A55 (1997), \texttt{arXiv:gr-qc/9602046}.

\bibitem{improvedmu} A. Ashtekar, T. Pawlowski and P. Singh,
``Quantum nature of the Big Bang: Improved dynamics'',
Phys. Rev. D \textbf{74} 084003 (2006), \texttt{arXiv:gr-qc/0607039}.

\bibitem{eteradeformed} E. R. Livine,
``Deformation operators of spin networks and coarse-graining'',
Class. Quant. Grav. \textbf{31} 075004 (2014), \texttt{arXiv:1310.3362 [gr-qc]}.
  
\bibitem{howmany} S. Ariwahjoedi, J. S. Kosasih, C. Rovelli and F. P. Zen,
``How many quanta are there in a quantum spacetime?'',
(2014), \texttt{arXiv:1404.1750 [gr-qc]}.

\bibitem{clement2} B. Dittrich, C. Delcamp,
``On the coarse graining of spin networks and entanglement entropy",
to appear.

\bibitem{lanery1} S. Lan\'ery and T. Thiemann,
``Projective limits of state spaces I. Classical formalism'',
(2014), \texttt{arXiv:1411.3589 [gr-qc]}.

\bibitem{lanery2} S. Lan\'ery and T. Thiemann,
``Projective limits of state spaces II. Quantum formalism'',
(2014), \texttt{arXiv:1411.3590 [gr-qc]}.

\bibitem{lanery3} S. Lan\'ery and T. Thiemann,
``Projective limits of state spaces III. Toy-models'',
(2014), \texttt{arXiv:1411.3591 [gr-qc]}.

\bibitem{lanery4} S. Lan\'ery and T. Thiemann,
``Projective loop quantum gravity I. State space'',
(2014), \texttt{arXiv:1411.3592 [gr-qc]}.

\bibitem{chaostoappear} B. Dittrich, P. Hoehn, T. Koslowski and M. Nelson,
``Can chaos be observed in quantum gravity?",
to appear.

\bibitem{BHentropyLQG} C. Rovelli,
``Black hole entropy from loop quantum gravity'',
Phys. Rev. Lett. \textbf{77} 3288-3291 (1996), \texttt{arXiv:gr-qc/9603063}.

\bibitem{AshKras} A. Ashtekar, J. Baez and K. Krasnov,
``Quantum geometry of isolated horizons and black hole entropy'',
Adv. Theor. Math. Phys. \textbf{4} 1 (2000), \texttt{arXiv:gr-qc/0005126}.

\bibitem{discretespectra?} B. Dittrich and T. Thiemann,
``Are the spectra of geometrical operators in loop quantum gravity really discrete?'',
J. Math. Phys \textbf{50} 012503 (2009), \texttt{arXiv:0708.1721 [gr-qc]}.

\bibitem{hannoBH} H. Sahlmann,
``Black hole horizons from within loop quantum gravity''
Phys. Rev. D \textbf{84} 044049 (2011), \texttt{arXiv:1104.4691 [gr-qc]}.

\bibitem{ihorizon} A. Ashtekar and B. Krishnan,
``Isolated and dynamical horizons and their applications'',
Living Rev. Rel. \textbf{7} 10 (2004), \texttt{arXiv:gr-qc/0407042}.

\bibitem{areaangle} B. Dittrich and S. Speziale,
``Area-angle variables for general relativity'',
New J. Phys. \textbf{10} 083006 (2008), \texttt{arXiv:0802.0864 [gr-qc]}.

\bibitem{bonzomarea} V. Bonzom,
``From lattice BF gauge theory to area-angle Regge calculus'',
Class. Quant. Grav. \textbf{26} 155020 (2009), \texttt{arXiv:0903.0267 [gr-qc]}.

\bibitem{thooftstudent} M. van de Meent,
``Piecewise flat gravitational waves",
Class. Quant. Grav. \textbf{28} 075005 (2011), \texttt{arXiv:1012.1991 [gr-qc]}.

\bibitem{aristideandco} A. Baratin, B. Dittrich, D. Oriti and J. Tambornino,
``Non-commutative flux representation for loop quantum gravity'',
Class. Quant. Grav. \textbf{28} 175011 (2011), \texttt{arXiv:1004.3450 [hep-th]}.

\bibitem{guedes} B. Dittrich, C. Guedes and D. Oriti,
``On the space of generalized fluxes for loop quantum gravity",
Class. Quant. Grav. \textbf{30} 055008 (2013), \texttt{arXiv:1205.6166 [gr-qc]}.

\bibitem{zapataBF} J. A. Zapata,
``Topological lattice gravity using selfdual variables'',
Class. Quant. Grav. \textbf{13} 2617 (1996), \texttt{arXiv:gr-qc/9603030}.

\bibitem{Lewandowskietal} R. Gambini, J. Lewandowski, D. Marolf and J. Pullin,
``On the consistency of the constraint algebra in spin network quantum gravity'',
Int. J. Mod. Phys. D \textbf{7} (1998) 97--109, \texttt{arXiv:gr-qc/9710018}.

\bibitem{bddiffeo} B. Dittrich,
``Diffeomorphism symmetry in quantum gravity models",
Adv. Sci. Lett. \textbf{2} (2009) 151, \texttt{arXiv:0810.3594 [gr-qc]}.

\bibitem{bdreview12} B. Dittrich,
``How to construct diffeomorphism symmetry on the lattice,"
PoS (QGQGS 2011) 012, \texttt{arXiv:1201.3840 [gr-qc]}.

\bibitem{wolfgangH}   W. M. Wieland,
``Hamiltonian spinfoam gravity'',
Class. Quant. Grav. \textbf{31} (2014) 025002, \texttt{arXiv:1301.5859 [gr-qc]}.

\bibitem{wolfgang} W. M. Wieland,
``New action for simplicial gravity in four dimensions'',
Class. Quant. Grav. \textbf{32} 015016 (2015), \texttt{arXiv:1407.0025 [gr-qc]}.

\bibitem{consistentdiscr} R. Gambini, J. Pullin,
``Consistent discretization and loop quantum geometry",
Phys. Rev. Lett. \textbf{94} (2005) 101302, \texttt{arXiv:gr-qc/0409057}.

\bibitem{bdhoehn2} B. Dittrich and P. A. Hohn,
``Canonical simplicial gravity'',
Class. Quant. Grav. \textbf{29} (2012) 115009, \texttt{arXiv:1108.1974 [gr-qc]}.

\bibitem{bdhoehn3} B. Dittrich and P. A. Hoehn,
``Constraint analysis for variational discrete systems'',
J. Math. Phys. \textbf{54} (2013) 093505, \texttt{arXiv:1303.4294 [math-ph]}.

\bibitem{uniform} M. Campiglia, C. Di Bartolo, R. Gambini, J. Pullin,
``Uniform discretizations: a new approach for the quantization of totally constrained systems",
Phys. Rev. \textbf{D74} (2006) 124012, \texttt{arXiv:gr-qc/0610023}.

\bibitem{cylconsis} B. Dittrich,
``From the discrete to the continuous: Towards a cylindrically consistent dynamics'',
New J. Phys. \textbf{14} (2012) 123004, \texttt{arXiv:1205.6127 [gr-qc]}.

\bibitem{phoehn2} P. A. Hoehn,
``Quantization of systems with temporally varying discretization II: Local evolution moves'',
J. Math. Phys. \textbf{55}, (2014) 103507, \texttt{arXiv:1401.7731 [gr-qc]}.

\bibitem{holonmysf} B. Bahr, B. Dittrich, F. Hellmann and W. Kaminski,
``Holonomy spin foam models: Definition and coarse graining'',
Phys. Rev. D \textbf{87} 044048 (2013), \texttt{arXiv:1208.3388 [gr-qc]}.

\bibitem{bahr14} B. Bahr,
``On background-independent renormalization of spin foam models'',
(2014), \texttt{arXiv:1407.7746 [gr-qc]}.

\bibitem{bd14} B. Dittrich,
``The continuum limit of loop quantum gravity - a framework for solving the theory'',
in A. Ashtekar and J. Pullin, ed., to be published in the World Scientific series ``100 Years of General Relativity'',
(2014), \texttt{arXiv:1409.1450 [gr-qc]}.

\bibitem{improved} B. Bahr and B. Dittrich, 
``Improved and Perfect Actions in Discrete Gravity'',
Phys. Rev. D \textbf{80} (2009) 124030, \texttt{arXiv:0907.4323 [gr-qc]}.

\bibitem{bahrdittrichsteinhaus} B. Bahr, B. Dittrich and S. Steinhaus,
``Perfect discretization of reparametrization invariant path integrals'',
Phys. Rev. D \textbf{83} (2011) 105026, \texttt{arXiv:1101.4775 [gr-qc]}.

\bibitem{bahrdittrich09b} B. Bahr and B. Dittrich,
``Breaking and restoring of diffeomorphism symmetry in discrete gravity'',
Proceedings of the XXV Max Born Symposium "The Planck Scale", Wroclaw, 29 June - 3 July, 2009, \texttt{arXiv:0909.5688 [gr-qc]}.

\bibitem{merce} B. Dittrich, M. Martin-Benito and E. Schnetter,
``Coarse graining of spin net models: dynamics of intertwiners'',
New J. Phys. \textbf{15} (2013) 103004, \texttt{arXiv:1306.2987 [gr-qc]}.

\bibitem{intertwiner} B. Dittrich and W. Kaminski,
``Topological lattice field theories from intertwiner dynamics'',
(2013), \texttt{arXiv:1311.1798 [gr-qc]}.

\bibitem{qgroupspinnet} B. Dittrich, M. Martin-Benito and S. Steinhaus,
``Quantum group spin nets: refinement limit and relation to spin foams'',
Phys. Rev. D \textbf{90} (2014) 024058, \texttt{arXiv:1312.0905 [gr-qc]}.

\bibitem{decoratedtnw} B. Dittrich, S. Mizera and S. Steinhaus,
``Decorated tensor network renormalization for lattice gauge theories and spin foam models'',
(2014), \texttt{arXiv:1409.2407 [gr-qc]}.  

\bibitem{eckert} B. Dittrich, F. C. Eckert and M. Martin-Benito,
``Coarse graining methods for spin net and spin foam models'',
New J. Phys. \textbf{14} 035008 (2012), \texttt{arXiv:1109.4927 [gr-qc]}.

\bibitem{clement} B. Dittrich and C. Delcamp,
``Continuum limit of 3D spin foam models",
to appear.

\bibitem{lee} S. Major and L. Smolin,
``Quantum deformation of quantum gravity'',
Nucl. Phys. B \textbf{473} 267-290, (1996), \texttt{arXiv:gr-qc/9512020}.
  
\bibitem{girelli1} V. Bonzom, M. Dupuis, F. Girelli and E. R. Livine,
``Deformed phase space for 3d loop gravity and hyperbolic discrete geometries'',
(2014), \texttt{arXiv:1402.2323 [gr-qc]}.
  
\bibitem{girelli2} V. Bonzom, M. Dupuis and F. Girelli,
``Towards the Turaev--Viro amplitudes from a Hamiltonian constraint'',
Phys. Rev. D \textbf{90} 104038 (2014), \texttt{arXiv:1403.7121 [gr-qc]}.

\bibitem{pranzetti} D. Pranzetti,
``Turaev--Viro amplitudes from 2+1 loop quantum gravity'',
Phys. Rev. D \textbf{89} 084058 (2014), \texttt{arXiv:1402.2384 [gr-qc]}.

\bibitem{rovvid} C. Rovelli and F. Vidotto,
``Compact phase space, cosmological constant, discrete time'',
Phys. Rev. D \textbf{91} 084037 (2015), \texttt{arXiv:1502.00278 [gr-qc]}.

\bibitem{newregge} B. Bahr and B. Dittrich,
``Regge calculus from a new angle'',
New J. Phys. \textbf{12} 033010 (2010), \texttt{arXiv:0907.4325 [gr-qc]}.
  
\bibitem{barrettmeus} J. W. Barrett, C. Meusburger, G. Schaumann,
``Gray categories with duals and their diagrams",
(2012), \texttt{arXiv:1211.0529 [math.QA]}.

\end{thebibliography}
\end{document}